\documentclass[pdflatex,sn-aps]{sn-jnl}

\jyear{2023}%

\theoremstyle{thmstyleone}%
\theoremstyle{thmstyletwo}%
\theoremstyle{thmstylethree}%

\newcommand{\dakshainst}{Paper~I}

\newcommand{\vbdone}{}

\newcommand{\daksha}{{\em Daksha}\xspace}
\newcommand{\asat}{{\em AstroSat}\xspace}
\newcommand{\swift}{{\em Swift}\xspace}
\newcommand{\fermi}{{\em Fermi}\xspace}
\newcommand{\integral}{{\em Integral}\xspace}

\newcommand{\fluence}{\ensuremath{\mathrm{erg~cm}^{-2}}}
\newcommand{\flux}{\ensuremath{\mathrm{erg~cm}^{-2}~\mathrm{s}^{-1}}}

\newcommand{\eu}[2]{\ensuremath{\times 10^{#1}~\mathrm{#2}}}

\newcommand{\ecs}{\ensuremath{\mathrm{erg~cm}^{-2}~\mathrm{s}^{-1}}}
\newcommand{\pcs}{\ensuremath{\mathrm{ph~cm}^{-2}~\mathrm{s}^{-1}}}

\newcommand{\update}[1]{ #1}

\newcommand{\degr}{\ensuremath{^\circ}}

\makeatletter
\let\jnl@style=\rm
\def\ref@jnl#1{{\jnl@style#1}}

\def\aj{\ref@jnl{AJ}}                   
\def\actaa{\ref@jnl{Acta Astron.}}      
\def\araa{\ref@jnl{ARA\&A}}             
\def\apj{\ref@jnl{ApJ}}                 
\def\apjl{\ref@jnl{ApJ}}                
\def\apjs{\ref@jnl{ApJS}}               
\def\ao{\ref@jnl{Appl.~Opt.}}           
\def\apss{\ref@jnl{Ap\&SS}}             
\def\aap{\ref@jnl{A\&A}}                
\def\aapr{\ref@jnl{A\&A~Rev.}}          
\def\aaps{\ref@jnl{A\&AS}}              
\def\azh{\ref@jnl{AZh}}                 
\def\baas{\ref@jnl{BAAS}}               
\def\bac{\ref@jnl{Bull. astr. Inst. Czechosl.}}
\def\caa{\ref@jnl{Chinese Astron. Astrophys.}}
\def\cjaa{\ref@jnl{Chinese J. Astron. Astrophys.}}
\def\icarus{\ref@jnl{Icarus}}           
\def\jcap{\ref@jnl{J. Cosmology Astropart. Phys.}}
\def\jrasc{\ref@jnl{JRASC}}             
\def\memras{\ref@jnl{MmRAS}}            
\def\mnras{\ref@jnl{MNRAS}}             
\def\na{\ref@jnl{New A}}                
\def\nar{\ref@jnl{New A Rev.}}          
\def\pra{\ref@jnl{Phys.~Rev.~A}}        
\def\prb{\ref@jnl{Phys.~Rev.~B}}        
\def\prc{\ref@jnl{Phys.~Rev.~C}}        
\def\prd{\ref@jnl{Phys.~Rev.~D}}        
\def\pre{\ref@jnl{Phys.~Rev.~E}}        
\def\prl{\ref@jnl{Phys.~Rev.~Lett.}}    
\def\pasa{\ref@jnl{PASA}}               
\def\pasp{\ref@jnl{PASP}}               
\def\pasj{\ref@jnl{PASJ}}               
\def\rmxaa{\ref@jnl{Rev. Mexicana Astron. Astrofis.}}%
\def\qjras{\ref@jnl{QJRAS}}             
\def\skytel{\ref@jnl{S\&T}}             
\def\solphys{\ref@jnl{Sol.~Phys.}}      
\def\sovast{\ref@jnl{Soviet~Ast.}}      
\def\ssr{\ref@jnl{Space~Sci.~Rev.}}     
\def\zap{\ref@jnl{ZAp}}                 
\def\nat{\ref@jnl{Nature}}              
\def\iaucirc{\ref@jnl{IAU~Circ.}}       
\def\aplett{\ref@jnl{Astrophys.~Lett.}} 
\def\apspr{\ref@jnl{Astrophys.~Space~Phys.~Res.}}
\def\bain{\ref@jnl{Bull.~Astron.~Inst.~Netherlands}} 
\def\fcp{\ref@jnl{Fund.~Cosmic~Phys.}}  
\def\gca{\ref@jnl{Geochim.~Cosmochim.~Acta}}   
\def\grl{\ref@jnl{Geophys.~Res.~Lett.}} 
\def\jcp{\ref@jnl{J.~Chem.~Phys.}}      
\def\jgr{\ref@jnl{J.~Geophys.~Res.}}    
\def\jqsrt{\ref@jnl{J.~Quant.~Spec.~Radiat.~Transf.}}
\def\memsai{\ref@jnl{Mem.~Soc.~Astron.~Italiana}}
\def\nphysa{\ref@jnl{Nucl.~Phys.~A}}   
\def\physrep{\ref@jnl{Phys.~Rep.}}   
\def\physscr{\ref@jnl{Phys.~Scr}}   
\def\planss{\ref@jnl{Planet.~Space~Sci.}}   
\def\procspie{\ref@jnl{Proc.~SPIE}}   

\makeatother

\usepackage{multirow}
\usepackage{multicol}
\usepackage{xspace}
\usepackage[utf8]{inputenc}
\newcommand{\orcid}[1]{\href{https://orcid.org/#1}{\includegraphics[scale=.05]{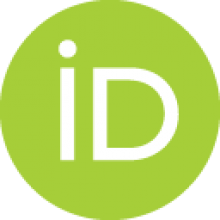}}}
\usepackage{devanagari}

\begin{document}

\title[Daksha Science]{Science with the Daksha High Energy Transients Mission}


\author*[1]{\fnm{Varun} \sur{Bhalerao}\orcid{0000-0002-6112-7609} } \email{varunb@iitb.ac.in}
\author[1]{\fnm{Disha} \sur{Sawant} \orcid{0000-0002-9702-6324}}
\author*[1]{\fnm{Archana} \sur{Pai}\orcid{0000-0003-3476-4589}}
\author[2]{\fnm{Shriharsh} \sur{Tendulkar}\orcid{0000-0003-2548-2926}}
\author[3]{\fnm{Santosh} \sur{Vadawale}\orcid{0000-0002-2050-0913}}
\author[4]{\fnm{Dipankar} \sur{Bhattacharya}\orcid{0000-0003-3352-3142  }}
\author[5]{\fnm{Vikram} \sur{Rana}\orcid{0000-0003-1703-8796}}

\author[3]{\fnm{Hitesh Kumar L.} \sur{Adalja} \orcid{0000-0002-5272-6386}}
\author[6]{\fnm{G} \sur{C Anupama} \orcid{0000-0003-3533-7183}}
\author[1]{\fnm{Suman} \sur{Bala} \orcid{0000-0002-6657-9022}}
\author[7,8]{\fnm{Smaranika} \sur{Banerjee} \orcid{0000-0001-6595-2238}}
\author[6]{\fnm{Judhajeet} \sur{Basu} \orcid{0000-0001-7570-545X}}
\author[9]{\fnm{Hrishikesh} \sur{Belatikar} \orcid{0000-0001-9954-7329}}
\author[10]{\fnm{Paz} \sur{Beniamini} \orcid{0000-0001-7833-1043}}
\author[9]{\fnm{Mahesh} \sur{Bhaganagare}}
\author[11]{\fnm{Ankush} \sur{Bhaskar} \orcid{0000-0003-4281-1744}}
\author[12]{\fnm{Soumyadeep} \sur{Bhattacharjee} \orcid{0000-0003-2071-2956}}
\author[13]{\fnm{Sukanta} \sur{Bose} \orcid{0000-0002-4151-1347}}
\author[14]{\fnm{Brad} \sur{Cenko} \orcid{0000-0003-1673-970X}}
\author[1]{\fnm{Mehul} \sur{Vijay Chanda} \orcid{0000-0001-6343-1674}}
\author[13]{\fnm{Gulab} \sur{Dewangan} \orcid{0000-0003-1589-2075}}
\author[15]{\fnm{Vishal} \sur{Dixit} \orcid{0000-0002-1468-5253}}
\author[5]{\fnm{Anirban} \sur{Dutta} \orcid{0000-0001-9526-7872}} 
\author[13]{\fnm{Priyanka} \sur{Gawade} \orcid{0000-0003-1775-4530}}
\author[1]{\fnm{Abhijeet} \sur{Ghodgaonkar} \orcid{0000-0001-6211-2209}}
\author[3]{\fnm{Shiv} \sur{Kumar Goyal} \orcid{0000-0002-3153-537X}}
\author[16]{\fnm{Suresh} \sur{Gunasekaran}}
\author[5]{\fnm{Manikantan} \sur{Hemanth} \orcid{0000-0001-9404-1601}}
\author[17]{\fnm{Kenta} \sur{Hotokezaka} \orcid{0000-0001-5023-6933}}
\author[18]{\fnm{Shabnam} \sur{Iyyani} \orcid{0000-0002-2525-3464}}
\author[16]{\fnm{P. J.} \sur{Guruprasad} \orcid{0000-0003-0162-0132}}
\author[19]{\fnm{Mansi} \sur{Kasliwal} \orcid{0000-0002-5619-4938}}
\author[2]{\fnm{Jayprakash} \sur{G. Koyande} \orcid{0000-0001-7829-3366}}
\author[20]{\fnm{Salil} \sur{Kulkarni}}
\author[2]{\fnm{APK} \sur{Kutty}}
\author[3]{\fnm{Tinkal} \sur{Ladiya} \orcid{0000-0001-6022-8283}}
\author[1]{\fnm{Suddhasatta} \sur{Mahapatra}}
\author[20]{\fnm{Deepak} \sur{Marla} \orcid{0000-0001-6829-7830}}
\author[2]{\fnm{Sujay} \sur{Mate} \orcid{0000-0001-5536-4635}}
\author[1]{\fnm{Advait} \sur{Mehla} \orcid{0000-0002-3155-6584}}
\author[3]{\fnm{N. P. S.} \sur{Mithun} \orcid{0000-0003-3431-6110}}
\author[13]{\fnm{Surhud} \sur{More} \orcid{0000-0002-2986-2371}}
\author[20]{\fnm{Rakesh} \sur{Mote} \orcid{0000-0001-7853-5150}}
\author[13]{\fnm{Dipanjan} \sur{Mukherjee} \orcid{0000-0003-0632-1000}}
\author[9]{\fnm{Sanjoli} \sur{Narang} \orcid{0000-0002-8723-2263}}
\author[21]{\fnm{Shyama} \sur{Narendranath}}
\author[2]{\fnm{Ayush} \sur{Nema} \orcid{0000-0003-0701-9639}}
\author[9]{\fnm{Sudhanshu} \sur{Nimbalkar} \orcid{0000-0002-2131-6720}}
\author[22,23]{\fnm{Samaya} \sur{Nissanke} \orcid{0000-0001-6573-7773}}
\author[1]{\fnm{Sourav} \sur{Palit} \orcid{0000-0003-2932-3666}}
\author[24,16]{\fnm{Jinaykumar} \sur{Patel} \orcid{0000-0002-6551-0963}}
\author[3]{\fnm{Arpit} \sur{Patel} \orcid{0000-0002-0929-1401}}
\author[5]{\fnm{Biswajit} \sur{Paul}}
\author[25]{\fnm{Priya} \sur{Pradeep}}
\author[16]{\fnm{Prabhu} \sur{Ramachandran} \orcid{0000-0001-6337-1720}}
\author[5]{\fnm{Kinjal} \sur{Roy} \orcid{0000-0002-7391-5776}}
\author[3]{\fnm{B.S. Bharath} \sur{Saiguhan} \orcid{0000-0001-7580-364X}}
\author[18]{\fnm{Joseph} \sur{Saji} \orcid{0000-0002-8066-8478}}
\author[26,27]{\fnm{M.} \sur{Saleem} \orcid{0000-0002-3836-7751}}
\author[1]{\fnm{Divita} \sur{Saraogi} \orcid{0000-0001-6332-1723}}
\author[1]{\fnm{Parth} \sur{Sastry} \orcid{0000-0001-9645-5453}}
\author[3]{\fnm{M.} \sur{Shanmugam} \orcid{0000-0002-5995-8681}}
\author[3]{\fnm{Piyush} \sur{Sharma} \orcid{0000-0001-9670-1511}}
\author[9]{\fnm{Amit} \sur{Shetye}}
\author[3]{\fnm{Nishant} \sur{Singh} \orcid{0000-0002-1975-0552}}
\author[28]{\fnm{Shreeya} \sur{Singh}}
\author[1]{\fnm{Akshat} \sur{Singhal} \orcid{0000-0003-1275-1904}}
\author[25]{\fnm{S.} \sur{Sreekumar}}
\author[9,29]{\fnm{Srividhya} \sur{Sridhar} \orcid{0000-0002-7972-169X}}
\author[1,30,31]{\fnm{Rahul} \sur{Srinivasan} \orcid{0000-0002-7176-6690}}
\author[9]{\fnm{Siddharth} \sur{Tallur} \orcid{0000-0003-1399-2187}}
\author[3]{\fnm{Neeraj} \sur{K. Tiwari} \orcid{0000-0003-4269-340X.}}
\author[16]{\fnm{Amrutha} \sur{Lakshmi Vadladi} \orcid{0000-0003-0814-9064}}
\author[3]{\fnm{C.} \sur{S. Vaishnava} \orcid{0000-0002-8096-5683}}
\author[2]{\fnm{Sandeep} \sur{Vishwakarma} \orcid{0000-0001-8159-5656}}
\author[1]{\fnm{Gaurav} \sur{Waratkar} \orcid{0000-0003-3630-9440}}

\affil*[1]{\orgdiv{Department of Physics}, \orgname{IIT Bombay}, \orgaddress{\street{Powai}, \city{Mumbai}, \postcode{400076}, \country{India}}}
\affil[2]{\orgname{Department of Astronomy and Astrophysics}, \orgaddress{Tata Institute of Fundamental Research}, \city{Mumbai}, \postcode{400005}, \country{India}}
\affil[3]{\orgname{Physical Research Laboratory}, \orgaddress{Navrangpura}, \city{Ahmedabad}, \postcode{380009}, \country{India}}
\affil[4]{\orgdiv{Ashoka University}, \orgname{Department of Physics}, \orgaddress{Sonepat}, \city{Haryana}, \postcode{131029}, \country{India}}
\affil[5]{\orgdiv{Raman Research Institute}, \orgname{C. V. Raman Avenue}, \orgaddress{Sadashivanagar}, \city{Bengaluru}, \postcode{560080}, \country{India}}

\affil[6]{\orgname{Indian Institute of Astrophysics}, \orgaddress{2nd Block 100 Feet Rd, Koramangala}, \city{Bengaluru}, \postcode{560034}, \country{India}}
\affil[7]{\orgname{Tohoku University}, \orgaddress{Aoba}, \city{Sendai}, \postcode{980-0845}, \country{Japan}}
\affil[8]{\orgname{Stockholm University}, \orgaddress{AlbaNova}, \city{Stockholm}, \postcode{SE 10691}, \country{Sweden}}
\affil[9]{\orgdiv{Department of Electrical Engineering}, \orgname{IIT Bombay}, \orgaddress{Powai}, \city{Mumbai}, \postcode{400076}, \country{India}}
\affil[10]{\orgdiv{Department of Natural Sciences}, \orgname{The Open University of Israel}, \orgaddress{P.O Box 808}, \city{Ra'anana}, \postcode{4353701}, \country{Israel}}
\affil[11]{\orgdiv{Space Physics Laboratory}, \orgname{ISRO/Vikram Sarabhai Space Centre}, \city{Thiruvananthapuram}, \postcode{695022}, \country{India}}
\affil[12]{\orgdiv{Department of Physics}, \orgname{Indian Institute of Science}, \city{Bengaluru}, \state{Karnataka}, \postcode{560012}, \country{India}}
\affil[13]{\orgname{Inter-University Center for Astronomy and Astrophysics}, \city{Pune}, \state{Maharashtra}, \postcode{411007}, \country{India}}
\affil[14]{\orgname{University of Maryland}, \orgaddress{College Park}, \country{USA}}
\affil[15]{\orgdiv{Interdisciplinary Program in Climate Studies}, \orgname{IIT Bombay}, \orgaddress{Powai}, \city{Mumbai}, \postcode{400076}, \country{India}}
\affil[16]{\orgdiv{Department of Aerospace Engineering}, \orgname{IIT Bombay}, \orgaddress{Powai}, \city{Mumbai}, \postcode{400076}, \country{India}}
\affil[17]{\orgname{University of Tokyo}, \orgaddress{Bunkyo City}, \city{Tokyo}, \postcode{113-8654}}
\affil[18]{\orgdiv{School of Physics}, \orgname{IISER Thiruvananthapuram}, \postcode{695551}, \state{Kerala}}
\affil[19]{\orgdiv{Division of Physics, Mathematics, and Astronomy}, \orgname{California Institute of Technology}, \city{Pasadena}, \state{CA} \postcode{91125}, \country{USA}}
\affil[20]{\orgdiv{Department of Mechanical Engineering}, \orgname{IIT Bombay}, \orgaddress{Powai}, \city{Mumbai}, \postcode{400076}, \country{India}}
\affil[21]{\orgdiv{Space Astronomy Group}, \orgname{U R Rao Satellite Centre, ISRO}, \city{Bengaluru} \postcode{560037}, \country{India}}
\affil[22]{\orgdiv{GRAPPA, Anton Pannekoek Institute for Astronomy and Institute of High-Energy Physics}, \orgname{University of Amsterdam}, \orgaddress{Science Park 904 1098 XH}, \city{Amsterdam}, \country{The Netherlands}}
\affil[23]{\orgname{Nikhef}, \orgaddress{Science Park 105, 1098 XG}, \city{Amsterdam}, \country{The Netherlands}}
\affil[24]{\orgdiv{Department of Mechanical and Aerospace Engineering}, \orgname{The University of Texas at Arlington}, \orgaddress{Arlington}, \city{TX}, \postcode{76019}, \country{USA}}
\affil[25]{\orgname{Vikram Sarabhai Space Centre}, \orgaddress{Kochuveli}, \city{Thiruvananthapuram}, \postcode{695022}, \country{India}}
\affil[26]{\orgname{Chennai Mathematical Institute}, \city{Siruseri}, \postcode{603103}, \state{Tamilnadu}, \country{India}}
\affil[27]{\orgdiv{School of Physics and Astronomy}, \orgname{University of Minnesota}, \city{Minneapolis}, \state{MN}, \postcode{55455}, \country{USA}}
\affil[28]{\orgdiv{Department of Chemical Engineering}, \orgname{IIT Bombay}, \orgaddress{Powai}, \city{Mumbai}, \postcode{400076}, \country{India}}
\affil[29]{\orgdiv{Department of Aerospace}, \orgname{University of Illinois Urbana-Champaign}, \postcode{61801}, \country{US}}
\affil[30]{\orgdiv{Universit\'{e} C\^{o}te d'Azur, Observatoire de la C\^{o}te d'Azur}, \orgname{CNRS, Laboratoire Lagrange, Bd de l'Observatoire}, \orgaddress{CS, 34229, 06304 Nice cedex 4}, \country{France}}
\affil[31]{\orgdiv{Artemis, Université Côte d’Azur, Observatoire de la Côte d’Azur}, \orgname{CNRS}, \postcode{F-06304}, \city{Nice}, \country{France}}

\abstract{We present the science case for the proposed \daksha\ high energy transients mission. \daksha\ will comprise of two satellites covering the entire sky from 1~keV to $>1$~MeV. The primary objectives of the mission are to discover and characterize electromagnetic counterparts to gravitational wave source; and to study Gamma Ray Bursts (GRBs). \daksha\ is a versatile all-sky monitor that can address a wide variety of science cases. With its broadband spectral response, high sensitivity, and continuous all-sky coverage, it will discover fainter and rarer sources than any other existing or proposed mission. \daksha\ can make key strides in GRB research with polarization studies, prompt soft spectroscopy, and fine time-resolved spectral studies. 
\daksha\ will provide continuous monitoring of X-ray pulsars. It will detect magnetar outbursts and high energy counterparts to Fast Radio Bursts. Using Earth occultation to measure source fluxes, the two satellites together will obtain daily flux measurements of bright hard X-ray sources including active galactic nuclei, X-ray binaries, and slow transients like Novae. Correlation studies between the two satellites can be used to probe primordial black holes through lensing. \daksha\ will have a set of detectors continuously pointing towards the Sun, providing excellent hard X-ray monitoring data. Closer to home, the high sensitivity and time resolution of \daksha\ can be leveraged for the characterization of Terrestrial Gamma-ray Flashes.}




\maketitle

\section{Introduction}\label{sec1}
In the past decade, transient astronomy has received a great boost due to the expansion from electromagnetic (EM) regime to multi-messenger astronomy by including gravitational waves (GW), neutrinos, and high-energy cosmic rays. The joint detection of a faint gamma-ray burst (GRB) coincident with the gravitational wave detection of a binary neutron star (BNS) merger GW170817 \citep{LIGOScientificCollaboration2017,gw170817} was the first direct proof of the long hypothesized link between short GRBs and BNS mergers. This has also provided us with a detailed understanding of the production of r-process elements in kilonovae \citep{whs+19,Growth170817} and independent measurements of the neutron star equation of state \citep{De2018}.

The GRB counterpart of GW170817, however, was several orders of magnitude fainter than expected than ``classical'' GRBs and was barely detected by most current space-based detectors \citep{zhang2018peculiar}. With the improved sensitivity of advanced GW detectors, the horizon for detecting compact binary mergers involving neutron stars (possibly with a stellar mass black hole companion) will reach farther distances. We will need comparable improvements in X-ray and gamma-ray telescope sensitivities to be able to fully leverage the increased number of GW detections. Indeed, after GW170817 (at a distance of 40\,Mpc), the BNS and neutron star black hole (NSBH) events detected in LIGO--Virgo observing run 3 (O3) where typically at distances $\gtrsim$ 100\,Mpc. Despite extensive searches, no EM counterparts were found ~\citep[see for instance][]{Hosseinzadeh2019,Coughlin2019}. Roughly scaling the flux from GW170817 to these distances, the non-detections are not unexpected with the current sensitivity of all-sky missions.

X-ray and gamma-ray transient astronomy has been a rich field with missions such as \emph{BATSE}~\citep{sdl+84,bmf+92}, \emph{BeppoSAX}~\citep{1997A&AS..122..299B}, \emph{HETE-2}~\citep{Ricker2003}, \swift~\citep{gcg+04,Barthelmy2005} and \fermi~\citep{Meegan2009} leading the exploration of GRBs, magnetar flares, and X-ray binary outbursts. More recently, instruments such as \asat\ CZTI~\citep{czti}, POLAR~\citep{2005NIMPA.550..616P}, and IKAROS-GAP~\citep{2011ApJ...743L..30Y} have characterized the polarization of GRBs. Yet, there remain open questions about the detailed emission mechanism, the jet launching physics, jet composition, the effects of magnetic fields, and the nature of the remnant and the afterglow \citep{Metzger2008,Kumar_2015}.

The next generation of transient detection telescopes need to have a much higher sensitivity and all-sky coverage to match the nearly isotropic visibility and detection horizons of GW detector networks --- aLIGO \citep{aLIGO}, adVirgo \citep{AdVirgo}, KAGRA \citep{kagra}, and LIGO-India \citep{indigo,2022CQGra..39b5004S}. These should also provide low-latency alerts in order to coordinate with new synoptic radio, optical and infrared telescopes that will coordinate the EM followup response such as the ZTF~\citep{bellm14}, Rubin Observatory~\citep{rubin}, Square Kilometer Array~\citep{ska}, uGMRT~\citep{ugmrt}, LOFAR~\citep{lofar}, ASKAP~\citep{askap_science}, and IceCube~\citep{icecube}. A greater sensitivity, especially at lower energies will improve detection rates of high-redshift GRBs, improving their use as probes of cosmology, star formation rates etc.

With these considerations, we have proposed \daksha\footnote{{\update \daksha ({\dn d\322w}) means ``alert'' in Marathi, Sanskrit, and related languages. It can also be interpreted as able or skillful.}}, a broadband high-energy all-sky mission dedicated to X-ray and gamma-ray transient astronomy. In this paper, we discuss the science cases enabled by \daksha. The technical details of \daksha are given in a companion paper (Bhalerao et al, hereafter \dakshainst), but we have included a brief summary in Section~\ref{sec:overview} for completeness. In the next sections, we discuss the primary and secondary science enabled by \daksha\ along with the detection rate estimates and simulations --- Section~\ref{sec:emgw} covers the impact of \daksha\ for coincident detections of prompt GRB-like counterparts of GW events expected from upcoming gravitational wave detector networks; Section~\ref{sec:grb} covers the detection rates, prompt soft spectra, and polarization of GRBs; Section~\ref{sec:other} covers the secondary science cases --- X-ray pulsars, magnetar flares, fast radio burst (FRB) counterparts, primordial blackholes (PBHs), Earth occultation studies of bright X-ray sources, solar physics, and terrestrial gamma-ray flares.

\section{Mission overview}\label{sec:overview}
\begin{figure}
    \centering
     \includegraphics[width=0.5\textwidth]{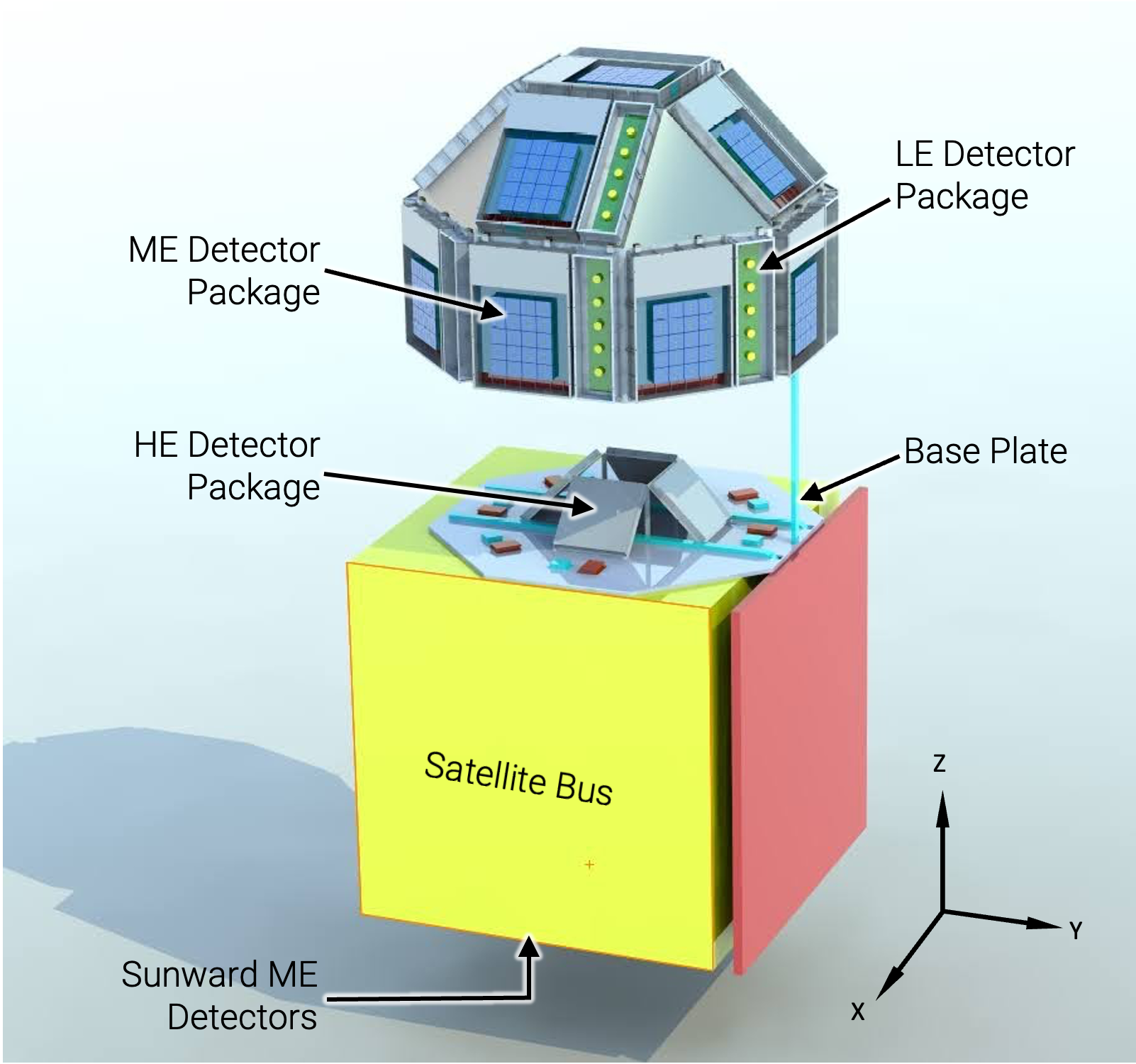}
    \caption{\update{Overall design of a \daksha\ satellite. The dome-shaped payload has 13 surfaces, each carrying one Low-energy detector package (LEP) and one Medium-energy detector packages (MEP). Four MEPs are mounted under the satellite bus, and always point directly to the sun. Four High-energy detector packages (HEP) are mounted inside the dome, along with processing electronics.}}
    \label{fig:config}
\end{figure}

\vbdone The mission comprises of two satellites launched in a near-equatorial low-earth orbit (LEO). The pair of satellites located opposite to each other in their orbit helps to gain all-sky coverage by mitigating the impact of the South Atlantic Anomaly and earth occultation in LEO.
\daksha satellites use three types of detectors to cover an energy range from 1~keV to $>1$~MeV (Figure~\ref{fig:config}). {\update The Low Energy Package (LEP) consists of five Silicon Drift Detectors (SDDs) and covers the 1--30~keV energy range. The Medium Energy Package (MEP) consists of 20 Cadmium Zinc Telluride (CZT) detectors and covers the 20--200~keV range. The High Energy Package (HEP) covers the 100~keV to $>1$~MeV range using NaI scintillator detectors readout by Silicon Photomultipliers (SiPM). \daksha is designed for a 5-year lifetime, but with no consumables on board it can be expected to operate for much longer. All the instruments serve as open all-sky monitors but do not have direct imaging capabilities.} Details of the instruments and their capabilities are given in \dakshainst.

\begin{figure}[!htbp]
    \centering
    \includegraphics[width=0.75\textwidth]{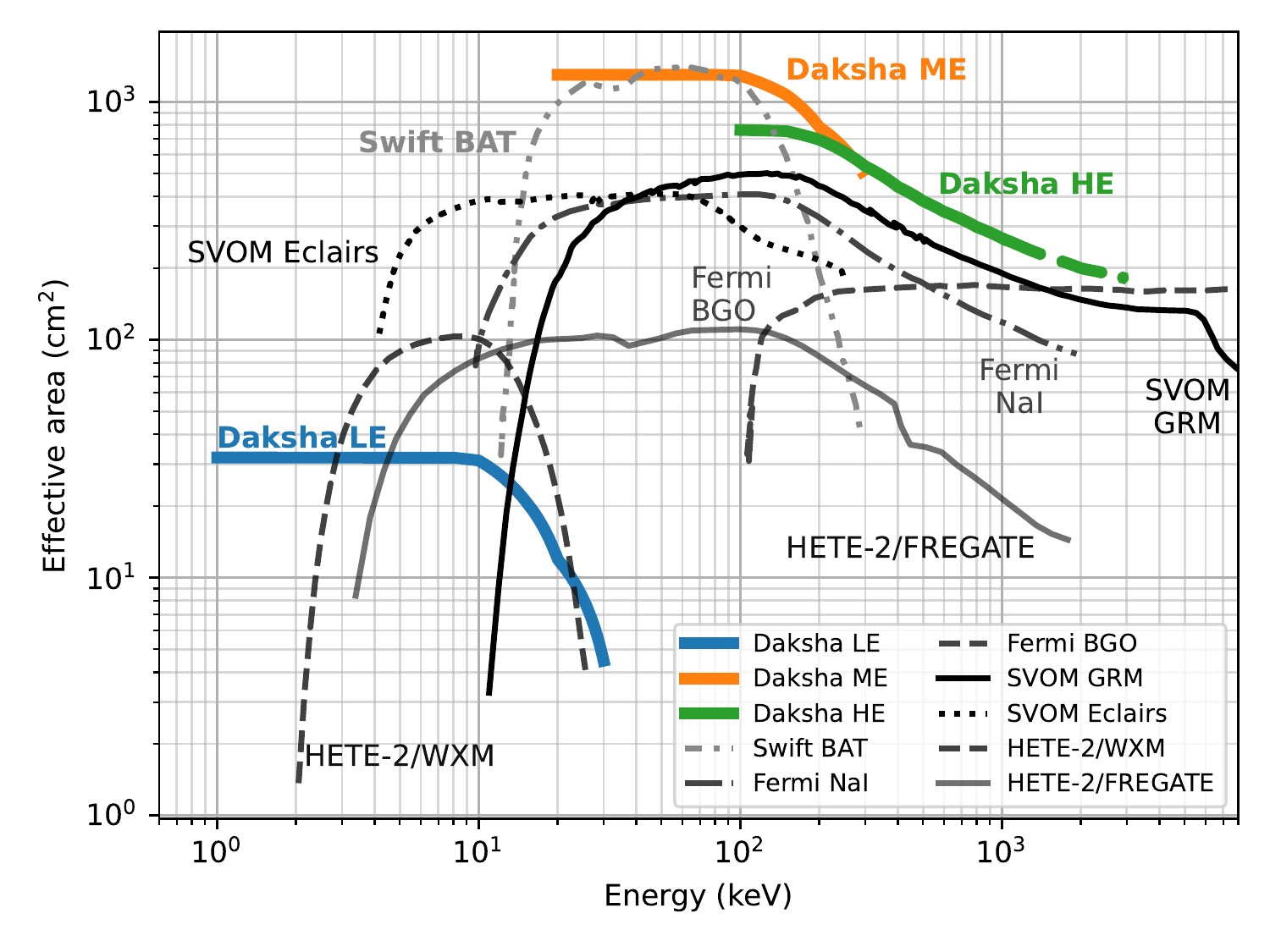}
    \caption{
    Effective area of \daksha\ detectors as a function of energy. Effective areas of \swift-BAT and both NaI and BGO detectors of \fermi-GBM are shown for comparison. These areas were adapted from \citep{2020JHEAp..27....1L} {\update and~\citep{Sakamoto2011}}. The NaI effective area is the averaged over the unocculted sky.}
    \label{fig:effectArea}
\end{figure}

The workhorse of \daksha are {\update MEPs}, which provide nearly uniform all-sky coverage with a median effective area of $\sim1300\mathrm{~cm}^2$ (Figure~\ref{fig:effectArea}), and a 5-$\sigma$ sensitivity of 4\eu{-8}{\flux} for 1-second bursts. Bursts with fluence of about 1\eu{-7}{\fluence} can be localized with an accuracy of 10\degr, with sub-degree localization for brighter bursts.

\section{Science Driver I: Electromagnetic Counterparts to Gravitational Wave Sources}\label{sec:emgw}

\vbdone On 2017 August 17, gravitational wave detectors had the first direct detection of coalescing neutron stars. 1.7~s later, a flash of gamma rays was detected by \fermi\ and \integral\ missions \citep{gw170817,LIGOScientificCollaboration2017}: {\update the first electromagnetic counterpart to a gravitational wave source (EMGW)} . The high energy detection spurred broadband follow-up observations {\update across various bands}, leading to the discovery of a kilonova \citep{Kilpatrick2017}, and eventually proving that the merger produced a successful structured jet \citep{Mooley2018,2018Natur.561..355M,teg+19}. Thanks to an extensive multi-wavelength data set, significant progress was made on many unsolved mysteries --- proving the connection between neutron star mergers and short GRBs ({\update sGRBs; }Section \ref{sec:grb}) \citep{Goldstein2017}, establishing that such mergers indeed are the sites of r-process nucleosynthesis \citep{Kasen2017,2017Sci...358.1559K}, providing an independent measurement of the Hubble constant \citep{2017Natur.551...85A}, and a measurement of the equation of state of ultra-dense matter \citep{2018PhRvL.121p1101A,2019PhRvX...9a1001A}.

GRB~170817 was peculiar in many ways. It was detected further off axis than typical GRBs \citep{2018Natur.554..207M,2018Natur.561..355M,2019MNRAS.482.5430B}. The prompt emission had soft spectral components which could not be adequately characterized or explained \citep{Goldstein2017}. While it is the closest GRB ever detected, it was still faint --- and intrinsically the most sub-luminous of all sGRBs detected to date \citep{zhang2018peculiar}. Had this event even been 30$\%$ fainter for instance, it would have been missed by several EM missions \citep{Saleem:2019wdv}. This is consistent with studies conducted in the third observing run of the advanced gravitational wave detectors (O3): while several neutron star merger events were detected in gravitational waves, no electromagnetic counterparts were found. A notable example is the lack of counterparts to BNS event GW190425 at a distance of $\sim150$~Mpc~\citep{S190425z,coughlin2020implications}. Besides sensitivity, GW170817 also underscored the importance of continuous all-sky coverage: sensitive missions like \swift\ \citep{Evans17} and \asat\ \citep{Growth170817} could not detect the source as it was occulted by the Earth at that instant.

The peculiarity GRB170817A of having low luminosity but comparable peak energy as any standard sGRBs \citep{Begue} posed additional questions: Does GRB170817A belong to a new, unexplored class of sGRBs? Do we need high soft X-ray sensitivity to detect and study this class of sources? 
How would they be different in terms of source distribution, energetics, spectral properties etc? Multi-wavelength and multi-messenger studies of more sources are critical for tackling these questions. Such observations can shed light on the sGRBs distribution as well as into the physics of sGRB jet structure, energetics \citep{kentaP,2019MNRAS.483..840B}.

The interferometric GW detector network is constantly getting upgraded. In future observing runs, the Advanced LIGO-Virgo detectors will achieve the sensitive distance reach of {$\sim$ 200 Mpc} \citep{Abbott2020}. Additional advanced detectors such as the Japanese detector KAGRA \citep{aso} and the LIGO-India \citep{2022CQGra..39b5004S} will make the network uniformly sensitive over the sky, with the localization capability up to ten square degrees \citep{Mills18,2020LRR....23....3A}. A growing number of BNS sources will be detected within a few hundred Mpc \citep{burns2019gamma,2022ApJ...924...54P}. To maximize the science returns from these detections, we need new higher sensitivity instruments with all--sky coverage \citep{gw170817,astro2020}. 

\subsection{Design considerations for EMGW}\label{sec:emgw_impact}
\vbdone \daksha has been designed keeping in mind the lessons learnt from the GW network observing runs and corresponding follow-up programs. Two key features of \daksha are the higher sensitivity and all-sky coverage, thanks to which we will detect far more events than other missions (Section~\ref{sec:emgw_rates}). On-board algorithms will detect bursts, localize them, and create coarse light curves and spectra to be broadcast globally within $\sim1$~minute of each event. This information will enable groups to prioritize their resources and start rapid follow-up observations.

All event-mode data will be downlinked on the next ground-station pass, to create improved data products. The broad spectral coverage will play a critical role in modeling spectral components of the prompt emission. A unique capability for \daksha\ is its low energy coverage for prompt soft emission, which we discuss in Section~\ref{sec:emgw_soft}. For bright events, \daksha will also be able to measure polarization of the events, which is discussed in greater detail in the context of GRBs in Section~\ref{sec:pol}. All bursts detected by \daksha can also be utilized for ``triggered'' searches in GW network data for corresponding GW signals. Our calculations show that this can give a significant boost to the number of binary neutron star events detected in GW (Section~\ref{sec:emgw_rates}; also see Bhattacharjee et al. in prep).

\subsection{Probing science of soft X-ray sources: the cocoon shock breakout model}\label{sec:emgw_soft}
\vbdone
{\update An} sGRB jet propagating inside the merger ejecta forms a cocoon, which eventually breaks out from the ejecta surface. The emission arising from the cocoon shock break-out is expected to be composed of an initial hard-spike followed by a soft-tail \citep{2018MNRAS.479..588G,2020ApJ...897..141B}. The luminosity, duration, and spectral peak of the hard spike and soft tail depend on the ejecta structure as well as the jet properties such as the power, opening angle and time delay between the merger and jet launch.

For instance, the emission properties of GRB170817A can be explained with the cocoon shock breakout model with a breakout radius of $\sim 2\times 10^{11}\,{\rm cm}$, a maximum ejecta velocity of $\sim 0.7c$ \citep{2018MNRAS.479..588G}.  Although these values are expected to be different for different events, the emission properties must roughly satisfy a closure relation between the duration, total energy, and the temperature \citep{2012ApJ...747...88N}. Thus, the cocoon breakout model can be tested by detecting more X-ray flashes associated with neutron star mergers. Furthermore, by modeling the emission properties, we will be able to obtain valuable information about the structure of merger ejecta and the jet formation in neutron star mergers. This in turn may allow us to constrain the equation of state (EOS) of neutron stars, since the ejecta profile in the fast tail is sensitive to the EOS \citep{2018ApJ...867...95H,2018ApJ...869..130R}.

\subsection{EMGW Rates}\label{sec:emgw_rates}
The rate of joint EMGW detections depends on the volumetric BNS merger rate $\mathcal{R}$, the source emission models, and sensitivities of the GW detector network and EM satellites. We estimated these rates by injecting BNS events uniformly randomly in co-moving volume, with random inclination, then calculating their detectability by the GW network as well as \daksha. A detailed set of calculations with a variety of emission models will be presented elsewhere (Bhattacharjee et al., {\update under review}), but we give a quick overview and present the results here.

We consider a configuration where five GW detectors are operating at their full sensitivity: ``A+'' sensitivity for LIGO Hanford, Livingston, and India, ``AdV'' sensitivity for Advanced Virgo, and the standard design sensitivity for Kagra. Following Petrov et al.\cite{2022ApJ...924...54P}, we assume that the detectors have a duty cycle of 70\%, and an event is considered to be detected in gravitational waves if the network signal-to-noise ratio (SNR) is $>8$. We take $\mathcal{R} = 320~\mathrm{Gpc^{-3}yr^{-1}}$ \cite{2021ApJ...913L...7A}. 

First, we assume a simplistic EM model where every BNS merger has the same intrinsic luminosity as GW170817, independent of viewing angle\footnote{This implies that there is no bright jet even for face-on observers.}.  We assume a ``Comptonized'' spectral model with a photon power-law index $\alpha = -0.62$, an exponential cut-off at $E_p = 185$~keV, duration $\Delta t = 0.576$~s \citep{Goldstein2017}. The luminosity is calculated to match the observed Fermi flux of 3.1\eu{-7}{~\ecs} in the 10--1000~keV band. Using the \daksha\ sensitivity discussed in \dakshainst, we calculate that the pair of \daksha\ satellites will detect about 0.5 neutron star merger events per year (after accounting for time lost due to SAA etc), almost all of which will have a high enough GW amplitude to be detected by GW networks.

Next, we consider a ``Gaussian jet'' model from Ioka \& Nakamura\cite{Ioka19}, where the jet energy $E_\gamma$ and the Lorentz factor $\Gamma$ are given by,
\begin{eqnarray}
E_\gamma(\theta) &=& \eta_\gamma E_0e^{-\theta^2/2\theta_0^2} \\
\Gamma(\theta) &= &\Gamma_{\rm max}/(1 + (\theta/\theta_{0})^{\lambda})
\end{eqnarray}
where $\theta_{0} = 0.059$ rad, $\Gamma_{\rm max} = 2000$ and $\eta_\gamma = 0.1$. The event duration was assumed to be 0.3~s for all the events, the typical sGRB duration \citep{Kouveliotou93}. Source energies ($E_0$) are drawn from the luminosity function given by Wanderman \& Piran\cite{Wanderman15}. For this model, we find that \daksha\ will obtain 12 joint detections with the GW network per year.
On the other hand, {\update if} very few events are found despite \daksha's high sensitivity, the upper limits from our data will imply that gamma-ray production efficiency drops beyond the core: thereby giving new insights into the underlying dissipation and emission mechanisms \citep{2019MNRAS.482.5430B}.

Another key observation in the Gaussian jet simulations is the presence of several events that are clearly detectable by \daksha, but just below the GW network SNR of 8. Since the direction and time of the bursts will be known from \daksha, we can undertake sub-threshold searches for coincident GW events \citep[see for instance][]{2021ApJ...915...86A,2022ApJ...928..186A}. Considering such sub-threshold events to be detectable if their GW network SNR is $>6.5$, we find that \daksha\ will enable the detection of gravitational waves from an additional seven sub-threshold neutron star mergers each year.

\section{Science Driver II: Gamma Ray Burst science}\label{sec:grb}

The observed GRBs fall into two broad categories. They are long GRBs (LGRBs) with duration $T_{90} > 2$ sec and short GRBs (sGRBs) with $T_{90} < 2 $ sec. Long GRBs are produced by the core collapse {\update supernovae} of giant stars  \citep{bloom02,hjorth2003very,hjorth2012grb}, while, the short GRBs are produced by the merger of binary compact objects such as binary neutron stars and neutron star - black hole. The observation of GW170817 confirmed that at least a class of short GRBs are produced by binary neutron star (BNS) mergers \citep{Main170817}.

A typical GRB emission consists of two main parts: the {\it prompt} phase which consists of the immediate $\gamma$-rays produced closed to origin of the burst, and the late time {\it afterglow} phase which is produced as the outflow interacts with the {\update interstellar medium (ISM)} surrounding the burst. 
{\update GRB afterglows have been studied in detail, and are reasonably well understood --- while many details of the prompt phase remain open questions.}

{\update Optical/Infra-red spectroscopy of GRB afterglows provide the redshift of the burst. Multi-wavelength observations of the afterglow convey the broad picture like constituents of the interstellar medium, source energetics, and timeline of the underlying processes \citep{Greiner09, Gendre05,Schady15}.}

During the main prompt GRB phase, the source invokes highly relativistic jets with bulk Lorentz factors of a few hundreds emitting highly energetic photons. The exact physical mechanism producing such powerful $\gamma$-rays still remains debated~\citep{Kumar_2015}. The composition of GRB jets, the radiative processes giving rise to the prompt $\gamma$-rays are some of the open ended questions~\citep{Kumar_2015}. Both in terms of spectral properties and physical mechanisms, prompt emissions are still comparatively poorly explored~\citep{Zhang_2011} as opposed to the afterglow phase due to the transient nature of the event and lack of observations in the soft X-ray band~\citep{Gehrels12,Oganesyan18}. \daksha  is expected to improve over both these aspects with its all-sky capability and ability to probe in soft X-ray band and hence improving the population of the GRBs. Owing to the high sensitivity, \daksha will help towards better understanding of the transition between the prompt and afterglow emission in GRBs. This is of high importance --- for instance, it will enable the determination of the Lorentz factor of the external shocked region, and whether the deceleration is effectively in the ``thick shell'' or ``thin shell'' regimes. Below, we highlight on the GRB science we can probe with the \daksha~mission especially in the prompt phase. \par

\subsection{GRB Rates}\label{sec:grb_rates}
The distribution of the rate of long gamma ray burst with respect to redshift, $z$, mainly depends on the star formation rate $R_{GRB}(z)$ and the luminosity function $\phi (L)$. In case of the distribution of the rate of short GRBs, an additional factor of time delay ($\Delta t$) relative to the star formation rate is also considered.  We adopt a functional form of broken power law for both $R_{GRB}(z)$ and $\phi (L)$ as mentioned in Wanderman \& Piran\cite{Wanderman2010}, whereas, a model of lognormal distribution is adopted for the time delay from Wanderman \& Piran\cite{Wanderman_2015}. The various parameters defining $R_{GRB}(z)$, $\phi (L)$ and $\Delta t$ are evaluated separately using the observed distribution of the redshift detected long and short GRBs by the \swift mission during the span of $17.5$ years\footnote{Note $\Delta t$ is estimated for the short GRB distribution only.}. During this period, \swift has detected a total of $1314$ long GRBs and $133$ short GRBs, out of which $350$ long GRBs and $26$ short GRBs are found to have redshift measurements.

\begin{figure}[htbp]
    \centering
    \includegraphics[width=0.75\textwidth,trim=0 0 25 0,clip=true]{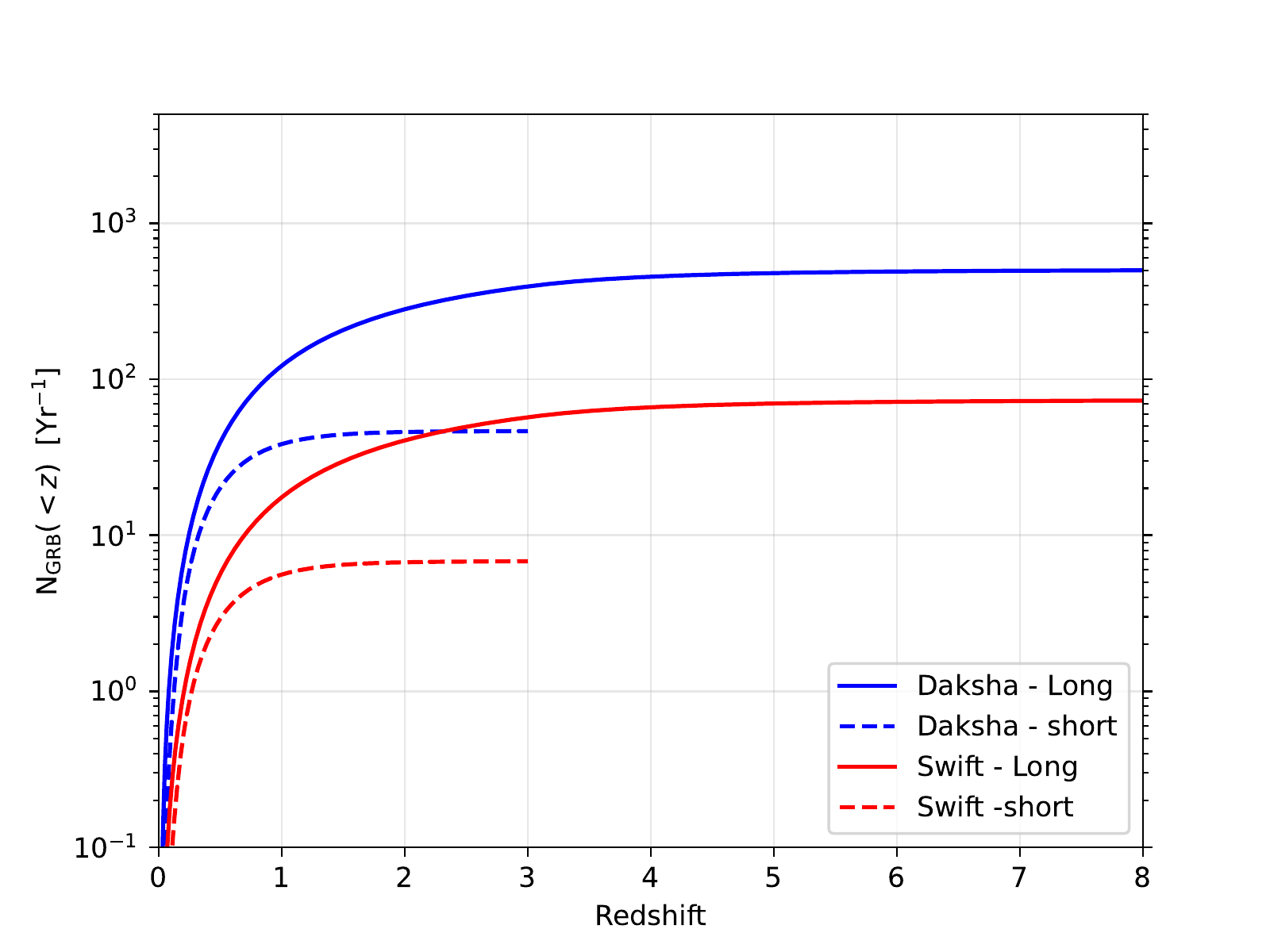}
    \caption{The cumulative plot showing the long and short GRB detection rate (per year) the by \daksha and \swift as a function of redshift.}
    \label{Daksha_swift}
\end{figure}

We use the {\it Swift} redshift distributions of long and short GRBs to predict GRB detection rate of \daksha. Using the obtained fit values and the \daksha  sensitivity of $S =4 \times 10^{-8}$~\ecs and spatiotemporal coverage of $\Omega \, T/(4 \pi) = 87\%$ (\dakshainst), we estimate that \daksha can detect nearly $500$ long GRBs and $46$ short GRBs per year (around $7$ and $2.5$ times greater than current \swift\ and \fermi\ rates respectively). Figure~\ref{Daksha_swift} shows the expected redshift distribution for long and short GRB that would be observed by \daksha.

\subsection{Broad-band spectroscopy of the prompt emission}
The physical origin of the prompt emission in GRBs is still a subject of intense debate. In absence of a clear understanding of the processes giving rise to the prompt emission, its spectrum is typically characterized by an empirical model consisting of two smoothly joining power-law \citep{bmf+92}, known as Band model or Band function. The typical GRB spectra in $\nu$F$_{\nu}$ form {\update (for instance in units of $\mathrm{keV~cm^{-2}~s^{-1}~keV^{-1}}$),} shows a peak around energy of $\sim$100 keV, characterized by the peak energy $E_{p}$, of the Band function. The spectra below and above $E_{p}$ are characterized by two spectral indices $\alpha$ and $\beta$ of the Band function. It has been observed that the peak energy is correlated to the isotropic equivalent energy of GRBs \citep{amati2002,Amati06}. Further, it has been found that the isotropic equivalent energy of the prompt emission is correlated with  the Lorentz factor of the outflow \citep{liang2010, ghirlanda2012}, which may define the other spectral parameters. Thus, accurate estimation of the GRB spectral parameters is important.

\subsection{Prompt emission spectral anomaly with soft X-ray studies}
A disagreement exists between the observed prompt spectral shape and the theoretical synchrotron predictions from a non-thermal population of ultra relativistic electrons in the soft X-ray band \citep{Derishev,Daigne}. The observed GRB prompt spectra consist of a photon index $\langle\alpha\rangle\sim-1$ which is on the harder side than the value $\alpha=-1.5$ expected from fast cooling synchrotron radiation. Unfortunately, the observational statistics for the existing missions in this low energy prompt phase are really poor.

Several explanations exist in literature to counter this discrepancy. One of the scenarios is either the cooling frequency $\nu_c$ or the self-absorption frequency $\nu_{sa}$ being comparable to the characteristic synchrotron frequency $\nu_m$ \citep{Daigne,2013ApJ...769...69B}. Another possibility points at the low-energy part of the synchrotron spectrum being modified by the energy-dependent inverse Compton scattering in the Klein-Nishina regime hence hardening spectral shape (up to $\alpha=-2/3$) \citep{Nakar2}. These models assume a constant magnetic field, while some other models explain the spectral hardening by invoking a non-uniform magnetic field that depends on the radius and/or the distance from the shock front \citep{Uhm}. Moreover, synchrotron spectra can have photon indices much harder than --1.5 if the pitch angles of the emitting electrons are distributed anisotropically \citep{Med}. It is then extremely important to explore whether these proposed scenarios represent a viable solution and are supported by observational evidence.

\begin{figure}[ht!]
    \centering
    \includegraphics[width=0.95\textwidth]{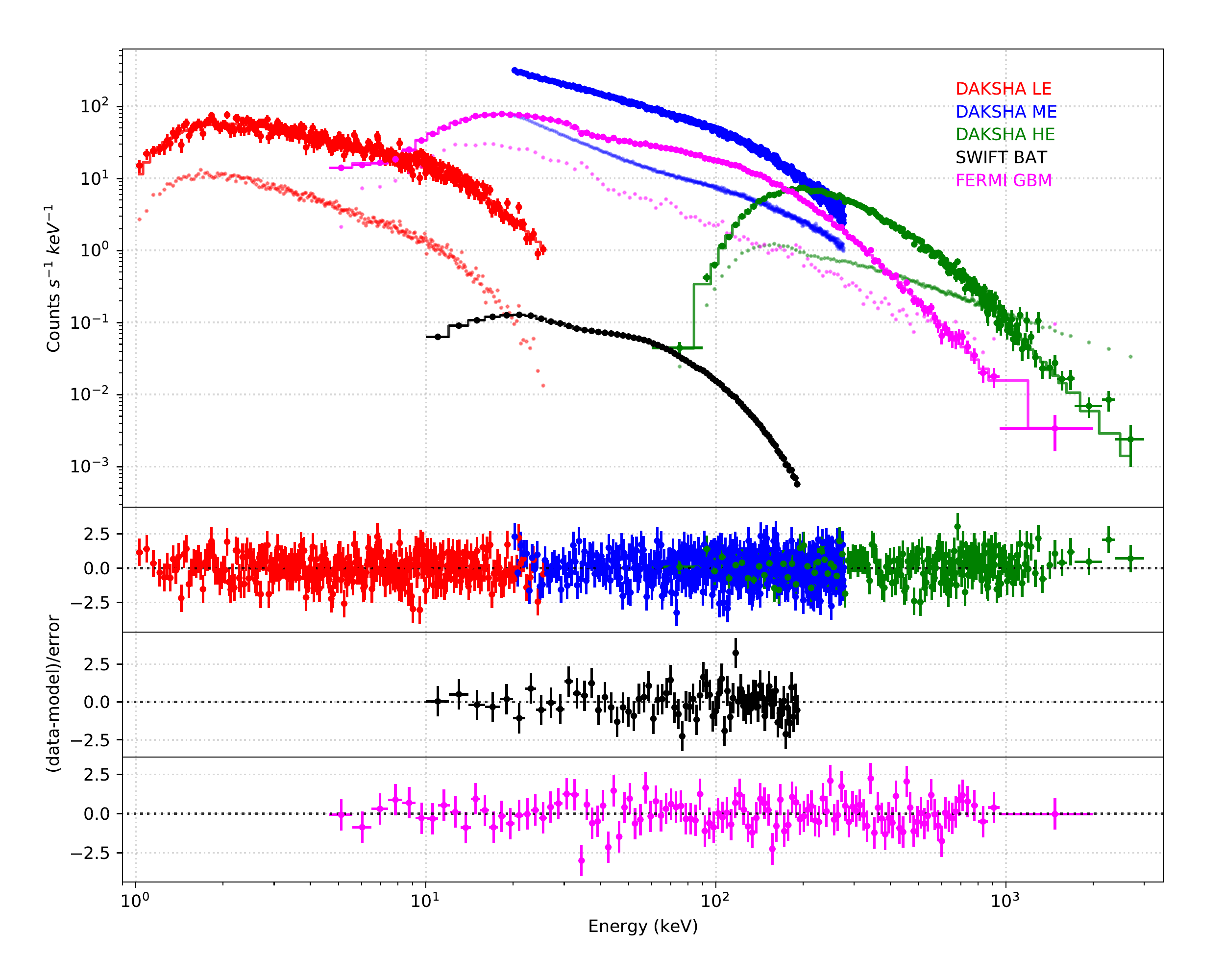}
    \caption{
    {\update Simulated spectra of a typical long GRB of $T_{90}$ 30 sec, assuming a  Band function spectrum \citep{band1993batse} with $\alpha = -1.0$, $\beta = -4.0$, $E_c = 300$ keV, and a 10 -- 1000 keV fluence of $\sim 1\times10^{-5}$ \fluence. The top panel shows the GRB spectra with \daksha, \swift-BAT and \fermi-GBM (one unit of NaI detector) along with the best fit models. 
    The background spectra for the corresponding instruments are also plotted as fainter points, except for \swift-BAT. 
    To simulate the GRB spectrum for \swift-BAT, we used \textit{batphasimerr} tool that takes into account the background statistics and hence  there is no separate background spectrum. 
    The rest of the three panels show residuals for \daksha, \swift-BAT, and \fermi-GBM, respectively. 
    }
    }
    \label{fig:bcpl}
\end{figure}

Oganesyan 2018\cite{Oganesyan2018} have characterized GRB prompt emission spectra down to soft X-rays; in their sample of about 34 GRBs, they detect a low energy break in about 62\% of the spectra where the spectra harden below the break energy (between 3~keV and 22~keV in their sample). However, this sample is small, and \daksha, with its dedicated detector to measure the soft X-ray spectra of the prompt emission, is ideally suited to determine the low-energy break for large number GRBs. 
{\update Figure~\ref{fig:bcpl} shows the simulated spectra of a GRB with 10--1000~keV fluence of 1~\eu{-5}\fluence\ as detected by \daksha, \swift-BAT and \fermi-GBM. This simulation is carried out considering on-axis incidence on just \textit{one each} of LEP, MEP, and HEP of \daksha. In practice, the effective area for \daksha is much higher when adding the signals from multiple detector packages, but this example suffices to show the high sensitivity of \daksha compared to other instruments.}

\subsection{Extended emission and long central engine activity} 
A fraction of sGRBs are observed to have softer extended emission on time scales of a few seconds to a hundred seconds. Amongst them, a good fraction of the sGRBs are followed by a kilonovae {\it e.g.} GRB 050724 \citep{Norris2010}, GRB 060614 \citep{Bostanci2012}, and GRB 080503 \citep{Perley2009}. The fluence of the extended emission of some of these sGRBs is higher than the prompt emission \citep{Norris2010}. It is also worth mentioning that the extended emission was not found in GW170817, suggesting either that NS mergers with masses similar to GW170817 do not produce extended emission or that it was produced in GW170817 but was significantly beamed away from the Earth \citep{LVCFERMI170817}. These observations pose few important questions: Are their progenitors different from sGRBs without extended emission? What is the energy source of long lasting emission if they
arise from binary NS mergers?

It has been suggested that the extended emission can be powered by the spin-down luminosity of a remnant magnetar or the Blandford-Znajek process of a remnant Kerr black hole in NS mergers \citep{Metzger2008}. The detection of an extended emission associated with the GW from confirmed BNS merger event will shine light on this as well as the nature of GRB central object. \daksha, with its high sensitivity and wide bandpass, particularly extending down to soft X-rays, is ideally suited to investigate these aspects.

\subsection{Time-resolved studies}
The time resolved prompt emission studies are governed by two distinct patterns in the evolution of the peak energy $E_{p}$: \textit{first} the spectral evolution from hard to soft energies. This can be explained by the matter dominant jet outflow first giving black body radiation from the collapsing material \citep{Shemi90} and then emitting non thermal photons resulting from interaction of internal shocks interacting with optically thin region above the photosphere \citep{Daigne98}. \textit{Second}, the luminosity correlation with temperature \citep{Golen83} resulting in flux driven spectra taking into account the local magnetic-energy dissipation in a Poynting flux dominant outflow fueling the acceleration of the flow to a high bulk Lorentz factor. For typical GRB parameters, the dissipation takes place mainly above the photosphere, producing non-thermal radiation \citep{McKinney12}. In order to understand the underlying emission mechanism and to reach to a unifying physical model, we need observations with fine time resolved prompt spectra as well as ample GRBs with soft energy detection.

Even for short GRBs, investigating temporal evolution is of great importance. The short GRB jet propagating inside the merger ejecta forms a cocoon, which eventually breaks out from the ejecta surface. The emission arising from the cocoon shock break-out is expected to be composed of an initial hard-spike followed by a soft-tail \citep{gottlieb170817, Beloborodov2020}. The luminosity, duration, and spectral peak of the hard spike and soft tail depend on the ejecta structure as well as the jet properties such as the power, opening angle and time delay between the merger and jet launch. By modeling the emission properties with time resolved spectroscopy of the prompt phase, we will be able to obtain valuable information about the structure of merger ejecta and the jet formation in neutron star mergers. This in turn may allow us to constrain the neutron star equation of state (EOS) since the ejecta profile in the fast tail is sensitive to the EOS \citep{Hotokezaka2018, Radice2018}. \daksha with micro-second time resolution capabilities and broadband energy coverage (including the soft X-ray energies) is well suited to probe the temporal evolution of the prompt phase for long as well as short GRBs.

\subsection{Polarization from GRBs}\label{sec:pol}
Polarization from GRB emission can be an important tool to probe the physics of emission mechanisms in GRBs, the geometry of the emission region, and the origin and nature of the magnetic field at the emission region \citep{Covino16,lazzati2004compton,waxman2003new}. Till date polarization has been detected in only a handful of sources \citep{gill20a,2022ApJ...936...12C,2011ApJ...743L..30Y}. There is significant debate in the literature regarding the source of the polarization e.g. synchrotron with ordered magnetic fields, synchrotron with random magnetic fields at shocks,  and inverse Compton interactions \citep{beloborodov11a,Toma_2009,Covino16,gill20a}. Different theoretical models predict varying degrees of maximum polarization, based on the magnetic field, geometry and the viewing angle \citep{ghisellini99a,lazzati2004compton,gill20a}. If the polarization is due to synchrotron processes, as is often conjectured \citep{burgess20a}, then a high polarization fraction would imply ordered magnetic fields within the jet structure. A disordered magnetic field structure that many theories propose may arise at the forward shock \citep{medvedev99a} would lower the polarization. Contemporary work on GRB spectra indicate the prompt emission resulting from synchrotron radiation(non-thermal) \citep{Burgess, Troja}. However, spectral modeling of photospheric emission(thermal) has also provided adequate fits to a subset of GRBs \citep{Vianello}. A combined study of the GRB spectrum and polarization will break the degeneracy between various such theoretical models.

Variation of the GRB polarization degree and the polarization angle has been observed in very few sources \citep{gotz,yonetoku11a,zhang19a,sharma19a,2022ApJ...936...12C}. Time-resolved polarization studies are challenging due to insufficient photon counts during a GRB outburst. However, such observations can be a valuable tool to understand the internal structure of the GRB jet and the nature of emission. Different theoretical models have put forth possible predictions of variation of the polarization angle, for example, the evolution of viewing angle cone resulting in observing different magnetic field geometries \citep{ghisellini99a} or inherently patchy, non-axisymmetric emission due to internal fluid inhomogeneities \citep{lazzati09a}. The strength of the observed polarization fraction and nature of the variation will help distinguish between such models \citep[e.g. as reported in ][]{yonetoku11a,sharma19a}, which will provide valuable constraints on the nature of GRB outflow. 

\begin{figure}[ht!]
    \centering
    \includegraphics[width=0.6\textwidth,trim=0 0 10 0,clip=true]{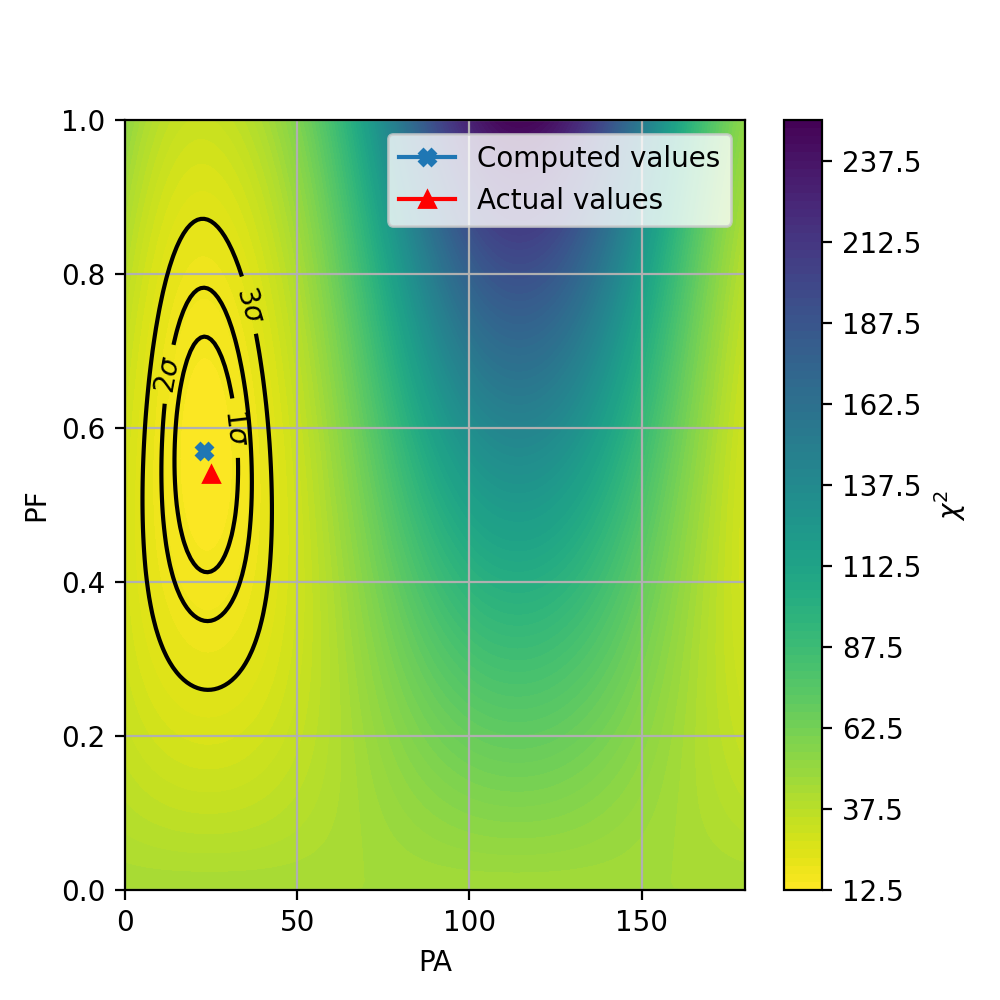}
    \caption{The plot shows hard X-ray polarization measurement capability of \daksha\ using GRB~160821A as an example. GRB~160821A had a fluence of $1.32\times10^{-4}$ \fluence\ in the energy range 10 -- 1000~keV ($3.64\times10^{-5}$ \fluence\ in 20 -- 200~keV). The red point shows the injected GRB with polarization angle (PA) = 25\textdegree\ and a polarization fraction (PF) of 0.54 while the blue point shows the recovered value of PA (=23\textdegree) and PF (=0.57). The recovered value lies well within 1-$\sigma$ contour.}
    \label{fig:pol_capability}
\end{figure}

\daksha, with its pixellated CZT detectors with large collecting area will be able to measure polarization of hard X-rays in the prompt phase for GRBs having sufficient brightness. Figure~\ref{fig:pol_capability} shows simulation results showing hard X-ray polarization measurement capabilities of \daksha. {\update Our study finds that the Minimum Detectable Polarisation (MDP) for Daksha will be 0.30 for a fluence of $10^{-4}$~\fluence, and we expect to measure polarization for at least 5 GRBs per year with a fluence of more than $10^{-4} $~\fluence. The details of the polarization sensitivity of \daksha (including the measurement method) are published in~\cite{Bala2023}.}

As an illustration, we consider the case of the polarized GRB~160821A. The GRB had a 10 -- 1000~keV fluence of 1.32\eu{-4}{~\fluence}, and a duration $T_{90} = 43$~s. We model the GRB spectrum as a Band function with parameters $\alpha = -1.05$, $\beta = -2.3$, and $E_\mathrm{peak} = 941$~keV.

\section{Additional science goals}\label{sec:other}
As a sensitive all-sky monitor, \daksha\ data can be used to probe many other scientific questions. Here, we discuss some of the science cases.

\subsection{Magnetar Bursts}
Magnetars are young (age $\lesssim 10^4$\,yrs), highly magnetized ($B_\mathrm{surf}\sim10^{13-16}$\,G) neutron stars, that are notorious for their wide variety of high-energy transient phenomena \citep[see][for an observational and theoretical review]{kaspi_beloborodov2017}. These include outbursts which last for months to giant flares that have been known to emit $10^{46}$\,erg in $\approx$100\,ms \citep{hurley2005}. With it's lower fluence threshold, wide energy range, and sky coverage \daksha\ will be a valuable tool for probing magnetar bursts and flares from the Milky Way magnetars as well as giant flares from magnetars in nearby galaxies ($d\lesssim 10^2\,\mathrm{Mpc}$). The increased volume probed by \daksha\ will allow for a better understanding of magnetar burst rates, luminosity distributions, and the birth-to-death cycle of magnetars.

For brighter bursts, \daksha's polarization capabilities will allow for the measurement of linear polarization which is expected to occur due to the propagation of photons through extremely strong magnetic fields \citep[$B\sim10^{15}$\,G,][]{tavernaturolla2017}. While other planned X-ray polarization missions aim to study the polarization of persistent emission from magnetars, \daksha's polarization measurement capabilities over a near-all-sky field of view are required for polarization measurements of magnetar bursts.

\subsection{Fast Radio Burst Counterparts}
FRBs are recently discovered millisecond timescale radio transients that are detected from cosmological distances ($\sim$Gpc). The isotropic burst energies of FRBs ($10^{38-42}$\,ergs) are almost a trillion times higher than the brightest radio pulses observed from Galactic pulsars. Due to the short timescale and the luminosity of FRBs, neutron stars, especially magnetars \citep[e.g. ][]{metzger2019, lyutikov2020}  are leading candidates for their origins. Similarly, many FRB models expect prompt radio counterparts to be emitted with BNS and NSBH mergers \citep{pshirkov2010, totani2013, mingarelli2015, paschalidis2019, rowlinson2019}. From the large volumetric rates of FRBs, compared to those of BNS and NSBH mergers, it is clear that these would contribute to a small fraction of observed FRBs. To date, while the observational data is rapidly increasing, the evidence is heterogenous with a multiple theories \citep[see e.g.][ for a review]{platts2019} and plenty of open questions remain about the origins of FRBs \citep{petroff2019, petroff2022}.

Most mechanisms expect that the radio emission of fast radio bursts is a small fraction of the total burst energy and prompt counterparts as well as afterglows are expected in different wavebands, including X-rays. Due to the extreme sensitivity of radio telescopes compared to X-ray and optical telescopes, it is expected that most models predict that typical FRBs will not have detectable high-energy counterparts. By significantly increasing the rate of X-ray and gamma-ray transients, {\update \daksha\ will accelerate the search for} X-ray and gamma-ray counterparts of FRBs, a search that has yet been unsuccessful from existing missions, see e.g. \cite{cunningham2019, anumarlapudi2020, guidorzi2020, curtin2022, principe2022}.

On 2020 April 28, an energetic radio burst with a total isotropic radio emission of $10^{34}$\,erg from the Galactic magnetar SGR\,1935+2154 \citep{chime_sgr1935_2020, bochenek2020b} accompanying an X-ray burst with an energy of $\approx10^{39}$\,erg. The X-ray burst was delayed by 6\,ms relative to the radio emission and was significantly harder ($E_\mathrm{peak} \sim 65\,$keV) compared to typical magnetar bursts \citep{mereghetti2020}. While the radio energy output of this burst was few orders of magnitude lower than that of typical FRBs, this burst was the brightest radio transient ever observed and partially bridges the energy gap between radio pulsars and FRBs. For a 10~ms burst, \daksha\ has a 5-$\sigma$ fluence sensitivity of 7\eu{-8}{~\fluence}. Given the scarcity of bright FRBs and the faintness of their corresponding high energy counterparts, \daksha's broad sky-coverage and low fluence threshold will be able to detect or rule out high energy counterparts to FRBs detected by ground-based instruments such as CHIME \citep{chime2018}, ASKAP \citep{bannister2017}, STARE2 \citep{bochenek2020a} and upcoming telescopes such as BURSTT \citep{lin2022}.

\subsection{X-ray pulsars}
The accretion powered X-ray pulsars are laboratories for study of various phenomena involving matter in strong magnetic fields\cite{1983ApJ...270..711W, 1989PASJ...41....1N}. The accreting matter imparts or draws out angular momentum from the neutron star. Both the accretion torque and the X-ray beaming from the neutron star are dependent on the mass accretion rate and structure of the accretion column on the magnetic poles of the neutron star. For a neutron star with a supergiant companion star, the mass accretion from the wind causes these sources to show random variation in accretion torque\cite{1997ApJS..113..367B, 2020ApJ...896...90M}. However, these sources occasionally show rapid spin-up phases, indicating a change in the mode of accretion from wind to a disk\cite{2012ApJ...759..124J, 2023MNRAS.520.1411M}.  Study of various pulsar characteristics like pulse shape, pulse fraction etc. at different accretion torque levels measured with \daksha\ will lead to better understanding of the relative importance of the two accretion mechanisms in the persistent accretion powered X-ray pulsars. The neutron stars with Be-Star companions undergo large outbursts during which the neutron star shows large spin-up followed by slow spin-down during the quiescence\cite{2011BASI...39..429P}. Measurement of the accretion torque as a function of the accretion power and the pulse profiles of X-ray pulsars will therefore be very useful in study of the accretion processes in high magnetic field neutron stars over a large range of mass accretion rate. Isolated X-ray pulsars and magnetars can also undergo discontinuous changes in their rotation rate (glitches and anti-glitches), some times accompanied by large scale pulse profile and emission mechanism changes.

As an all sky X-ray sensitive satellite, \daksha\ will be able to gather time-tagged photons from all sources in the sky. The {\update absolute precision of \daksha\ time-tagging} will be sub-ms. Since individual \daksha\ detectors lack directionality, {\update individual photons will not be identified to be from any specific sources}. However, by barycentering the arrival times and optimally combining photon counts from different detector facets of \daksha, we can search for periodic astrophysical signals in a frequency--frequency derivative--sky-position phase space. For known {\update accretion powered} pulsars, we can use \daksha\ photons to continuously monitor the spin periods for variations with accretion rates, and refine their ephemeris. This monitoring can simultaneously cover all the X-ray bright pulsars in the sky, unlike the monitoring with radio telescopes. In addition to barycentering the data and combining the signals from different detectors, \daksha\ can also use Earth occultation (see Section~\ref{sec:eo}) as a method to distinguish photons from bright sources near the Earth limb.

We calculated the expected source count rates for these sources with {\update \daksha\ MEPs} based on the \emph{NuSTAR} 20--50\,keV spectra of several accretion powered pulsars. We created lightcurves for these pulsars based on the \emph{NuSTAR} pulse profiles in the same energy band, scaled appropriately and {\update with a background count rate} of $\sim7400\,\mathrm{cts\,s^{-1}}$ (for seven surfaces). We simulated observations of these pulsars with 
a duty cycle appropriate for the low-earth orbit of \daksha. A Lomb-Scargle periodogram search was used to determine the pulsed flux level required to detect the pulsations above a 3-$\sigma$ and 5-$\sigma$ threshold.

\begin{figure}[ht!]
    \centering
    \includegraphics[width=0.75\textwidth]{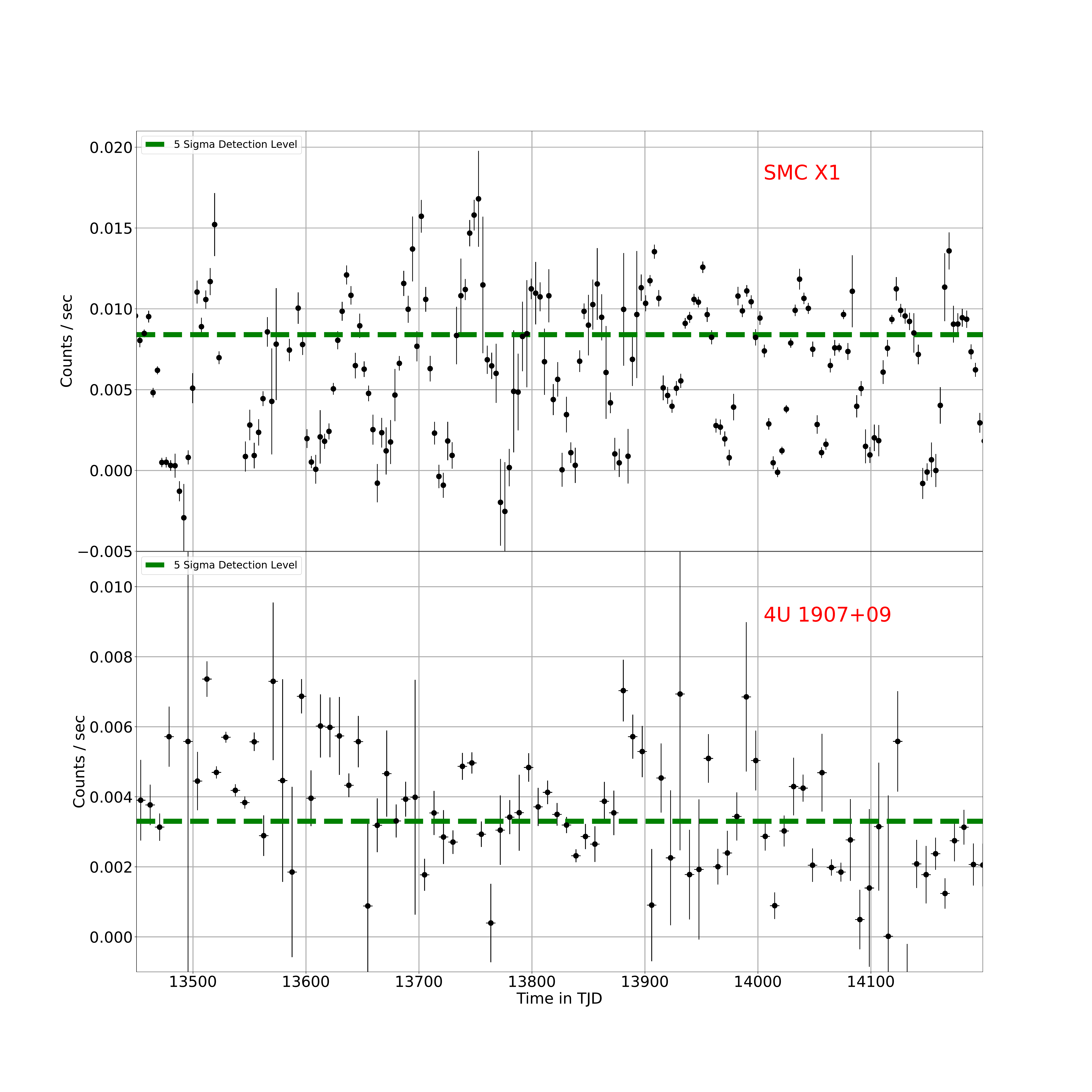}
    \caption{{\update Long term light curves of SMC X-1 and 4U 1907+09 from \emph{Swift}-BAT compared to the \daksha-MEPs 5-$\sigma$ detection threshold for integration times of 3.9 days and 8.4 days respectively.}
    }
    \label{fig:pulsar_obs}
\end{figure}

The simulations show that accretion powered pulsars with a {\update total} flux level above $2.6\times10^{-10} (3.1\times10^{-10})\,\mathrm{ergs\,s^{-1}\,cm^{-2}}$ in the 20--50 keV band will be detected with {\update \daksha-MEPs} at 3-$\sigma$ (5-$\sigma$) level respectively in every 2.5 day interval. 
{\update Even fainter sources will be detected with longer integration times.
Figure~\ref{fig:pulsar_obs} shows detectability of two accretion powered pulsars SMC X-1 and 4U 1907+09 with \daksha-MEPs. 
These two sources are persistent but have considerable variability and are not detected with \emph{CGRO}-BATSE and \emph{Fermi}-GBM.
The Swift-BAT light curves are shown here for a two year period along with the 5 sigma detection thresholds for these two sources with
\daksha-MEPs for pulsation search carried out on a light curve with a time span that is the same as their respective orbital periods 
(3.9 days and 8.4 days respectively).} The detection threshold  depends partly on the pulsed fraction and the pulse shape of each source. \daksha-MEPs will therefore be able to carry out continuous monitoring of the pulsar frequency and pulsed flux history of about 15 persistent X-ray pulsars. A large number of transient pulsars (about six transient pulsars every year) will also be monitored during their outbursts. About two new transient pulsars are expected to be discovered every year. For study of accretion powered X-ray pulsars, \daksha-MEPs 
{\update will be more effective than}
\emph{CGRO}-BATSE and \emph{Fermi}-GBM which have been extremely useful for study of this type of sources in the last three {\update decades\cite{1997ApJS..113..367B, 2017PASJ...69..100S}.}

\subsection{Primordial black hole abundance}
The nature of dark matter remains one of the biggest puzzles in cosmology today.  Primordial black holes (PBHs) that formed very early in the history of the Universe have been suggested as possible candidates to make up a large fraction of dark matter \citep{ZeldovichNovikov1966, Hawking1971}.  Cosmological observations including the cosmic microwave background, as well as various microlensing surveys carried out from both ground and space have ruled out a vast parameter range for the possible masses of such primordial black holes \citep[see e.g.,][]{Carr2020}.  However, there exists a window in the mass range {\update $[10^{-15}-10^{-11}]M_\odot$} where such PBHs remain unexplored thus far.  Such PBHs could even make up the entirety of the dark matter, thus solving the dark matter puzzle without recourse to any new particle physics.  Such mass scales cannot be explored with light in the optical wavelength range, as the Schwarzschild radii of such black holes are smaller than the wavelength of optical light \citep[see e.g.][]{Niikura2019}.  Due to its wide wavelength coverage, \daksha\ will probe this mass range in its high energy band.  The two \daksha\ satellites will enable a unique parallax microlensing experiment for the first time, where the same GRB can be observed with two different lines-of-sight {\update \citep{JungKim2020, Gawade2023}}.  The PBH will be at a different impact parameter compared to the GRB for the two different satellites, thus causing a difference in the measured flux of the GRB. A null detection of a difference in flux (i.e.  no detection of lensed GRBs) can constrain the PBH abundance in the unexplored mass range. 
{\update If the two Daksha satellites observe 1000 (10000) GRBs simultaneously during its nominal (extended\footnote{GRB missions in the past have  well out-lived their nominal mission lifetimes.}) lifetime and the entirety of dark matter is made up of $[10^{-15}-10^{-12}]M_\odot$ black holes, Daksha will detect non-zero lensing events with a decreasing probability which ranges from $16$ $(80)$ to $6$ $(50)$ per cent, respectively, at the edges of the bins. However, due to Poisson fluctuations, non-detections will not be able to conclusively constrain primordial black holes as dark matter just with the Daksha satellites alone. \citep{Gawade2023} show the synergy of \daksha\ when combined with a GRB mission kept at the Earth-Moon distance for this purpose. }
Any such experiment will require excellent cross-calibration between the detectors on board the twin \daksha\ satellites.

\subsection{Earth Occultation Studies}
\label{sec:eo}
\begin{figure}[tbhp]
    \centering
     \includegraphics[width=\columnwidth]{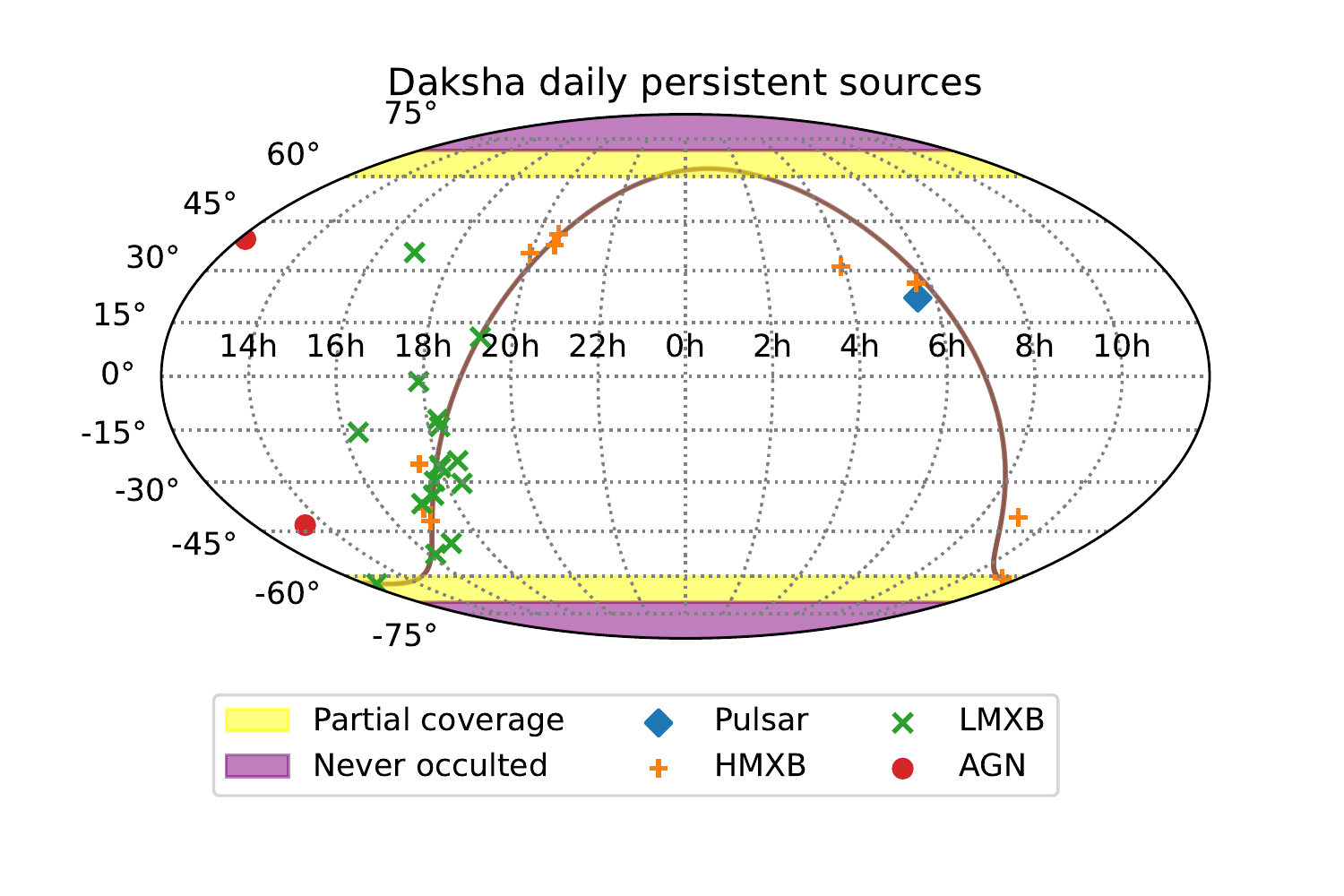}
    \caption{Sources whose flux can be monitored daily with \daksha\ by the Earth Occultation technique. {\update The curved brown line shows the galactic plane.} In addition to these persistent sources, bright transient sources like novae, X-ray binaries outbursts, etc can also be detected with this technique.    \label{fig:eo}}
\end{figure}

The background for \daksha\ arises from the diffuse cosmic X-ray background, a relatively small contribution from electronic noise, and counts from bright point sources in the sky. The occultation of these bright sources by the Earth causes changes in the net count rate, which can be used to monitor the brightness of these sources \citep{2002ApJS..138..149H}. We follow the method defined in  Singhal et al.\cite{2021JApA...42...64S} to calculate the sensitivity of \daksha\ for such measurements. The instant of the occultation event (ingress or egress) is known from the coordinates of the source and the orbital position of the satellite. We measure the flux levels in 100~s pre-- and post-- event windows, and estimate the change in flux from the change in counts. We have to account for one subtlety in these calculations: source photons are incident on the projected area of the satellite, while the background rates are determined by the total physical area of the relevant surfaces. We do these calculations for the MEPs considering a nominal case with the median effective area (1310~cm$^2$) for the source and seven surfaces for the background. We calculate that for a single ingress or egress, the 3-$\sigma$ sensitivity of \daksha\ is 0.048~\pcs\ (186~mCrab) in the 20--200~keV ME band. Each satellite completes about 14 orbits per day, with two measurements per orbit. Combining data from both satellites, the daily averaged sensitivity is 0.0064~\pcs\ (25~mCrab).

We used the \swift-BAT catalogue \citep{2018ApJS..235....4O} to estimate the number of sources that can be detected every day by \daksha. Using the reported hard X-ray fluxes and power-law indices, we calculate that \daksha\ can provide daily flux measurements of 29 sources: two active galactic nuclei (Cen~A, NGC~4151), the Crab Pulsar, ten High Mass X-ray Binaries, and sixteen Low Mass X-ray Binaries. As seen in Figure~\ref{fig:eo}, most sources are along the galactic plane. Note that sources in the $\pm$(60\degr~--~70\degr) declination range are not occulted in every satellite orbit, owing to the 6\degr\ orbital inclination. Sources close to the poles are never occulted by the Earth for \daksha, and cannot be monitored by this method.

{\update Similar monitoring of low-energy sources can potentially be undertaken with the LEPs. The sky coverage in this energy band is limited as the LEPs only cover half the sky, with highest effective area in a direction opposite to the sun. While source fluxes are higher in this energy range, so is the contribution of the Cosmic X-ray Background \citep{2009A&A...493..501M}. The final number of detectable sources in this band depends sensitively on our effective area as a function of energy. As a result, we do not consider LE earth occultation measurements further in this work.}

\subsection{Novae}
In addition to the study of persistent sources, the Earth occultation technique can also be used to detect bright transient sources like novae, outbursts of X-ray binaries, etc. As an illustration, we discuss the hard X-ray detectability of novae. Soft X-rays have been commonly detected in novae, but hard X-ray detections remain few in number --- and the emission mechanisms are poorly understood \citep{2008ASPC..401..323S,2021ApJ...910..134G}. Theoretical models predict that the early flash may be as bright as 0.1~\pcs\ in the 100--200~keV band, with emission fading rapidly to lower levels over the timescale of a day \citep[and references therein]{2008A&A...485..223S}. However, some novae have been bright in hard X-rays: for instance, the 2006 outburst of recurrent nova RS~Oph was detected by \swift/BAT \citep{2006ApJ...652..629B}, and the 2015 outburst of GK~Per was detected by \integral\ \citep{2015ATel.7244....1T}. Pointed hard X-ray observations of novae have led to more detections, with steep power-law spectral slopes:  $\Gamma = -3.6$ for V5855~Sgr \citep{2019ApJ...872...86N}, $-3.9$ for V906~Car \citep{2020MNRAS.497.2569S}, and $-3.3$ for YZ~Ret \citep{2022MNRAS.514.2239S}.

Predictions for hard X-ray fluxes of novae are often made by extrapolating the low-energy models \citep[see for instance][]{2006ApJ...652..629B}. To estimate if RS~Oph would be detectable by \daksha, we follow the approach of Page et. al. 2022~\citep{2022MNRAS.514.1557P} by fitting the \swift-XRT data with an APEC model, and calculating the expected flux in the \daksha\ energy band (20--200 keV). We find that the source would have been detectable to \daksha\ from day 3--6 of the outburst.

A direct comparison with \swift-BAT lightcurves shows that \daksha\ can also detect X-ray novae like MAXI~J1828$-$249 \citep{2016AstL...42...69G}, GRS~1739--278 \citep{2017AstL...43..167M} and MAXI~J1535$-$571 \citep{2018AstL...44..378M}.

\subsection{Terrestrial Gamma-ray Flashes}
The Earth's atmosphere is populated by processes like thunderstorms, lightning, and related electrical phenomena which emit X-rays and gamma-rays. Despite their global occurrences, the underlying science of these critical electromagnetic phenomena is still poorly understood. One such phenomenon is Terrestrial gamma-ray flashes (TGFs)  which were first detected in 1994 from space by Compton Gamma Ray Observatory \citep{tgf_fishman94}. TGFs are millisecond-duration sudden bursts of gamma-ray radiation having energies reaching as high as 100  MeV \citep{tgf_tavani11}. They are observed over thunderstorms and originate at around 20~km altitude \citep{tgf_carlson07}. On average 500 TGFs/day are expected to occur globally but not all get detected. Moreover, TGFs also produce energetic particles known as Terrestrial Electron Beams (TEBs) that are composed of mainly secondary electrons and positrons, which can be observed by spacecraft orbiting in the inner magnetosphere \citep{tgf_dwyer12}. Their impact on the inner magnetosphere is unknown and warrants detailed investigation. The simultaneous conjugate studies with existing radiation belt missions like the Exploration of energization and Radiation in Geospace (ERG) /Arase \citep{2018EP&S...70..101M} will help to decipher the linkage of the lower atmosphere and inner magnetosphere coupling.

High energy atmospheric physics is a new domain and evolving, the space-based continuous observations of gamma rays will assist in deciphering the unresolved problems of TGFs and TEBs. The microsecond time resolution and high sensitivity of \daksha\ in the hard X-ray band makes it a well-suited instrument for the study of TGFs. At an altitude of 650~km, \daksha\ can monitor a $\pm 25 \degr$ band on the Earth's surface, corresponding to {\update an instantaneous footprint of $\sim$2700~km (5\% of Earth's surface)}. Combined with the 6\degr\ inclination, the satellites will be able to detect activity in the latitude range from $-19\degr$ to $+19\degr$ in each orbit, reaching upto $\pm31\degr$ in various parts of the Earth on successive passes {\update --- covering a total of $\sim 50$\% of the surface.}

Key instruments for the study of TGFs inclued \fermi\ and the Atmosphere-Space Interactions Monitor ({\em ASIM}) instrument on the International Space Station (ISS). We can compare the TGF capabilities of \daksha\ directly with the ``Low Energy Detector'' (LED) of the Modular X-ray and Gamma-ray Sensor (MXGS) on {\em ASIM}. The LED comprises of CZT detectors covering the energy range from 50~keV to 400~keV, with an effective area of 400~cm$^2$ at 100~keV and a time resolution of 1~$\mu$s \citep{2019JGRD..12414024O}. \daksha\ has similar time resolution, and a much higher effective area: hence will be more sensitive to TGFs. The higher altitude of \daksha\ (650~km instead of 400~km) will also allow each \daksha\ satellite to view a much larger part of the Earth, while the near-equatorial orbit (as opposed to 55\degr\ orbit of the ISS) ensures that \daksha\ spends more time over areas with active thunderstorms, and avoids background magnetospheric particle fluxes as well as auroral X-ray emission which could have added to background noise. The orbital inclination is very important, and has been suggested to be the main reason why {\em ASIM} detects only about 260~TGFs/year while \fermi\ detects close to 800~TGFs/year \citep{2018JGRA..123.4381R,2019JGRD..12414024O}.The peak count rates reach tens of counts in 10~$\mu$s intervals, with only a fraction of events yielding over 100~counts. These rates are easily detectable in \daksha, where individual MEPs will have an average of 0.03~counts in 10~$\mu$s bins, and are designed to handle might higher peak count rates. TGFs are expected to have a few percent degrees of polarization \citep{tgf_bagheri19} which varies with source altitude. \daksha\ will be able to put stringent observational constraints on the polarization of the brightest TGFs.

There are several ground-based facilities studying the atmosphere: electric field measurements, cosmic ray observations, mesosphere--stratosphere--troposphere (MST) radars, etc. Simultaneous data from \daksha and such facilities can give great insight in understanding association of TGFs with convective activities and their evolutions. A particular example of such a synergy will be joint studies with \daksha, the GRAPES-3 muon telescope \citep{hariharan19}, Equatorial secondary cosmic ray Observatory, Tirunelveli, India \citep{vichare18}, and Indian MST radars \citep{uma09,subrahmanyam22}.

\subsection{Solar Flares}

Solar flares are sudden releases of energy in the solar atmosphere leading to emission across the entire electromagnetic spectrum, and often the release of energetic particles into the interplanetary medium. According to the standard flare model, the underlying mechanism powering the flares is magnetic reconnection that leads to acceleration of particles into non-thermal distributions and also heating of the plasma to temperatures often exceeding 10~MK~\citep{2017LRSP...14....2B}. While the standard flare model picture explains the observations in a broader context, several details such as the acceleration mechanism are still not well understood. As the accelerated 
electrons emit in hard X-rays by non-thermal bremsstrahlung, observations of the hard X-ray spectrum provide the most direct diagnostics of the non-thermal electron population~\citep{2008A&ARv..16..155K}. By modeling the observed hard X-ray spectrum, the distribution of the non-thermal electron population as well as quantitative estimates of their total energy content can be obtained. Reuven Ramaty High-Energy Solar Spectroscopic Imager (RHESSI, \citealp{2002SoPh..210....3L}) that observed the Sun in hard X-rays for 16 years until 2018 provided wealth of information on particle acceleration in solar flares with its broad band spectroscopic and imaging observations. RHESSI could observe non-thermal emission up to few tens of keV for flares down to GOES B-class intensities~\citep{2008ApJ...677..704H}; however, it was not possible to extend this to lower intensity flares.

Daksha, with its Sunward MEPs, will provide measurements of hard X-ray 
spectra of solar flares in 20--200 keV energy band. With 4 MEPs in the Sunward 
direction, Daksha will have about an order of magnitude larger effective area than that of RHESSI in this energy range. With the added advantage of simultaneous background measurements from other faces, Daksha is expected to have much better sensitivity than RHESSI for solar flare spectra. Simultaneous observations of flares by Daksha with instruments at other vantage points such as the Spectrometer/Telescope for Imaging X-rays (STIX, \citealp{2020A&A...642A..15K}) on Solar Orbiter or various instruments of {\em Aditya-L1} \citep{2017CSci..113..610S} also provides the opportunity to probe hard X-ray directivity.

\subsection{Earth-Sun interaction}
X-ray fluorescence (XRF) emission is triggered by solar X-rays from planetary atmospheres along with a scattered continuum. The scattered X-ray spectrum from Earth's atmosphere in the soft X-ray regime is a representation of the incident solar spectrum from which solar coronal abundances have been derived by Katsuda et. al~\citep{2020ApJ...891..126K}. The LEP on Daksha would measure the scattered solar X-ray spectra over a dynamic range enabling solar coronal studies. The reduction in intensity by scattering especially during strong flares would be an advantage here where often sun pointing spectrometers reach a saturation. In addition, a mapping of the Ar elemental abundance in Earth's atmosphere would be possible from its X-ray fluorescence line.

\section{Summary: the Impact of \daksha}
\daksha\ has unprecedented coverage of the transient high energy sky. As discussed in \dakshainst, the overall ``grasp'' of the mission, defined as the product of effective area and sky coverage, is higher than any current or proposed missions. Thanks to this, \daksha\ will discover the highest number of high energy counterparts to gravitational wave sources. It will boost our understanding of GRBs with prompt soft X-ray spectroscopy, highly time-resolved spectroscopic studies, and polarization measurements. 

In addition, \daksha\ covers a large number of secondary science cases. Compact object studies will benefit from continuous monitoring of accreting X-ray pulsars, detection and characterization of magnetar bursts, and the search for counterparts to FRBs. \daksha\ can monitor persistent sources and slower timescale transients like Novae to a sensitivity of 25~mCrab by the Earth Occultation Technique. The two identical satellites in the mission will allow us to probe primordial black hole abundance through microlensing.
Closer in, \daksha\ will provide excellent data for the study the Sun, TGFs, and the atmospheric response to solar activity.

As mentioned in \dakshainst, \daksha\ was proposed in response to the Indian Space Research Organisation's Announcement of Opportunity (AO) for Astronomy missions in 2018. The proposal was shortlisted for further studies, and was awarded seed funding by the Space Science Program Office for demonstration of a proof-of-concept. The team has completed the construction and testing of a laboratory model of the Medium Energy detector Package, as required. \daksha\ builds heavily on the legacy of various Indian space science missions, giving a high technology readiness level to all subsystems. The mission will be reviewed for full approval, after which we target a development timeline of three years to launch.

\section*{Acknowledgments}
We thank the Space Program Office (SPO) of the Indian Space Research Organisation for its Announcement of Opportunity for space astrophysics missions, under which \daksha\ was proposed. Development of the \daksha\ MEP laboratory model was started with funding support from SPO. We thank the administrative and support staff at all partner institutes for their help in all \daksha-related matters.

DS was supported under the CEFIPRA Grant No. IFC/5404-1. AP acknowledges the SERB Matrics grant MTR/2019/001096 and SERB-Power-fellowship grant SPF/2021/000036 of Department of Science and Technology, India for support.

\subsection*{Author Contributions}
Varun Bhalerao, Disha Sawant, Archana Pai, Shriharsh Tendulkar, Santosh Vadawale, Dipankar Bhattacharya, Vikram Rana, Gulab Dewangan contributed to overall definition and development of the mission science goals. 
The mission technical team that includes APK Kutty, Abhijeet Ghodgaonkar, Amit Shetye, Amrutha Lakshmi Vadladi, Arpit Patel, Ayush Nema, B.S. Bharath Saiguhan, Deepak Marla, Guruprasad P J, Hitesh Kumar L. Adalja, Hrishikesh Belatikar, Jayprakash G. Koyande, Jinaykumar Patel, M. Shanmugam, Mahesh Bhaganagare, Neeraj K. Tiwari, Nishant Singh, Piyush Sharma, Prabhu Ramachandran, Priya Pradeep, Rakesh Mote, S. Sreekumar, Salil Kulkarni, Sandeep Vishwakarma, Sanjoli Narang, Shiv Kumar Goyal, Shreeya Singh, Siddharth Tallur, Srividhya Sridhar, Suddhasatta Mahapatra, Sudhanshu Nimbalkar, Suresh Gunasekaran, Tinkal Ladiya contributed to developing and testing various hardware aspects of the mission, and their results were key for calculating sensitivity of the detectors for various science cases.
Akshat Singhal, Brad Cenko, Gaurav Waratkar, Kenta Hotokezaka, Mansi Kasliwal, Paz Beniamini, Rahul Srinivasan, Samaya Nissanke, Smaranika Banerjee, Soumyadeep Bhattacharjee, Sukanta Bose contributed to discussions, calculations and verification pertaining to the electromagnetic counterparts of gravitational wave sources.
Advait Mehla, C. S. Vaishnava, Dipanjan Mukherjee, Divita Saraogi, Joseph Saji, M. Saleem, Mehul Vijay Chanda, N. P. S. Mithun, Parth Sastry, Shabnam Iyyani, Sourav Palit, Sujay Mate, Suman Bala contributed to the science of Gamma-Ray Bursts and polarization measurements.
On broader science cases, Priyanka Gawade and Surhud More contributed to studies of primordial black holes; Vishal Dixit, Ankush Bhaskar, Shyama Narendranath contributed to to terrestrial studies; Biswajit Paul, Manikantan Hemanth, Kinjal Roy contributed to pulsars and X-ray binaries; G C Anupama, Judhajeet Basu, Anirban Dutta contributed to novae and slow transients.

\subsection*{Conflict of Interest}
The authors declare that there are no conflicts of interest in this manuscript.

\subsection*{Funding}
Development of Daksha was supported by a seed grant awarded to IIT Bombay from the Indian Space Research Organization. Further expenses were also supported by the partner institutes including direct and in-kind contributions.

\subsection*{Software}
Numpy~\citep{numpy}, Matplotlib~\citep{matplotlib}, Astropy \citep[\url{http://www.astropy.org}]{2013A&A...558A..33A,2018AJ....156..123A}, 
  HealPIX~\citep{ghb+05}, Healpy (\url{https://healpy.readthedocs.org/}), Ephem (\url{https://pypi.python.org/pypi/pyephem/}), WebPlot Digitizer \url{https://automeris.io/WebPlotDigitizer}, GEANT4~\citep[\url{https://geant4.web.cern.ch/}]{Agostinelli2003}

\bibliography{daksha}

\begin{thebibliography}{100}
\providecommand{\url}[1]{{#1}}
\providecommand{\urlprefix}{URL }
\providecommand{\doi}[1]{\url{https://doi.org/#1}}
\bibcommenthead

\bibitem{LIGOScientificCollaboration2017}
{LIGO Scientific Collaboration}, {Virgo Collaboration}, {Fermi Gamma-Ray Burst
  Monitor}, INTEGRAL, Gravitational waves and gamma-rays from a binary neutron
  star merger: Gw170817 and grb 170817a.
\newblock The Astrophysical Journal \textbf{848}, 13 (2017).
\newblock \doi{10.3847/2041-8213/aa920c}.
\newblock \urlprefix\url{http://arxiv.org/abs/1710.05834
  http://dx.doi.org/10.3847/2041-8213/aa920c}

\bibitem{gw170817}
B.~Abbott, R.~Abbott, T.~Abbott, F.~Acernese, K.~Ackley, C.~Adams, T.~Adams,
  P.~Addesso, R.~Adhikari, V.~Adya, C.~Affeldt, M.~Afrough, B.~Agarwal,
  M.~Agathos, K.~Agatsuma, N.~Aggarwal, O.~Aguiar, L.~Aiello, A.~Ain, P.~Ajith,
  O.~Smirnov, R.~Fender, P.~Woudt, {Multi-messenger Observations of a Binary
  Neutron Star Merger}.
\newblock \apjl \textbf{848}(2), L12 (2017).
\newblock \doi{10.3847/2041-8213/aa91c9}.
\newblock {\href{https://arxiv.org/abs/1710.05833}{{arXiv:1710.05833}}}
  {[astro-ph.HE]}

\bibitem{whs+19}
D.~Watson, C.J. Hansen, J.~Selsing, A.~Koch, D.B. Malesani, A.C. Andersen,
  J.P.U. Fynbo, A.~Arcones, A.~Bauswein, S.~Covino, A.~Grado, K.E. Heintz,
  L.~Hunt, C.~Kouveliotou, G.~Leloudas, A.J. Levan, P.~Mazzali, E.~Pian,
  {Identification of strontium in the merger of two neutron stars}.
\newblock Nature \textbf{574}(7779), 497--500 (2019).
\newblock \doi{10.1038/s41586-019-1676-3}.
\newblock \urlprefix\url{http://arxiv.org/abs/1910.10510
  http://dx.doi.org/10.1038/s41586-019-1676-3
  http://www.nature.com/articles/s41586-019-1676-3}.
\newblock {\href{https://arxiv.org/abs/1910.10510}{{arXiv:1910.10510}}}

\bibitem{Growth170817}
M.M. Kasliwal, E.~Nakar, L.P. Singer, D.L. Kaplan, D.O. Cook, A.~Van~Sistine,
  R.M. Lau, C.~Fremling, O.~Gottlieb, J.E. Jencson, S.M. Adams, U.~Feindt,
  K.~Hotokezaka, S.~Ghosh, D.A. Perley, P.C. Yu, T.~Piran, J.R. Allison, G.C.
  Anupama, A.~Balasubramanian, K.W. Bannister, J.~Bally, J.~Barnes, S.~Barway,
  E.~Bellm, V.~Bhalerao, D.~Bhattacharya, N.~Blagorodnova, J.S. Bloom, P.R.
  Brady, C.~Cannella, D.~Chatterjee, S.B. Cenko, B.E. Cobb, C.~Copperwheat,
  A.~Corsi, K.~De, D.~Dobie, S.W.K. Emery, P.A. Evans, O.D. Fox, D.A. Frail,
  C.~Frohmaier, A.~Goobar, G.~Hallinan, F.~Harrison, G.~Helou, T.~Hinderer,
  A.Y.Q. Ho, A.~Horesh, W.H. Ip, R.~Itoh, D.~Kasen, H.~Kim, N.P.M. Kuin,
  T.~Kupfer, C.~Lynch, K.~Madsen, P.A. Mazzali, A.A. Miller, K.~Mooley,
  T.~Murphy, C.C. Ngeow, D.~Nichols, S.~Nissanke, P.~Nugent, E.O. Ofek, H.~Qi,
  R.M. Quimby, S.~Rosswog, F.~Rusu, E.M. Sadler, P.~Schmidt, J.~Sollerman,
  I.~Steele, A.R. Williamson, Y.~Xu, L.~Yan, Y.~Yatsu, C.~Zhang, W.~Zhao,
  Illuminating gravitational waves: A concordant picture of photons from a
  neutron star merger.
\newblock Science \textbf{358}(6370), 1559--1565 (2017).
\newblock \doi{10.1126/science.aap9455}.
\newblock \urlprefix\url{https://science.sciencemag.org/content/358/6370/1559}.
\newblock
  {\href{https://arxiv.org/abs/https://science.sciencemag.org/content/358/6370/1559.full.pdf}{{https://science.sciencemag.org/content/358/6370/1559.full.pdf}}}

\bibitem{De2018}
S.~{De}, D.~{Finstad}, J.M. {Lattimer}, D.A. {Brown}, E.~{Berger}, C.M.
  {Biwer}, {Tidal Deformabilities and Radii of Neutron Stars from the
  Observation of GW170817}.
\newblock \prl \textbf{121}(9), 091102 (2018).
\newblock \doi{10.1103/PhysRevLett.121.091102}.
\newblock {\href{https://arxiv.org/abs/1804.08583}{{arXiv:1804.08583}}}
  {[astro-ph.HE]}

\bibitem{zhang2018peculiar}
B.B. {Zhang}, B.~{Zhang}, H.~{Sun}, W.H. {Lei}, H.~{Gao}, Y.~{Li}, L.~{Shao},
  Y.~{Zhao}, Y.D. {Hu}, H.J. {L{\"u}}, X.F. {Wu}, X.L. {Fan}, G.~{Wang}, A.J.
  {Castro-Tirado}, S.~{Zhang}, B.Y. {Yu}, Y.Y. {Cao}, E.W. {Liang}, {A peculiar
  low-luminosity short gamma-ray burst from a double neutron star merger
  progenitor}.
\newblock Nature Communications \textbf{9}, 447 (2018).
\newblock \doi{10.1038/s41467-018-02847-3}.
\newblock {\href{https://arxiv.org/abs/1710.05851}{{arXiv:1710.05851}}}
  {[astro-ph.HE]}

\bibitem{Hosseinzadeh2019}
G.~Hosseinzadeh, P.S. Cowperthwaite, S.~Gomez, V.A. Villar, M.~Nicholl,
  R.~Margutti, E.~Berger, R.~Chornock, K.~Paterson, W.~Fong, V.~Savchenko,
  P.~Short, K.D. Alexander, P.K. Blanchard, J.~Braga, M.L. Calkins, R.~Cartier,
  D.L. Coppejans, T.~Eftekhari, T.~Laskar, C.~Ly, L.~Patton, I.~Pelisoli, D.E.
  Reichart, G.~Terreran, P.K.G. Williams, {Follow-up of the Neutron Star
  Bearing Gravitational-wave Candidate Events S190425z and S190426c with MMT
  and SOAR}.
\newblock The Astrophysical Journal \textbf{880}(1), L4 (2019).
\newblock \doi{10.3847/2041-8213/ab271c}.
\newblock {\href{https://arxiv.org/abs/1905.02186}{{arXiv:1905.02186}}}

\bibitem{Coughlin2019}
M.W. Coughlin, T.~Ahumada, S.~Anand, K.~De, M.J. Hankins, M.M. Kasliwal, L.P.
  Singer, E.C. Bellm, I.~Andreoni, S.B. Cenko, J.~Cooke, C.M. Copperwheat, A.M.
  Dugas, J.E. Jencson, D.A. Perley, P.C. Yu, V.~Bhalerao, H.~Kumar, J.S. Bloom,
  G.C. Anupama, M.C.B. Ashley, A.~Bagdasaryan, R.~Biswas, D.A.H. Buckley, K.B.
  Burdge, D.O. Cook, J.~Cromer, V.~Cunningham, A.~D'A{\`{i}}, R.G. Dekany,
  A.~Delacroix, S.~Dichiara, D.A. Duev, A.~Dutta, M.~Feeney, S.~Frederick,
  P.~Gatkine, S.~Ghosh, D.A. Goldstein, V.Z. Golkhou, A.~Goobar, M.J. Graham,
  H.~Hanayama, T.~Horiuchi, T.~Hung, S.W. Jha, A.K.H. Kong, M.~Giomi, D.L.
  Kaplan, V.R. Karambelkar, M.~Kowalski, S.R. Kulkarni, T.~Kupfer, V.~{La
  Parola}, F.J. Masci, P.~Mazzali, A.M. Moore, M.~Mogotsi, J.D. Neill, C.C.
  Ngeow, J.~Mart{\'{i}}nez-Palomera, M.~Pavana, E.O. Ofek, A.S. Patil,
  R.~Riddle, M.~Rigault, B.~Rusholme, E.~Serabyn, D.L. Shupe, Y.~Sharma,
  J.~Sollerman, J.~Soon, K.~Staats, K.~Taggart, H.~Tan, T.~Travouillon,
  E.~Troja, G.~Waratkar, Y.~Yatsu, {GROWTH on S190425z: Searching thousands of
  square degrees to identify an optical or infrared counterpart to a binary
  neutron star merger with the Zwicky Transient Facility and Palomar Gattini
  IR}.
\newblock The Astrophysical Journal Letters \textbf{885}, L19 (2019).
\newblock \urlprefix\url{http://arxiv.org/abs/1907.12645}.
\newblock {\href{https://arxiv.org/abs/1907.12645}{{arXiv:1907.12645}}}

\bibitem{sdl+84}
V.~Schönfelder, R.~Diehl, G.G. Lichti, H.~Steinle, B.N. Swanenburg, A.J.M.
  Deerenberg, H.~Aarts, J.~Lockwood, W.~Webber, J.~Macri, J.~Ryan, G.~Simpson,
  B.G. Taylor, K.~Bennett, M.~Snelling, The imaging compton telescope comptel
  on the gamma ray observatory.
\newblock IEEE Transactions on Nuclear Science \textbf{1}, 766--770 (1984).
\newblock \doi{10.1109/TNS.1984.4333363}

\bibitem{bmf+92}
D.~Band, J.~Matteson, L.~Ford, B.~Schaefer, D.~Palmer, B.~Teegarden, T.~Cline,
  M.~Briggs, W.~Paciesas, G.~Pendleton, G.~Fishman, C.~Kouveliotou, C.~Meegan,
  R.~Wilson, P.~Lestrade, Batse observations of gamma-ray burst spectra. i -
  spectral diversity.
\newblock The Astrophysical Journal \textbf{413}, 281 (1993).
\newblock \doi{10.1086/172995}

\bibitem{1997A&AS..122..299B}
G.~{Boella}, R.C. {Butler}, G.C. {Perola}, L.~{Piro}, L.~{Scarsi}, J.A.M.
  {Bleeker}, {BeppoSAX, the wide band mission for X-ray astronomy}.
\newblock \aaps \textbf{122}, 299--307 (1997).
\newblock \doi{10.1051/aas:1997136}

\bibitem{Ricker2003}
G.R. {Ricker}, J.L. {Atteia}, G.B. {Crew}, J.P. {Doty}, E.E. {Fenimore},
  M.~{Galassi}, C.~{Graziani}, K.~{Hurley}, J.G. {Jernigan}, N.~{Kawai}, D.Q.
  {Lamb}, M.~{Matsuoka}, G.~{Pizzichini}, Y.~{Shirasaki}, T.~{Tamagawa},
  R.~{Vanderspek}, G.~{Vedrenne}, J.~{Villasenor}, S.E. {Woosley},
  A.~{Yoshida}, in \emph{Gamma-Ray Burst and Afterglow Astronomy 2001: A
  Workshop Celebrating the First Year of the HETE Mission}, \emph{American
  Institute of Physics Conference Series}, vol. 662, ed. by G.R. {Ricker}, R.K.
  {Vanderspek} (2003), pp. 3--16.
\newblock \doi{10.1063/1.1579291}

\bibitem{gcg+04}
N.~Gehrels, G.~Chincarini, P.~Giommi, K.O. Mason, J.A. Nousek, A.A. Wells, N.E.
  White, S.D. Barthelmy, D.N. Burrows, L.R. Cominsky, K.C. Hurley, F.E.
  Marshall, P.~M{\' e}sz{\' a}ros, P.W.A. Roming, L.~Angelini, L.M. Barbier,
  T.~Belloni, S.~Campana, P.A. Caraveo, M.M. Chester, O.~Citterio, T.L. Cline,
  M.S. Cropper, J.R. Cummings, A.J. Dean, E.D. Feigelson, E.E. Fenimore, D.A.
  Frail, A.S. Fruchter, G.P. Garmire, K.~Gendreau, G.~Ghisellini, J.~Greiner,
  J.E. Hill, S.D. Hunsberger, H.A. Krimm, S.R. Kulkarni, P.~Kumar, F.~Lebrun,
  N.M. Lloyd-Ronning, C.B. Markwardt, B.J. Mattson, R.F. Mushotzky, J.P.
  Norris, J.~Osborne, B.~Paczynski, D.M. Palmer, H.S. Park, A.M. Parsons,
  J.~Paul, M.J. Rees, C.S. Reynolds, J.E. Rhoads, T.P. Sasseen, B.E. Schaefer,
  A.T. Short, A.P. Smale, I.A. Smith, L.~Stella, G.~Tagliaferri, T.~Takahashi,
  M.~Tashiro, L.K. Townsley, J.~Tueller, M.J.L. Turner, M.~Vietri, W.~Voges,
  M.J. Ward, R.~Willingale, F.M. Zerbi, W.W. Zhang, {The Swift Gamma-Ray Burst
  Mission}.
\newblock The Astrophysical Journal \textbf{611}(2), 1005--1020 (2004).
\newblock \doi{10.1086/422091}.
\newblock \urlprefix\url{http://adsabs.harvard.edu/abs/2004ApJ...611.1005G}

\bibitem{Barthelmy2005}
S.D. Barthelmy, L.M. Barbier, J.R. Cummings, E.E. Fenimore, N.~Gehrels,
  D.~Hullinger, H.A. Krimm, C.B. Markwardt, D.M. Palmer, A.~Parsons, G.~Sato,
  M.~Suzuki, T.~Takahashi, M.~Tashiro, J.~Tueller, {The Burst Alert Telescope
  (BAT) on the Swift MIDEX Mission}.
\newblock Space Science Reviews, Volume 120, Issue 3-4, pp. 143-164
  \textbf{120}, 143--164 (2005).
\newblock \doi{10.1007/s11214-005-5096-3}.
\newblock \urlprefix\url{http://arxiv.org/abs/astro-ph/0507410
  http://dx.doi.org/10.1007/s11214-005-5096-3}.
\newblock {\href{https://arxiv.org/abs/0507410}{{arXiv:0507410}}} {[astro-ph]}

\bibitem{Meegan2009}
C.~Meegan, G.~Lichti, P.N. Bhat, E.~Bissaldi, M.S. Briggs, V.~Connaughton,
  R.~Diehl, G.~Fishman, J.~Greiner, A.S. Hoover, A.J. van~der Horst, A.~von
  Kienlin, R.M. Kippen, C.~Kouveliotou, S.~McBreen, W.S. Paciesas, R.~Preece,
  H.~Steinle, M.S. Wallace, R.B. Wilson, C.~Wilson-Hodge, The <i>fermi</i>
  gamma-ray burst monitor.
\newblock The Astrophysical Journal \textbf{702}, 791--804 (2009).
\newblock \doi{10.1088/0004-637X/702/1/791}.
\newblock
  \urlprefix\url{http://stacks.iop.org/0004-637X/702/i=1/a=791?key=crossref.55ed2ad645b9c8942db3a7bafcc85a32}

\bibitem{czti}
V.~Bhalerao, D.~Bhattacharya, A.~Vibhute, P.~Pawar, A.~Rao, M.~Hingar,
  R.~Khanna, A.~Kutty, J.~Malkar, M.~Patil, Y.~Arora, S.~Sinha, P.~Priya,
  E.~Samuel, S.~Sreekumar, P.~Vinod, N.~Mithun, S.~Vadawale, N.~Vagshette,
  K.~Navalgund, K.~Sarma, R.~Pandiyan, S.~Seetha, K.~Subbarao, {The Cadmium
  Zinc Telluride Imager on AstroSat}.
\newblock Journal of Astrophysics and Astronomy \textbf{38}, 31 (2017).
\newblock \doi{10.1007/s12036-017-9447-8}.
\newblock \urlprefix\url{http://arxiv.org/abs/1608.03408}.
\newblock {\href{https://arxiv.org/abs/1608.03408}{{arXiv:1608.03408}}}

\bibitem{2005NIMPA.550..616P}
N.~{Produit}, F.~{Barao}, S.~{Deluit}, W.~{Hajdas}, C.~{Leluc}, M.~{Pohl},
  D.~{Rapin}, J.P. {Vialle}, R.~{Walter}, C.~{Wigger}, {POLAR, a compact
  detector for gamma-ray bursts photon polarization measurements}.
\newblock Nuclear Instruments and Methods in Physics Research A
  \textbf{550}(3), 616--625 (2005).
\newblock \doi{10.1016/j.nima.2005.05.066}.
\newblock
  {\href{https://arxiv.org/abs/astro-ph/0504605}{{arXiv:astro-ph/0504605}}}
  {[astro-ph]}

\bibitem{2011ApJ...743L..30Y}
D.~{Yonetoku}, T.~{Murakami}, S.~{Gunji}, T.~{Mihara}, K.~{Toma},
  T.~{Sakashita}, Y.~{Morihara}, T.~{Takahashi}, N.~{Toukairin}, H.~{Fujimoto},
  Y.~{Kodama}, S.~{Kubo}, {IKAROS Demonstration Team}, {Detection of Gamma-Ray
  Polarization in Prompt Emission of GRB 100826A}.
\newblock \apjl \textbf{743}(2), L30 (2011).
\newblock \doi{10.1088/2041-8205/743/2/L30}.
\newblock {\href{https://arxiv.org/abs/1111.1779}{{arXiv:1111.1779}}}
  {[astro-ph.HE]}

\bibitem{Metzger2008}
B.D. Metzger, E.~Quataert, T.A. Thompson, Short-duration gamma-ray bursts with
  extended emission from protomagnetar spin-down.
\newblock Monthly Notices of the Royal Astronomical Society \textbf{385}(3),
  1455--1460 (2008).
\newblock \doi{10.1111/j.1365-2966.2008.12923.x}.
\newblock \urlprefix\url{http://dx.doi.org/10.1111/j.1365-2966.2008.12923.x}

\bibitem{Kumar_2015}
P.~Kumar, B.~Zhang, The physics of gamma-ray bursts relativistic jets.
\newblock Physics Reports \textbf{561}, 1--109 (2015).
\newblock \doi{10.1016/j.physrep.2014.09.008}.
\newblock \urlprefix\url{http://dx.doi.org/10.1016/j.physrep.2014.09.008}

\bibitem{aLIGO}
{LIGO Scientific Collaboration}, J.~{Aasi}, B.P. {Abbott}, R.~{Abbott},
  T.~{Abbott}, M.R. {Abernathy}, K.~{Ackley}, C.~{Adams}, T.~{Adams},
  P.~{Addesso}, R.X. {Adhikari}, V.~{Adya}, C.~{Affeldt}, N.~{Aggarwal},
  C.~{Wilkinson}, L.~{Williams}, R.~{Williams}, A.R. {Williamson}, J.L.
  {Willis}, B.~{Willke}, M.~{Wimmer}, W.~{Winkler}, C.C. {Wipf}, H.~{Wittel},
  G.~{Woan}, J.~{Worden}, S.~{Xie}, J.~{Yablon}, I.~{Yakushin}, W.~{Yam},
  H.~{Yamamoto}, C.C. {Yancey}, Q.~{Yang}, M.~{Zanolin}, F.~{Zhang},
  L.~{Zhang}, M.~{Zhang}, Y.~{Zhang}, C.~{Zhao}, M.~{Zhou}, X.J. {Zhu}, M.E.
  {Zucker}, S.~{Zuraw}, J.~{Zweizig}, {Advanced LIGO}.
\newblock Classical and Quantum Gravity \textbf{32}(7), 074001 (2015).
\newblock \doi{10.1088/0264-9381/32/7/074001}.
\newblock {\href{https://arxiv.org/abs/1411.4547}{{arXiv:1411.4547}}} {[gr-qc]}

\bibitem{AdVirgo}
F.~{Acernese}, M.~{Agathos}, K.~{Agatsuma}, D.~{Aisa}, N.~{Allemandou},
  A.~{Allocca}, J.~{Amarni}, P.~{Astone}, G.~{Balestri}, G.~{Ballardin},
  F.~{Barone}, J.P. {Baronick}, M.~{Barsuglia}, A.~{Basti}, F.~{Basti}, T.S.
  {Bauer}, V.~{Bavigadda}, M.~{Bejger}, M.G. {Beker}, C.~{Belczynski},
  D.~{Bersanetti}, A.~{Bertolini}, M.~{Bitossi}, M.A. {Bizouard}, S.~{Bloemen},
  M.~{Blom}, M.~{Boer}, G.~{Bogaert}, D.~{Bondi}, G.~{Vajente}, N.~{van Bakel},
  M.~{van Beuzekom}, J.F.J. {van den Brand}, C.~{Van Den Broeck}, M.V. {van der
  Sluys}, J.~{van Heijningen}, M.~{Vas{\'u}th}, G.~{Vedovato}, J.~{Veitch},
  D.~{Verkindt}, F.~{Vetrano}, A.~{Vicer{\'e}}, J.Y. {Vinet}, G.~{Visser},
  H.~{Vocca}, R.~{Ward}, M.~{Was}, L.W. {Wei}, M.~{Yvert}, A.Z. {{\.z}ny}, J.P.
  {Zendri}, {Advanced Virgo: a second-generation interferometric gravitational
  wave detector}.
\newblock Classical and Quantum Gravity \textbf{32}(2), 024001 (2015).
\newblock \doi{10.1088/0264-9381/32/2/024001}.
\newblock {\href{https://arxiv.org/abs/1408.3978}{{arXiv:1408.3978}}} {[gr-qc]}

\bibitem{kagra}
{KAGRA Collaboration}, T.~{Akutsu}, M.~{Ando}, K.~{Arai}, Y.~{Arai},
  S.~{Araki}, A.~{Araya}, N.~{Aritomi}, H.~{Asada}, Y.~{Aso}, S.~{Bae},
  Y.~{Bae}, L.~{Baiotti}, R.~{Bajpai}, M.A. {Barton}, K.~{Cannon}, Z.~{Cao},
  E.~{Capocasa}, M.~{Chan}, C.~{Chen}, K.~{Chen}, Y.~{Chen}, C.Y. {Chiang},
  H.~{Chu}, Y.K. {Chu}, S.~{Eguchi}, Y.~{Enomoto}, R.~{Flaminio}, Y.~{Fujii},
  F.~{Fujikawa}, M.~{Fukunaga}, M.~{Fukushima}, D.~{Gao}, G.~{Ge}, S.~{Ha},
  A.~{Hagiwara}, S.~{Haino}, W.B. {Han}, K.~{Hasegawa}, K.~{Hattori},
  T.~{Ushiba}, M.H.P.M. {van Putten}, H.~{Vocca}, J.~{Wang}, T.~{Washimi},
  C.~{Wu}, H.~{Wu}, S.~{Wu}, W.R. {Xu}, T.~{Yamada}, K.~{Yamamoto},
  K.~{Yamamoto}, T.~{Yamamoto}, K.~{Yamashita}, R.~{Yamazaki}, Y.~{Yang},
  K.~{Yokogawa}, J.~{Yokoyama}, T.~{Yokozawa}, T.~{Yoshioka}, H.~{Yuzurihara},
  S.~{Zeidler}, M.~{Zhan}, H.~{Zhang}, Y.~{Zhao}, Z.H. {Zhu}, {Overview of
  KAGRA : KAGRA science}.
\newblock arXiv e-prints arXiv:2008.02921 (2020).
\newblock {\href{https://arxiv.org/abs/2008.02921}{{arXiv:2008.02921}}}
  {[gr-qc]}

\bibitem{indigo}
B.~Iyer, et~al., Ligo-india. tech. rep. m1100296-v2, indigo, india  (2011).
\newblock \urlprefix\url{https://dcc.ligo.org/ligo-M1100296/public}

\bibitem{2022CQGra..39b5004S}
M.~{Saleem}, J.~{Rana}, V.~{Gayathri}, A.~{Vijaykumar}, S.~{Goyal},
  S.~{Sachdev}, J.~{Suresh}, S.~{Sudhagar}, A.~{Mukherjee}, G.~{Gaur},
  B.~{Sathyaprakash}, A.~{Pai}, R.X. {Adhikari}, P.~{Ajith}, S.~{Bose}, {The
  science case for LIGO-India}.
\newblock Classical and Quantum Gravity \textbf{39}(2), 025004 (2022).
\newblock \doi{10.1088/1361-6382/ac3b99}.
\newblock {\href{https://arxiv.org/abs/2105.01716}{{arXiv:2105.01716}}}
  {[gr-qc]}

\bibitem{bellm14}
E.~Bellm, The zwicky transient facility.
\newblock The Third Hot-wiring the Transient Universe Workshop (HTU-III) pp.
  27--33 (2014).
\newblock \urlprefix\url{http://adsabs.harvard.edu/abs/2014arXiv1410.8185B
  http://adsabs.harvard.edu/abs/2014htu..conf...27B}

\bibitem{rubin}
{LSST Science Collaboration}, P.A. {Abell}, J.~{Allison}, S.F. {Anderson}, J.R.
  {Andrew}, J.R.P. {Angel}, L.~{Armus}, D.~{Arnett}, S.J. {Asztalos}, T.S.
  {Axelrod}, S.~{Bailey}, D.R. {Ballantyne}, J.R. {Bankert}, W.A. {Barkhouse},
  J.D. {Barr}, L.F. {Barrientos}, A.J. {Barth}, J.G. {Bartlett}, A.C. {Becker},
  J.~{Becla}, T.C. {Beers}, J.P. {Bernstein}, R.~{Biswas}, M.R. {Blanton},
  P.~{Young}, A.~{Zentner}, H.~{Zhan}, {LSST Science Book, Version 2.0}.
\newblock arXiv e-prints arXiv:0912.0201 (2009).
\newblock {\href{https://arxiv.org/abs/0912.0201}{{arXiv:0912.0201}}}
  {[astro-ph.IM]}

\bibitem{ska}
P.E. {Dewdney}, P.J. {Hall}, R.T. {Schilizzi}, T.J.L.W. {Lazio}, {The Square
  Kilometre Array}.
\newblock IEEE Proceedings \textbf{97}(8), 1482--1496 (2009).
\newblock \doi{10.1109/JPROC.2009.2021005}

\bibitem{ugmrt}
Y.~{Gupta}, B.~{Ajithkumar}, H.S. {Kale}, S.~{Nayak}, S.~{Sabhapathy},
  S.~{Sureshkumar}, R.V. {Swami}, J.N. {Chengalur}, S.K. {Ghosh}, C.H.
  {Ishwara-Chandra}, B.C. {Joshi}, N.~{Kanekar}, D.V. {Lal}, S.~{Roy}, {The
  upgraded GMRT: opening new windows on the radio Universe}.
\newblock Current Science \textbf{113}(4), 707--714 (2017).
\newblock \doi{10.18520/cs/v113/i04/707-714}

\bibitem{lofar}
M.P. {van Haarlem}, M.W. {Wise}, A.W. {Gunst}, G.~{Heald}, J.P. {McKean},
  J.W.T. {Hessels}, A.G. {de Bruyn}, R.~{Nijboer}, J.~{Swinbank}, R.~{Fallows},
  M.~{Brentjens}, A.~{Nelles}, R.~{Beck}, H.~{Falcke}, R.~{Fender},
  J.~{H{\"o}randel}, L.V.E. {Koopmans}, G.~{Mann}, G.~{Miley},
  H.~{R{\"o}ttgering}, B.W. {Stappers}, R.A.M.J. {Wijers}, S.~{Zaroubi},
  M.~{van den Akker}, A.~{Alexov}, J.~{Anderson}, K.~{Anderson}, A.~{van
  Ardenne}, M.~{Arts}, A.~{Asgekar}, I.M. {Avruch}, F.~{Batejat},
  L.~{B{\"a}hren}, M.E. {Bell}, M.R. {Bell}, I.~{van Bemmel}, P.~{Bennema},
  M.J. {Bentum}, G.~{Bernardi}, P.~{Best}, L.~{B{\^\i}rzan}, A.~{Bonafede},
  A.J. {Boonstra}, R.~{Braun}, J.~{Bregman}, F.~{Breitling}, R.H. {van de
  Brink}, J.~{Broderick}, P.C. {Broekema}, W.N. {Brouw}, M.~{Br{\"u}ggen}, H.R.
  {Butcher}, W.~{van Cappellen}, B.~{Ciardi}, T.~{Coenen}, J.~{Conway},
  A.~{Coolen}, A.~{Corstanje}, S.~{Damstra}, O.~{Davies}, A.T. {Deller}, R.J.
  {Dettmar}, G.~{van Diepen}, K.~{Dijkstra}, P.~{Donker}, A.~{Doorduin},
  J.~{Dromer}, M.~{Drost}, A.~{van Duin}, J.~{Eisl{\"o}ffel}, J.~{van Enst},
  C.~{Ferrari}, W.~{Frieswijk}, H.~{Gankema}, M.A. {Garrett}, F.~{de Gasperin},
  M.~{Gerbers}, E.~{de Geus}, J.M. {Grie{\ss}meier}, T.~{Grit}, P.~{Gruppen},
  J.P. {Hamaker}, T.~{Hassall}, M.~{Hoeft}, H.A. {Holties}, A.~{Horneffer},
  A.~{van der Horst}, A.~{van Houwelingen}, A.~{Huijgen}, M.~{Iacobelli},
  H.~{Intema}, N.~{Jackson}, V.~{Jelic}, A.~{de Jong}, E.~{Juette}, D.~{Kant},
  A.~{Karastergiou}, A.~{Koers}, H.~{Kollen}, V.I. {Kondratiev}, E.~{Kooistra},
  Y.~{Koopman}, A.~{Koster}, M.~{Kuniyoshi}, M.~{Kramer}, G.~{Kuper},
  P.~{Lambropoulos}, C.~{Law}, J.~{van Leeuwen}, J.~{Lemaitre}, M.~{Loose},
  P.~{Maat}, G.~{Macario}, S.~{Markoff}, J.~{Masters}, R.A. {McFadden},
  D.~{McKay-Bukowski}, H.~{Meijering}, H.~{Meulman}, M.~{Mevius},
  E.~{Middelberg}, R.~{Millenaar}, J.C.A. {Miller-Jones}, R.N. {Mohan}, J.D.
  {Mol}, J.~{Morawietz}, R.~{Morganti}, D.D. {Mulcahy}, E.~{Mulder}, H.~{Munk},
  L.~{Nieuwenhuis}, R.~{van Nieuwpoort}, J.E. {Noordam}, M.~{Norden},
  A.~{Noutsos}, A.R. {Offringa}, H.~{Olofsson}, A.~{Omar}, E.~{Orr{\'u}},
  R.~{Overeem}, H.~{Paas}, M.~{Pandey-Pommier}, V.N. {Pandey}, R.~{Pizzo},
  A.~{Polatidis}, D.~{Rafferty}, S.~{Rawlings}, W.~{Reich}, J.P. {de Reijer},
  J.~{Reitsma}, G.A. {Renting}, P.~{Riemers}, E.~{Rol}, J.W. {Romein},
  J.~{Roosjen}, M.~{Ruiter}, A.~{Scaife}, K.~{van der Schaaf}, B.~{Scheers},
  P.~{Schellart}, A.~{Schoenmakers}, G.~{Schoonderbeek}, M.~{Serylak},
  A.~{Shulevski}, J.~{Sluman}, O.~{Smirnov}, C.~{Sobey}, H.~{Spreeuw},
  M.~{Steinmetz}, C.G.M. {Sterks}, H.J. {Stiepel}, K.~{Stuurwold}, M.~{Tagger},
  Y.~{Tang}, C.~{Tasse}, I.~{Thomas}, S.~{Thoudam}, M.C. {Toribio}, B.~{van der
  Tol}, O.~{Usov}, M.~{van Veelen}, A.J. {van der Veen}, S.~{ter Veen}, J.P.W.
  {Verbiest}, R.~{Vermeulen}, N.~{Vermaas}, C.~{Vocks}, C.~{Vogt}, M.~{de Vos},
  E.~{van der Wal}, R.~{van Weeren}, H.~{Weggemans}, P.~{Weltevrede},
  S.~{White}, S.J. {Wijnholds}, T.~{Wilhelmsson}, O.~{Wucknitz},
  S.~{Yatawatta}, P.~{Zarka}, A.~{Zensus}, J.~{van Zwieten}, {LOFAR: The
  LOw-Frequency ARray}.
\newblock \aap \textbf{556}, A2 (2013).
\newblock \doi{10.1051/0004-6361/201220873}.
\newblock {\href{https://arxiv.org/abs/1305.3550}{{arXiv:1305.3550}}}
  {[astro-ph.IM]}

\bibitem{askap_science}
S.~{Johnston}, M.~{Bailes}, N.~{Bartel}, C.~{Baugh}, M.~{Bietenholz},
  C.~{Blake}, R.~{Braun}, J.~{Brown}, S.~{Chatterjee}, J.~{Darling},
  A.~{Deller}, R.~{Dodson}, P.G. {Edwards}, R.~{Ekers}, S.~{Ellingsen},
  I.~{Feain}, B.M. {Gaensler}, M.~{Haverkorn}, G.~{Hobbs}, A.~{Hopkins},
  C.~{Jackson}, C.~{James}, G.~{Joncas}, V.~{Kaspi}, V.~{Kilborn},
  B.~{Koribalski}, R.~{Kothes}, T.L. {Landecker}, E.~{Lenc}, J.~{Lovell}, J.P.
  {Macquart}, R.~{Manchester}, D.~{Matthews}, N.M. {McClure-Griffiths},
  R.~{Norris}, U.L. {Pen}, C.~{Phillips}, C.~{Power}, R.~{Protheroe},
  E.~{Sadler}, B.~{Schmidt}, I.~{Stairs}, L.~{Staveley-Smith}, J.~{Stil},
  R.~{Taylor}, S.~{Tingay}, A.~{Tzioumis}, M.~{Walker}, J.~{Wall},
  M.~{Wolleben}, {Science with the Australian Square Kilometre Array
  Pathfinder}.
\newblock \pasa \textbf{24}(4), 174--188 (2007).
\newblock \doi{10.1071/AS07033}.
\newblock {\href{https://arxiv.org/abs/0711.2103}{{arXiv:0711.2103}}}
  {[astro-ph]}

\bibitem{icecube}
M.G. {Aartsen}, M.~{Ackermann}, J.~{Adams}, J.A. {Aguilar}, M.~{Ahlers},
  M.~{Ahrens}, D.~{Altmann}, K.~{Andeen}, T.~{Anderson}, I.~{Ansseau},
  G.~{Anton}, M.~{Archinger}, C.~{Arg{\"u}elles}, R.~{Auer}, J.~{Auffenberg},
  S.~{Axani}, J.~{Baccus}, X.~{Bai}, S.~{Barnet}, S.W. {Barwick}, V.~{Baum},
  R.~{Bay}, K.~{Beattie}, J.J. {Beatty}, J.~{Becker Tjus}, K.H. {Becker},
  T.~{Bendfelt}, S.~{BenZvi}, D.~{Berley}, E.~{Bernardini}, A.~{Bernhard}, D.Z.
  {Besson}, G.~{Binder}, D.~{Bindig}, M.~{Bissok}, E.~{Blaufuss}, S.~{Blot},
  D.~{Boersma}, C.~{Bohm}, M.~{B{\"o}rner}, F.~{Bos}, D.~{Bose},
  S.~{B{\"o}ser}, O.~{Botner}, A.~{Bouchta}, J.~{Braun}, L.~{Brayeur}, H.P.
  {Bretz}, S.~{Bron}, A.~{Burgman}, C.~{Burreson}, T.~{Carver}, M.~{Casier},
  E.~{Cheung}, D.~{Chirkin}, A.~{Christov}, K.~{Clark}, L.~{Classen},
  S.~{Coenders}, G.H. {Collin}, J.M. {Conrad}, D.F. {Cowen}, R.~{Cross},
  C.~{Day}, M.~{Day}, J.P.A.M. {de Andr{\'e}}, C.~{De Clercq}, E.~{del Pino
  Rosendo}, H.~{Dembinski}, S.~{De Ridder}, F.~{Descamps}, P.~{Desiati}, K.D.
  {de Vries}, G.~{de Wasseige}, M.~{de With}, T.~{DeYoung}, J.C.
  {D{\'\i}az-V{\'e}lez}, V.~{di Lorenzo}, H.~{Dujmovic}, J.P. {Dumm},
  M.~{Dunkman}, B.~{Eberhardt}, W.R. {Edwards}, T.~{Ehrhardt}, B.~{Eichmann},
  P.~{Eller}, S.~{Euler}, P.A. {Evenson}, S.~{Fahey}, A.R. {Fazely},
  J.~{Feintzeig}, J.~{Felde}, K.~{Filimonov}, C.~{Finley}, S.~{Flis}, C.C.
  {F{\"o}sig}, A.~{Franckowiak}, M.~{Fr{\`e}re}, E.~{Friedman}, T.~{Fuchs},
  T.K. {Gaisser}, J.~{Gallagher}, L.~{Gerhardt}, K.~{Ghorbani}, W.~{Giang},
  L.~{Gladstone}, T.~{Glauch}, D.~{Glowacki}, T.~{Gl{\"u}senkamp},
  A.~{Goldschmidt}, J.G. {Gonzalez}, D.~{Grant}, Z.~{Griffith},
  L.~{Gustafsson}, C.~{Haack}, A.~{Hallgren}, F.~{Halzen}, E.~{Hansen},
  T.~{Hansmann}, K.~{Hanson}, J.~{Haugen}, D.~{Hebecker}, D.~{Heereman},
  K.~{Helbing}, R.~{Hellauer}, R.~{Heller}, S.~{Hickford}, J.~{Hignight}, G.C.
  {Hill}, K.D. {Hoffman}, R.~{Hoffmann}, K.~{Hoshina}, F.~{Huang}, M.~{Huber},
  P.O. {Hulth}, K.~{Hultqvist}, S.~{In}, M.~{Inaba}, A.~{Ishihara},
  E.~{Jacobi}, J.~{Jacobsen}, G.S. {Japaridze}, M.~{Jeong}, K.~{Jero},
  A.~{Jones}, B.J.P. {Jones}, J.~{Joseph}, W.~{Kang}, A.~{Kappes}, T.~{Karg},
  A.~{Karle}, U.~{Katz}, M.~{Kauer}, A.~{Keivani}, J.L. {Kelley}, J.~{Kemp},
  A.~{Kheirandish}, J.~{Kim}, M.~{Kim}, T.~{Kintscher}, J.~{Kiryluk},
  N.~{Kitamura}, T.~{Kittler}, S.R. {Klein}, S.~{Kleinfelder}, M.~{Kleist},
  G.~{Kohnen}, R.~{Koirala}, H.~{Kolanoski}, R.~{Konietz}, L.~{K{\"o}pke},
  C.~{Kopper}, S.~{Kopper}, D.J. {Koskinen}, M.~{Kowalski}, M.~{Krasberg},
  K.~{Krings}, M.~{Kroll}, G.~{Kr{\"u}ckl}, C.~{Kr{\"u}ger}, J.~{Kunnen},
  S.~{Kunwar}, N.~{Kurahashi}, T.~{Kuwabara}, M.~{Labare}, K.~{Laihem},
  H.~{Landsman}, J.L. {Lanfranchi}, M.J. {Larson}, F.~{Lauber}, A.~{Laundrie},
  D.~{Lennarz}, H.~{Leich}, M.~{Lesiak-Bzdak}, M.~{Leuermann}, L.~{Lu},
  J.~{Ludwig}, J.~{L{\"u}nemann}, C.~{Mackenzie}, J.~{Madsen}, G.~{Maggi},
  K.B.M. {Mahn}, S.~{Mancina}, M.~{Mandelartz}, R.~{Maruyama}, K.~{Mase},
  H.~{Matis}, R.~{Maunu}, F.~{McNally}, C.P. {McParland}, P.~{Meade},
  K.~{Meagher}, M.~{Medici}, M.~{Meier}, A.~{Meli}, T.~{Menne}, G.~{Merino},
  T.~{Meures}, S.~{Miarecki}, R.H. {Minor}, T.~{Montaruli}, M.~{Moulai},
  T.~{Murray}, R.~{Nahnhauer}, U.~{Naumann}, G.~{Neer}, M.~{Newcomb},
  H.~{Niederhausen}, S.C. {Nowicki}, D.R. {Nygren}, A.~{Obertacke Pollmann},
  A.~{Olivas}, A.~{O'Murchadha}, T.~{Palczewski}, H.~{Pandya}, D.V. {Pankova},
  S.~{Patton}, P.~{Peiffer}, {\"O}.~{Penek}, J.A. {Pepper}, C.~{P{\'e}rez de
  los Heros}, C.~{Pettersen}, D.~{Pieloth}, E.~{Pinat}, P.B. {Price}, G.T.
  {Przybylski}, M.~{Quinnan}, C.~{Raab}, L.~{R{\"a}del}, M.~{Rameez},
  K.~{Rawlins}, R.~{Reimann}, B.~{Relethford}, M.~{Relich}, E.~{Resconi},
  W.~{Rhode}, M.~{Richman}, B.~{Riedel}, S.~{Robertson}, M.~{Rongen},
  C.~{Roucelle}, C.~{Rott}, T.~{Ruhe}, D.~{Ryckbosch}, D.~{Rysewyk},
  L.~{Sabbatini}, S.E. {Sanchez Herrera}, A.~{Sandrock}, J.~{Sandroos},
  P.~{Sandstrom}, S.~{Sarkar}, K.~{Satalecka}, P.~{Schlunder}, T.~{Schmidt},
  S.~{Schoenen}, S.~{Sch{\"o}neberg}, A.~{Schukraft}, L.~{Schumacher},
  D.~{Seckel}, S.~{Seunarine}, M.~{Solarz}, D.~{Soldin}, M.~{Song}, G.M.
  {Spiczak}, C.~{Spiering}, T.~{Stanev}, A.~{Stasik}, J.~{Stettner},
  A.~{Steuer}, T.~{Stezelberger}, R.G. {Stokstad}, A.~{St{\"o}{\ss}l},
  R.~{Str{\"o}m}, N.L. {Strotjohann}, K.H. {Sulanke}, G.W. {Sullivan},
  M.~{Sutherland}, H.~{Taavola}, I.~{Taboada}, J.~{Tatar}, F.~{Tenholt},
  S.~{Ter-Antonyan}, A.~{Terliuk}, G.~{Te{\v{s}}i{\'c}}, L.~{Thollander},
  S.~{Tilav}, P.A. {Toale}, M.N. {Tobin}, S.~{Toscano}, D.~{Tosi},
  M.~{Tselengidou}, A.~{Turcati}, E.~{Unger}, M.~{Usner}, J.~{Vandenbroucke},
  N.~{van Eijndhoven}, S.~{Vanheule}, M.~{van Rossem}, J.~{van Santen},
  M.~{Vehring}, M.~{Voge}, E.~{Vogel}, M.~{Vraeghe}, D.~{Wahl}, C.~{Walck},
  A.~{Wallace}, M.~{Wallraff}, N.~{Wandkowsky}, C.~{Weaver}, M.J. {Weiss},
  C.~{Wendt}, S.~{Westerhoff}, D.~{Wharton}, B.J. {Whelan}, S.~{Wickmann},
  K.~{Wiebe}, C.H. {Wiebusch}, L.~{Wille}, D.R. {Williams}, L.~{Wills},
  P.~{Wisniewski}, M.~{Wolf}, T.R. {Wood}, E.~{Woolsey}, K.~{Woschnagg}, D.L.
  {Xu}, X.W. {Xu}, Y.~{Xu}, J.P. {Yanez}, G.~{Yodh}, S.~{Yoshida}, M.~{Zoll},
  {The IceCube Neutrino Observatory: instrumentation and online systems}.
\newblock Journal of Instrumentation \textbf{12}(3), P03,012 (2017).
\newblock \doi{10.1088/1748-0221/12/03/P03012}.
\newblock {\href{https://arxiv.org/abs/1612.05093}{{arXiv:1612.05093}}}
  {[astro-ph.IM]}

\bibitem{2020JHEAp..27....1L}
Q.~{Luo}, J.Y. {Liao}, X.F. {Li}, G.~{Li}, J.~{Zhang}, C.Z. {Liu}, X.B. {Li},
  Y.~{Zhu}, C.K. {Li}, Y.~{Huang}, M.Y. {Ge}, Y.P. {Xu}, Z.W. {Li}, C.~{Cai},
  S.~{Xiao}, Q.B. {Yi}, Y.F. {Zhang}, S.L. {Xiong}, S.~{Zhang}, S.N. {Zhang},
  {Calibration of the instrumental response of Insight-HXMT/HE CsI detectors
  for gamma-ray monitoring}.
\newblock Journal of High Energy Astrophysics \textbf{27}, 1--13 (2020).
\newblock \doi{10.1016/j.jheap.2020.04.004}.
\newblock {\href{https://arxiv.org/abs/2005.01367}{{arXiv:2005.01367}}}
  {[astro-ph.IM]}

\bibitem{Sakamoto2011}
T.~{Sakamoto}, S.D. {Barthelmy}, W.H. {Baumgartner}, J.R. {Cummings}, E.E.
  {Fenimore}, N.~{Gehrels}, H.A. {Krimm}, C.B. {Markwardt}, D.M. {Palmer}, A.M.
  {Parsons}, G.~{Sato}, M.~{Stamatikos}, J.~{Tueller}, T.N. {Ukwatta},
  B.~{Zhang}, {The Second Swift Burst Alert Telescope Gamma-Ray Burst Catalog}.
\newblock \apjs \textbf{195}(1), 2 (2011).
\newblock \doi{10.1088/0067-0049/195/1/2}.
\newblock {\href{https://arxiv.org/abs/1104.4689}{{arXiv:1104.4689}}}
  {[astro-ph.HE]}

\bibitem{Kilpatrick2017}
C.D. Kilpatrick, R.J. Foley, D.~Kasen, A.~Murguia-Berthier, E.~Ramirez-Ruiz,
  D.A. Coulter, M.R. Drout, A.L. Piro, B.J. Shappee, K.~Boutsia, C.~Contreras,
  F.D. Mille, B.F. Madore, N.~Morrell, Y.C. Pan, J.X. Prochaska, A.~Rest,
  C.~Rojas-Bravo, M.R. Siebert, J.D. Simon, N.~Ulloa, Electromagnetic evidence
  that sss17a is the result of a binary neutron star merger.
\newblock Science \textbf{358}, 1583--1587 (2017).
\newblock \doi{10.1126/science.aaq0073}.
\newblock
  \urlprefix\url{http://www.sciencemag.org/lookup/doi/10.1126/science.aaq0073
  http://arxiv.org/abs/1710.05434 http://dx.doi.org/10.1126/science.aaq0073}

\bibitem{Mooley2018}
K.P. Mooley, D.A. Frail, D.~Dobie, E.~Lenc, A.~Corsi, K.~De, A.J. Nayana,
  S.~Makhathini, I.~Heywood, T.~Murphy, D.L. Kaplan, P.~Chandra, O.~Smirnov,
  E.~Nakar, G.~Hallinan, F.~Camilo, R.~Fender, S.~Goedhart, P.~Groot, M.M.
  Kasliwal, S.R. Kulkarni, P.A. Woudt, A strong jet signature in the late-time
  light curve of gw170817.
\newblock The Astrophysical Journal \textbf{868}, L11 (2018).
\newblock \doi{10.3847/2041-8213/aaeda7}

\bibitem{2018Natur.561..355M}
K.P. Mooley, A.T. Deller, O.~Gottlieb, E.~Nakar, G.~Hallinan, S.~Bourke, D.A.
  Frail, A.~Horesh, A.~Corsi, K.~Hotokezaka, Superluminal motion of a
  relativistic jet in the neutron-star merger gw170817.
\newblock \nat \textbf{561}, 355--359 (2018).
\newblock \doi{10.1038/s41586-018-0486-3}

\bibitem{teg+19}
E.~Troja, H.~van Eerten, G.~Ryan, R.~Ricci, J.M. Burgess, M.~Wieringa, L.~Piro,
  S.B. Cenko, T.~Sakamoto, A year in the life of gw170817: the rise and fall of
  a structured jet from a binary neutron star merger.
\newblock Monthly Notices of the Royal Astronomical Society p. 2169 (2018).
\newblock \doi{10.1093/mnras/stz2248}.
\newblock
  \urlprefix\url{https://academic.oup.com/mnras/advance-article/doi/10.1093/mnras/stz2248/5549529
  http://arxiv.org/abs/1808.06617 http://dx.doi.org/10.1093/mnras/stz2248}

\bibitem{Goldstein2017}
A.~Goldstein, P.~Veres, E.~Burns, M.S. Briggs, R.~Hamburg, D.~Kocevski, C.A.
  Wilson-Hodge, R.D. Preece, S.~Poolakkil, O.J. Roberts, C.M. Hui,
  V.~Connaughton, J.~Racusin, A.~von Kienlin, T.D. Canton, N.~Christensen,
  T.~Littenberg, K.~Siellez, L.~Blackburn, J.~Broida, E.~Bissaldi, W.H.
  Cleveland, M.H. Gibby, M.M. Giles, R.M. Kippen, S.~McBreen, J.~McEnery, C.A.
  Meegan, W.S. Paciesas, M.~Stanbro, An ordinary short gamma-ray burst with
  extraordinary implications: Fermi -gbm detection of grb 170817a.
\newblock The Astrophysical Journal \textbf{848}, L14 (2017).
\newblock \doi{10.3847/2041-8213/aa8f41}.
\newblock
  \urlprefix\url{https://iopscience.iop.org/article/10.3847/2041-8213/aa8f41}

\bibitem{Kasen2017}
D.~Kasen, B.~Metzger, J.~Barnes, E.~Quataert, E.~Ramirez-Ruiz, Origin of the
  heavy elements in binary neutron-star mergers from a gravitational wave
  event.
\newblock Nature \textbf{551}, 80--84 (2017).
\newblock \doi{10.1038/nature24453}.
\newblock \urlprefix\url{http://arxiv.org/abs/1710.05463
  http://dx.doi.org/10.1038/nature24453}

\bibitem{2017Sci...358.1559K}
M.M. Kasliwal, E.~Nakar, L.P. Singer, D.L. Kaplan, D.O. Cook, A.V. Sistine,
  R.M. Lau, C.~Fremling, O.~Gottlieb, J.E. Jencson, S.M. Adams, U.~Feindt,
  K.~Hotokezaka, S.~Ghosh, D.A. Perley, P.C. Yu, T.~Piran, J.R. Allison, G.C.
  Anupama, A.~Balasubramanian, K.W. Bannister, J.~Bally, J.~Barnes, S.~Barway,
  E.~Bellm, V.~Bhalerao, D.~Bhattacharya, N.~Blagorodnova, J.S. Bloom, P.R.
  Brady, C.~Cannella, D.~Chatterjee, S.B. Cenko, B.E. Cobb, C.~Copperwheat,
  A.~Corsi, K.~De, D.~Dobie, S.W.K. Emery, P.A. Evans, O.D. Fox, D.A. Frail,
  C.~Frohmaier, A.~Goobar, G.~Hallinan, F.~Harrison, G.~Helou, T.~Hinderer,
  A.Y.Q. Ho, A.~Horesh, W.H. Ip, R.~Itoh, D.~Kasen, H.~Kim, N.P.M. Kuin,
  T.~Kupfer, C.~Lynch, K.~Madsen, P.A. Mazzali, A.A. Miller, K.~Mooley,
  T.~Murphy, C.C. Ngeow, D.~Nichols, S.~Nissanke, P.~Nugent, E.O. Ofek, H.~Qi,
  R.M. Quimby, S.~Rosswog, F.~Rusu, E.M. Sadler, P.~Schmidt, J.~Sollerman,
  I.~Steele, A.R. Williamson, Y.~Xu, L.~Yan, Y.~Yatsu, C.~Zhang, W.~Zhao,
  Illuminating gravitational waves: A concordant picture of photons from a
  neutron star merger.
\newblock Science \textbf{358}, 1559--1565 (2017).
\newblock \doi{10.1126/science.aap9455}

\bibitem{2017Natur.551...85A}
B.P. Abbott, R.~Abbott, T.D. Abbott, F.~Acernese, K.~Ackley, C.~Adams,
  T.~Adams, P.~Addesso, R.X. Adhikari, V.B. Adya, C.~Affeldt, M.~Afrough,
  B.~Agarwal, M.~Agathos, K.~Agatsuma, N.~Aggarwal, O.D. Aguiar, L.~Aiello,
  A.~Ain, P.~Ajith, R.~Podesta, H.~Levato, C.~Saffe, D.A.H. Buckley, N.M.
  Budnev, O.~Gress, V.~Yurkov, R.~Rebolo, M.~Serra-Ricart, A gravitational-wave
  standard siren measurement of the hubble constant.
\newblock Nature \textbf{551}, 85--88 (2017).
\newblock \doi{10.1038/nature24471}

\bibitem{2018PhRvL.121p1101A}
B.P. Abbott, R.~Abbott, T.D. Abbott, F.~Acernese, K.~Ackley, C.~Adams,
  T.~Adams, P.~Addesso, R.X. Adhikari, V.B. Adya, C.~Affeldt, B.~Agarwal,
  M.~Agathos, K.~Agatsuma, N.~Aggarwal, O.D. Aguiar, L.~Aiello, A.~Ain,
  P.~Ajith, B.~Allen, G.~Allen, A.~Allocca, M.A. Aloy, P.A. Altin, A.~Amato,
  A.~Ananyeva, S.B. Anderson, W.G. Anderson, S.V. Angelova, S.~Antier,
  S.~Appert, K.~Arai, M.C. Araya, J.S. Areeda, M.~Arène, N.~Arnaud, K.G. Arun,
  J.~Zweizig, {GW170817: Measurements of Neutron Star Radii and Equation of
  State}.
\newblock \prl \textbf{121}(16), 161101 (2018).
\newblock \doi{10.1103/PhysRevLett.121.161101}.
\newblock {\href{https://arxiv.org/abs/1805.11581}{{arXiv:1805.11581}}}
  {[gr-qc]}

\bibitem{2019PhRvX...9a1001A}
B.P. Abbott, R.~Abbott, T.D. Abbott, F.~Acernese, K.~Ackley, C.~Adams,
  T.~Adams, P.~Addesso, R.X. Adhikari, V.B. Adya, C.~Affeldt, B.~Agarwal,
  M.~Agathos, K.~Agatsuma, N.~Aggarwal, O.D. Aguiar, L.~Aiello, A.~Ain,
  P.~Ajith, B.~Allen, G.~Allen, A.~Allocca, M.A. Aloy, P.A. Altin, A.~Amato,
  A.~Ananyeva, S.B. Anderson, J.~Zweizig, Properties of the binary neutron star
  merger gw170817.
\newblock Physical Review X \textbf{9}, 011,001 (2019).
\newblock \doi{10.1103/PhysRevX.9.011001}.
\newblock
  \urlprefix\url{https://ui.adsabs.harvard.edu/abs/2019PhRvX...9a1001A/abstract}

\bibitem{2018Natur.554..207M}
K.P. Mooley, E.~Nakar, K.~Hotokezaka, G.~Hallinan, A.~Corsi, D.A. Frail,
  A.~Horesh, T.~Murphy, E.~Lenc, D.L. Kaplan, K.~de, D.~Dobie, P.C. ra,
  A.~Deller, O.~Gottlieb, M.M. Kasliwal, S.R. Kulkarni, S.T. Myers,
  S.~Nissanke, T.~Piran, C.~Lynch, V.~Bhalerao, S.~Bourke, K.W. Bannister, L.P.
  Singer, A mildly relativistic wide-angle outflow in the neutron-star merger
  event gw170817.
\newblock \nat \textbf{554}, 207--210 (2018).
\newblock \doi{10.1038/nature25452}

\bibitem{2019MNRAS.482.5430B}
P.~{Beniamini}, E.~{Nakar}, {Observational constraints on the structure of
  gamma-ray burst jets}.
\newblock \mnras \textbf{482}(4), 5430--5440 (2019).
\newblock \doi{10.1093/mnras/sty3110}.
\newblock {\href{https://arxiv.org/abs/1808.07493}{{arXiv:1808.07493}}}
  {[astro-ph.HE]}

\bibitem{Saleem:2019wdv}
M.~{Saleem}, L.~{Resmi}, K.G. {Arun}, S.~{Mohan}, {On the Energetics of a
  Possible Relativistic Jet Associated with the Binary Neutron Star Merger
  Candidate S190425z}.
\newblock \apj \textbf{891}(2), 130 (2020).
\newblock \doi{10.3847/1538-4357/ab6731}.
\newblock {\href{https://arxiv.org/abs/1905.00337}{{arXiv:1905.00337}}}
  {[astro-ph.HE]}

\bibitem{S190425z}
G.~{Hosseinzadeh}, P.S. {Cowperthwaite}, S.~{Gomez}, V.A. {Villar},
  M.~{Nicholl}, R.~{Margutti}, E.~{Berger}, R.~{Chornock}, K.~{Paterson},
  W.~{Fong}, V.~{Savchenko}, P.~{Short}, K.D. {Alexander}, P.K. {Blanchard},
  J.~{Braga}, M.L. {Calkins}, R.~{Cartier}, D.L. {Coppejans}, T.~{Eftekhari},
  T.~{Laskar}, C.~{Ly}, L.~{Patton}, I.~{Pelisoli}, D.E. {Reichart},
  G.~{Terreran}, P.K.G. {Williams}, {Follow-up of the Neutron Star Bearing
  Gravitational-wave Candidate Events S190425z and S190426c with MMT and SOAR}.
\newblock \apjl \textbf{880}(1), L4 (2019).
\newblock \doi{10.3847/2041-8213/ab271c}.
\newblock {\href{https://arxiv.org/abs/1905.02186}{{arXiv:1905.02186}}}
  {[astro-ph.HE]}

\bibitem{coughlin2020implications}
M.W. {Coughlin}, T.~{Dietrich}, S.~{Antier}, M.~{Almualla}, S.~{Anand},
  M.~{Bulla}, F.~{Foucart}, N.~{Guessoum}, K.~{Hotokezaka}, V.~{Kumar},
  G.~{Raaijmakers}, S.~{Nissanke}, {Implications of the search for optical
  counterparts during the second part of the Advanced LIGO's and Advanced
  Virgo's third observing run: lessons learned for future follow-up
  observations}.
\newblock \mnras \textbf{497}(1), 1181--1196 (2020).
\newblock \doi{10.1093/mnras/staa1925}.
\newblock {\href{https://arxiv.org/abs/2006.14756}{{arXiv:2006.14756}}}
  {[astro-ph.HE]}

\bibitem{Evans17}
P.A. {Evans}, S.B. {Cenko}, J.A. {Kennea}, S.W.K. {Emery}, N.P.M. {Kuin},
  O.~{Korobkin}, R.T. {Wollaeger}, C.L. {Fryer}, K.K. {Madsen}, F.A.
  {Harrison}, Y.~{Xu}, E.~{Nakar}, K.~{Hotokezaka}, A.~{Lien}, S.~{Campana},
  S.R. {Oates}, E.~{Troja}, A.A. {Breeveld}, F.E. {Marshall}, S.D. {Barthelmy},
  A.P. {Beardmore}, D.N. {Burrows}, G.~{Cusumano}, A.~{D'A{\`\i}},
  P.~{D'Avanzo}, V.~{D'Elia}, M.~{de Pasquale}, W.P. {Even}, C.J. {Fontes},
  K.~{Forster}, J.~{Garcia}, P.~{Giommi}, B.~{Grefenstette}, C.~{Gronwall},
  D.H. {Hartmann}, M.~{Heida}, A.L. {Hungerford}, M.M. {Kasliwal}, H.A.
  {Krimm}, A.J. {Levan}, D.~{Malesani}, A.~{Melandri}, H.~{Miyasaka}, J.A.
  {Nousek}, P.T. {O'Brien}, J.P. {Osborne}, C.~{Pagani}, K.L. {Page}, D.M.
  {Palmer}, M.~{Perri}, S.~{Pike}, J.L. {Racusin}, S.~{Rosswog}, M.H. {Siegel},
  T.~{Sakamoto}, B.~{Sbarufatti}, G.~{Tagliaferri}, N.R. {Tanvir},
  A.~{Tohuvavohu}, {Swift and NuSTAR observations of GW170817: Detection of a
  blue kilonova}.
\newblock Science \textbf{358}(6370), 1565--1570 (2017).
\newblock \doi{10.1126/science.aap9580}.
\newblock {\href{https://arxiv.org/abs/1710.05437}{{arXiv:1710.05437}}}
  {[astro-ph.HE]}

\bibitem{Begue}
D.~{B{\'e}gu{\'e}}, J.M. {Burgess}, J.~{Greiner}, {The Peculiar Physics of GRB
  170817A and Their Implications for Short GRBs}.
\newblock \apjl \textbf{851}(1), L19 (2017).
\newblock \doi{10.3847/2041-8213/aa9d85}.
\newblock {\href{https://arxiv.org/abs/1710.07987}{{arXiv:1710.07987}}}
  {[astro-ph.HE]}

\bibitem{kentaP}
K.~Hotokezaka, P.~Beniamini, T.~Piran, Neutron star mergers as sites of
  r-process nucleosynthesis and short gamma-ray bursts.
\newblock International Journal of Modern Physics D  (2018).
\newblock \doi{10.1142/S0218271818420051}

\bibitem{2019MNRAS.483..840B}
P.~{Beniamini}, M.~{Petropoulou}, R.~{Barniol Duran}, D.~{Giannios}, {A lesson
  from GW170817: most neutron star mergers result in tightly collimated
  successful GRB jets}.
\newblock \mnras \textbf{483}(1), 840--851 (2019).
\newblock \doi{10.1093/mnras/sty3093}.
\newblock {\href{https://arxiv.org/abs/1808.04831}{{arXiv:1808.04831}}}
  {[astro-ph.HE]}

\bibitem{Abbott2020}
B.P. Abbott, R.~Abbott, T.D. Abbott, M.R. Abernathy, F.~Acernese, K.~Ackley,
  C.~Adams, T.~Adams, P.~Addesso, et~al., Prospects for observing and
  localizing gravitational-wave transients with advanced ligo, advanced virgo
  and kagra.
\newblock Living Reviews in Relativity \textbf{21}(1) (2018).
\newblock \doi{10.1007/s41114-018-0012-9}.
\newblock \urlprefix\url{http://dx.doi.org/10.1007/s41114-018-0012-9}

\bibitem{aso}
Y.~{Aso}, Y.~{Michimura}, K.~{Somiya}, M.~{Ando}, O.~{Miyakawa},
  T.~{Sekiguchi}, D.~{Tatsumi}, H.~{Yamamoto}, {Interferometer design of the
  KAGRA gravitational wave detector}.
\newblock \prd \textbf{88}(4), 043007 (2013).
\newblock \doi{10.1103/PhysRevD.88.043007}.
\newblock {\href{https://arxiv.org/abs/1306.6747}{{arXiv:1306.6747}}} {[gr-qc]}

\bibitem{Mills18}
C.~Mills, V.~Tiwari, S.~Fairhurst, Localization of binary neutron star mergers
  with second and third generation gravitational-wave detectors.
\newblock Physical Review D \textbf{97}(10) (2018).
\newblock \doi{10.1103/physrevd.97.104064}.
\newblock \urlprefix\url{http://dx.doi.org/10.1103/PhysRevD.97.104064}

\bibitem{2020LRR....23....3A}
B.P. {Abbott}, R.~{Abbott}, T.D. {Abbott}, S.~{Abraham}, F.~{Acernese},
  K.~{Ackley}, C.~{Adams}, V.B. {Adya}, C.~{Affeldt}, M.~{Agathos},
  K.~{Agatsuma}, N.~{Aggarwal}, O.D. {Aguiar}, L.~{Aiello}, A.~{Ain},
  P.~{Ajith}, T.~{Akutsu}, G.~{Allen}, A.~{Allocca}, M.A. {Aloy}, P.A. {Altin},
  A.~{Amato}, A.~{Ananyeva}, S.B. {Anderson}, W.G. {Anderson}, M.~{Ando}, S.V.
  {Angelova}, S.~{Antier}, S.~{Appert}, K.~{Arai}, K.~{Arai}, Y.~{Arai},
  S.~{Araki}, A.~{Araya}, M.C. {Araya}, J.S. {Areeda}, M.~{Ar{\`e}ne},
  N.~{Aritomi}, N.~{Arnaud}, K.G. {Arun}, S.~{Ascenzi}, G.~{Ashton}, Y.~{Aso},
  S.M. {Aston}, P.~{Astone}, F.~{Aubin}, P.~{Aufmuth}, J.~{Worden}, J.L.
  {Wright}, C.M. {Wu}, D.S. {Wu}, H.C. {Wu}, S.R. {Wu}, D.M. {Wysocki},
  L.~{Xiao}, W.R. {Xu}, T.~{Yamada}, H.~{Yamamoto}, K.~{Yamamoto},
  K.~{Yamamoto}, T.~{Yamamoto}, C.C. {Yancey}, L.~{Yang}, M.J. {Yap},
  M.~{Yazback}, D.W. {Yeeles}, K.~{Yokogawa}, J.~{Yokoyama}, T.~{Yokozawa},
  T.~{Yoshioka}, H.~{Yu}, H.~{Yu}, S.H.R. {Yuen}, H.~{Yuzurihara}, M.~{Yvert},
  A.K. {Zadro{\.z}ny}, M.~{Zanolin}, S.~{Zeidler}, T.~{Zelenova}, J.P.
  {Zendri}, M.~{Zevin}, J.~{Zhang}, L.~{Zhang}, T.~{Zhang}, C.~{Zhao},
  Y.~{Zhao}, M.~{Zhou}, Z.~{Zhou}, X.J. {Zhu}, Z.H. {Zhu}, A.B. {Zimmerman},
  M.E. {Zucker}, J.~{Zweizig}, L.S.C. {Kagra Collaboration}, {VIRGO
  Collaboration}, {Prospects for observing and localizing gravitational-wave
  transients with Advanced LIGO, Advanced Virgo and KAGRA}.
\newblock Living Reviews in Relativity \textbf{23}(1), 3 (2020).
\newblock \doi{10.1007/s41114-020-00026-9}

\bibitem{burns2019gamma}
E.~{Burns}, S.~{Zhu}, C.M. {Hui}, S.~{Ansoldi}, S.~{Barthelmy}, S.~{Boggs},
  S.B. {Cenko}, N.~{Christensen}, C.~{Fryer}, A.~{Goldstein}, A.~{Harding},
  D.~{Hartmann}, A.~{Joens}, G.~{Kanbach}, M.~{Kerr}, C.~{Kierans},
  J.~{McEnery}, B.~{Patricelli}, J.~{Perkins}, J.~{Racusin}, P.~{Ray},
  J.~{Schlieder}, H.~{Schoorlemmer}, F.~{Schussler}, A.~{Stamerra},
  J.~{Tomsick}, Z.~{Wadiasingh}, C.~{Wilson-Hodge}, G.~{Younes}, B.~{Zhang},
  A.~{Zoglauer}, {Gamma Rays and Gravitational Waves}.
\newblock \baas \textbf{51}(3), 260 (2019).
\newblock {\href{https://arxiv.org/abs/1903.04472}{{arXiv:1903.04472}}}
  {[astro-ph.HE]}

\bibitem{2022ApJ...924...54P}
P.~{Petrov}, L.P. {Singer}, M.W. {Coughlin}, V.~{Kumar}, M.~{Almualla},
  S.~{Anand}, M.~{Bulla}, T.~{Dietrich}, F.~{Foucart}, N.~{Guessoum},
  {Data-driven Expectations for Electromagnetic Counterpart Searches Based on
  LIGO/Virgo Public Alerts}.
\newblock \apj \textbf{924}(2), 54 (2022).
\newblock \doi{10.3847/1538-4357/ac366d}.
\newblock {\href{https://arxiv.org/abs/2108.07277}{{arXiv:2108.07277}}}
  {[astro-ph.HE]}

\bibitem{astro2020}
{National Academies of Sciences Engineering and Medicine}, \emph{Pathways to
  Discovery in Astronomy and Astrophysics for the 2020s} (The National
  Academies Press, Washington, DC, 2021).
\newblock \doi{10.17226/26141}.
\newblock
  \urlprefix\url{https://nap.nationalacademies.org/catalog/26141/pathways-to-discovery-in-astronomy-and-astrophysics-for-the-2020s}

\bibitem{2018MNRAS.479..588G}
O.~{Gottlieb}, E.~{Nakar}, T.~{Piran}, K.~{Hotokezaka}, {A cocoon shock
  breakout as the origin of the {\ensuremath{\gamma}}-ray emission in
  GW170817}.
\newblock \mnras \textbf{479}(1), 588--600 (2018).
\newblock \doi{10.1093/mnras/sty1462}.
\newblock {\href{https://arxiv.org/abs/1710.05896}{{arXiv:1710.05896}}}
  {[astro-ph.HE]}

\bibitem{2020ApJ...897..141B}
A.M. {Beloborodov}, C.~{Lundman}, Y.~{Levin}, {Relativistic Envelopes and
  Gamma-Rays from Neutron Star Mergers}.
\newblock \apj \textbf{897}(2), 141 (2020).
\newblock \doi{10.3847/1538-4357/ab86a0}.
\newblock {\href{https://arxiv.org/abs/1812.11247}{{arXiv:1812.11247}}}
  {[astro-ph.HE]}

\bibitem{2012ApJ...747...88N}
E.~{Nakar}, R.~{Sari}, {Relativistic Shock Breakouts{\textemdash}A Variety of
  Gamma-Ray Flares: From Low-luminosity Gamma-Ray Bursts to Type Ia
  Supernovae}.
\newblock \apj \textbf{747}(2), 88 (2012).
\newblock \doi{10.1088/0004-637X/747/2/88}.
\newblock {\href{https://arxiv.org/abs/1106.2556}{{arXiv:1106.2556}}}
  {[astro-ph.HE]}

\bibitem{2018ApJ...867...95H}
K.~{Hotokezaka}, K.~{Kiuchi}, M.~{Shibata}, E.~{Nakar}, T.~{Piran},
  {Synchrotron Radiation from the Fast Tail of Dynamical Ejecta of Neutron Star
  Mergers}.
\newblock \apj \textbf{867}(2), 95 (2018).
\newblock \doi{10.3847/1538-4357/aadf92}.
\newblock {\href{https://arxiv.org/abs/1803.00599}{{arXiv:1803.00599}}}
  {[astro-ph.HE]}

\bibitem{2018ApJ...869..130R}
D.~{Radice}, A.~{Perego}, K.~{Hotokezaka}, S.A. {Fromm}, S.~{Bernuzzi}, L.F.
  {Roberts}, {Binary Neutron Star Mergers: Mass Ejection, Electromagnetic
  Counterparts, and Nucleosynthesis}.
\newblock \apj \textbf{869}(2), 130 (2018).
\newblock \doi{10.3847/1538-4357/aaf054}.
\newblock {\href{https://arxiv.org/abs/1809.11161}{{arXiv:1809.11161}}}
  {[astro-ph.HE]}

\bibitem{2021ApJ...913L...7A}
R.~{Abbott}, T.D. {Abbott}, S.~{Abraham}, F.~{Acernese}, K.~{Ackley},
  A.~{Adams}, C.~{Adams}, R.X. {Adhikari}, V.B. {Adya}, C.~{Affeldt},
  M.~{Agathos}, C.~{Zhao}, G.~{Zhao}, M.~{Zhou}, Z.~{Zhou}, X.J. {Zhu}, A.B.
  {Zimmerman}, M.E. {Zucker}, J.~{Zweizig}, {LIGO Scientific Collaboration},
  {Virgo Collaboration}, {Population Properties of Compact Objects from the
  Second LIGO-Virgo Gravitational-Wave Transient Catalog}.
\newblock \apjl \textbf{913}(1), L7 (2021).
\newblock \doi{10.3847/2041-8213/abe949}.
\newblock {\href{https://arxiv.org/abs/2010.14533}{{arXiv:2010.14533}}}
  {[astro-ph.HE]}

\bibitem{Ioka19}
K.~{Ioka}, T.~{Nakamura}, {Spectral puzzle of the off-axis gamma-ray burst in
  GW170817}.
\newblock \mnras \textbf{487}(4), 4884--4889 (2019).
\newblock \doi{10.1093/mnras/stz1650}.
\newblock {\href{https://arxiv.org/abs/1903.01484}{{arXiv:1903.01484}}}
  {[astro-ph.HE]}

\bibitem{Kouveliotou93}
C.~{Kouveliotou}, C.A. {Meegan}, G.J. {Fishman}, N.P. {Bhat}, M.S. {Briggs},
  T.M. {Koshut}, W.S. {Paciesas}, G.N. {Pendleton}, {Identification of Two
  Classes of Gamma-Ray Bursts}.
\newblock \apjl \textbf{413}, L101 (1993).
\newblock \doi{10.1086/186969}

\bibitem{Wanderman15}
D.~Wanderman, T.~Piran, The rate, luminosity function and time delay of
  non-collapsar short grbs.
\newblock Monthly Notices of the Royal Astronomical Society \textbf{448}(4),
  3026--3037 (2015).
\newblock \doi{10.1093/mnras/stv123}.
\newblock \urlprefix\url{https://doi.org/10.1093/mnras/stv123}.
\newblock
  {\href{https://arxiv.org/abs/https://academic.oup.com/mnras/article-pdf/448/4/3026/2810182/stv123.pdf}{{https://academic.oup.com/mnras/article-pdf/448/4/3026/2810182/stv123.pdf}}}

\bibitem{2021ApJ...915...86A}
R.~{Abbott}, T.D. {Abbott}, S.~{Abraham}, F.~{Acernese}, K.~{Ackley},
  C.~{Adams}, R.X. {Adhikari}, V.B. {Adya}, C.~{Affeldt}, M.~{Agathos},
  K.~{Agatsuma}, N.~{Aggarwal}, O.D. {Aguiar}, A.~{Aich}, L.~{Aiello},
  A.~{Ain}, P.~{Ajith}, G.~{Allen}, A.~{Allocca}, P.A. {Altin}, A.~{Amato},
  S.~{Anand}, A.~{Ananyeva}, S.B. {Anderson}, W.G. {Anderson}, S.V. {Angelova},
  S.~{Ansoldi}, S.~{Antier}, S.~{Appert}, K.~{Arai}, M.C. {Araya}, J.S.
  {Areeda}, M.~{Ar{\`e}ne}, N.~{Arnaud}, S.M. {Aronson}, Y.~{Asali},
  S.~{Ascenzi}, G.~{Ashton}, M.~{Assiduo}, S.M. {Aston}, P.~{Astone},
  F.~{Aubin}, P.~{Aufmuth}, K.~{AultONeal}, C.~{Austin}, V.~{Avendano},
  S.~{Babak}, P.~{Bacon}, F.~{Badaracco}, M.K.M. {Bader}, S.~{Bae}, A.M.
  {Baer}, J.~{Baird}, F.~{Baldaccini}, G.~{Ballardin}, S.W. {Ballmer},
  A.~{Bals}, A.~{Balsamo}, M.E. {Zucker}, J.~{Zweizig}, {LIGO Scientific
  Collaboration}, {Virgo Collaboration}, {Search for Gravitational Waves
  Associated with Gamma-Ray Bursts Detected by Fermi and Swift during the
  LIGO-Virgo Run O3a}.
\newblock \apj \textbf{915}(2), 86 (2021).
\newblock \doi{10.3847/1538-4357/abee15}.
\newblock {\href{https://arxiv.org/abs/2010.14550}{{arXiv:2010.14550}}}
  {[astro-ph.HE]}

\bibitem{2022ApJ...928..186A}
R.~{Abbott}, T.D. {Abbott}, F.~{Acernese}, K.~{Ackley}, C.~{Adams},
  N.~{Adhikari}, R.X. {Adhikari}, V.B. {Adya}, C.~{Affeldt}, D.~{Agarwal},
  M.~{Agathos}, K.~{Agatsuma}, N.~{Aggarwal}, O.D. {Aguiar}, L.~{Aiello},
  A.~{Ain}, P.~{Ajith}, T.~{Akutsu}, S.~{Albanesi}, A.~{Allocca}, P.A. {Altin},
  A.~{Amato}, C.~{Anand}, S.~{Anand}, A.~{Ananyeva}, S.B. {Anderson}, W.G.
  {Anderson}, M.~{Ando}, T.~{Andrade}, N.~{Andres}, T.~{Andri{\'c}}, S.V.
  {Angelova}, S.~{Ansoldi}, J.M. {Antelis}, S.~{Antier}, S.~{Appert},
  K.~{Arai}, K.~{Arai}, Y.~{Arai}, S.~{Araki}, A.~{Araya}, M.C. {Araya}, J.S.
  {Areeda}, M.~{Ar{\`e}ne}, N.~{Aritomi}, N.~{Arnaud}, S.M. {Aronson}, K.G.
  {Arun}, H.~{Asada}, Y.~{Asali}, G.~{Ashton}, Y.~{Aso}, M.~{Assiduo}, S.M.
  {Aston}, P.~{Astone}, F.~{Aubin}, C.~{Austin}, S.~{Babak}, F.~{Badaracco},
  M.K.M. {Bader}, C.~{Badger}, S.~{Bae}, Y.~{Bae}, A.M. {Baer}, S.~{Bagnasco},
  Y.~{Bai}, A.B. {Zimmerman}, M.E. {Zucker}, J.~{Zweizig}, {Ligo Scientific
  Collaboration}, {VIRGO Collaboration}, {Kagra Collaboration}, {Search for
  Gravitational Waves Associated with Gamma-Ray Bursts Detected by Fermi and
  Swift during the LIGO-Virgo Run O3b}.
\newblock \apj \textbf{928}(2), 186 (2022).
\newblock \doi{10.3847/1538-4357/ac532b}.
\newblock {\href{https://arxiv.org/abs/2111.03608}{{arXiv:2111.03608}}}
  {[astro-ph.HE]}

\bibitem{bloom02}
J.S. {Bloom}, S.R. {Kulkarni}, S.G. {Djorgovski}, The observed offset
  distribution of gamma-ray bursts from their host galaxies: A robust clue to
  the nature of the progenitors.
\newblock \aj \textbf{123}(3), 1111--1148 (2002).
\newblock \doi{10.1086/338893}.
\newblock
  {\href{https://arxiv.org/abs/astro-ph/0010176}{{arXiv:astro-ph/0010176}}}
  {[astro-ph]}

\bibitem{hjorth2003very}
J.~{Hjorth}, J.~{Sollerman}, P.~{M{\o}ller}, J.P.U. {Fynbo}, S.E. {Woosley},
  C.~{Kouveliotou}, N.R. {Tanvir}, J.~{Greiner}, M.I. {Andersen}, A.J.
  {Castro-Tirado}, J.M. {Castro Cer{\'o}n}, A.S. {Fruchter}, J.~{Gorosabel},
  P.~{Jakobsson}, L.~{Kaper}, S.~{Klose}, N.~{Masetti}, H.~{Pedersen},
  K.~{Pedersen}, E.~{Pian}, E.~{Palazzi}, J.E. {Rhoads}, E.~{Rol}, E.P.J. {van
  den Heuvel}, P.M. {Vreeswijk}, D.~{Watson}, R.A.M.J. {Wijers}, {A very
  energetic supernova associated with the {\ensuremath{\gamma}}-ray burst of 29
  March 2003}.
\newblock \nat \textbf{423}(6942), 847--850 (2003).
\newblock \doi{10.1038/nature01750}.
\newblock
  {\href{https://arxiv.org/abs/astro-ph/0306347}{{arXiv:astro-ph/0306347}}}
  {[astro-ph]}

\bibitem{hjorth2012grb}
J.~{Hjorth}, J.S. {Bloom}, in \emph{Chapter 9 in ''Gamma-Ray Bursts} (Cambridge
  University Press, 2012), pp. 169--190

\bibitem{Main170817}
B.P. Abbott, et~al., {GW170817: Observation of Gravitational Waves from a
  Binary Neutron Star Inspiral}.
\newblock Phys. Rev. Lett. \textbf{119}, 161,101 (2017).
\newblock \doi{10.1103/PhysRevLett.119.161101}.
\newblock
  \urlprefix\url{https://link.aps.org/doi/10.1103/PhysRevLett.119.161101}

\bibitem{Greiner09}
{Greiner, J.}, {Clemens, C.}, {Kr\"uhler, T.}, {von Kienlin, A.}, {Rau, A.},
  {Sari, R.}, {Fox, D. B.}, {Kawai, N.}, {Afonso, P.}, {Ajello, M.}, {Berger,
  E.}, {Cenko, S. B.}, {Cucchiara, A.}, {Filgas, R.}, {Klose, S.}, {K\"upc\"u
  Yoldas, A.}, {Lichti, G. G.}, {L\"ow, S.}, {McBreen, S.}, {Nagayama, T.},
  {Rossi, A.}, {Sato, S.}, {Szokoly, G.}, {Yoldas, A.}, {Zhang, X.-L.}, The
  redshift and afterglow of the extremely energetic gamma-ray burst grb
  080916c.
\newblock A\&A \textbf{498}(1), 89--94 (2009).
\newblock \doi{10.1051/0004-6361/200811571}.
\newblock \urlprefix\url{https://doi.org/10.1051/0004-6361/200811571}

\bibitem{Gendre05}
{Gendre, B.}, {Bo\"er, M.}, Decay properties of the x-ray afterglows of
  gamma-ray bursts.
\newblock A\&A \textbf{430}(2), 465--470 (2005).
\newblock \doi{10.1051/0004-6361:20042031}.
\newblock \urlprefix\url{https://doi.org/10.1051/0004-6361:20042031}

\bibitem{Schady15}
P.~Schady, Gamma-ray burst afterglows as probes of the ism.
\newblock Journal of High Energy Astrophysics \textbf{7}, 56 -- 63 (2015).
\newblock \doi{https://doi.org/10.1016/j.jheap.2015.05.001}.
\newblock
  \urlprefix\url{http://www.sciencedirect.com/science/article/pii/S2214404815000191}.
\newblock Swift 10 Years of Discovery, a novel approach to Time Domain
  Astronomy

\bibitem{Zhang_2011}
B.~Zhang, Open questions in grb physics.
\newblock Comptes Rendus Physique \textbf{12}(3), 206--225 (2011).
\newblock \doi{10.1016/j.crhy.2011.03.004}.
\newblock \urlprefix\url{http://dx.doi.org/10.1016/j.crhy.2011.03.004}

\bibitem{Gehrels12}
N.~Gehrels, P.~M{\' e}sz{\' a}ros, Gamma-ray bursts.
\newblock Science \textbf{337}(6097), 932--936 (2012).
\newblock \doi{10.1126/science.1216793}.
\newblock \urlprefix\url{http://dx.doi.org/10.1126/science.1216793}

\bibitem{Oganesyan18}
{Oganesyan, G.}, {Nava, L.}, {Ghirlanda, G.}, {Celotti, A.}, Characterization
  of gamma-ray burst prompt emission spectra down to soft x-rays.
\newblock A\&A \textbf{616}, A138 (2018).
\newblock \doi{10.1051/0004-6361/201732172}.
\newblock \urlprefix\url{https://doi.org/10.1051/0004-6361/201732172}

\bibitem{Wanderman2010}
D.~{Wanderman}, T.~{Piran}, {The luminosity function and the rate of Swift's
  gamma-ray bursts}.
\newblock \mnras \textbf{406}(3), 1944--1958 (2010).
\newblock \doi{10.1111/j.1365-2966.2010.16787.x}.
\newblock {\href{https://arxiv.org/abs/0912.0709}{{arXiv:0912.0709}}}
  {[astro-ph.HE]}

\bibitem{Wanderman_2015}
D.~Wanderman, T.~Piran, The rate, luminosity function and time delay of
  non-collapsar short {GRBs}.
\newblock Monthly Notices of the Royal Astronomical Society \textbf{448}(4),
  3026--3037 (2015).
\newblock \doi{10.1093/mnras/stv123}.
\newblock \urlprefix\url{https://doi.org/10.1093%2Fmnras%2Fstv123}

\bibitem{amati2002}
L.~{Amati}, F.~{Frontera}, M.~{Tavani}, J.J.M. {in't Zand}, A.~{Antonelli},
  E.~{Costa}, M.~{Feroci}, C.~{Guidorzi}, J.~{Heise}, N.~{Masetti},
  E.~{Montanari}, L.~{Nicastro}, E.~{Palazzi}, E.~{Pian}, L.~{Piro},
  P.~{Soffitta}, {Intrinsic spectra and energetics of BeppoSAX Gamma-Ray Bursts
  with known redshifts}.
\newblock \aap \textbf{390}, 81--89 (2002).
\newblock \doi{10.1051/0004-6361:20020722}.
\newblock
  {\href{https://arxiv.org/abs/astro-ph/0205230}{{arXiv:astro-ph/0205230}}}
  {[astro-ph]}

\bibitem{Amati06}
L.~{Amati}, {The E$_{p,i}$-E$_{iso}$ correlation in gamma-ray bursts: updated
  observational status, re-analysis and main implications}.
\newblock \mnras \textbf{372}(1), 233--245 (2006).
\newblock \doi{10.1111/j.1365-2966.2006.10840.x}.
\newblock
  {\href{https://arxiv.org/abs/astro-ph/0601553}{{arXiv:astro-ph/0601553}}}
  {[astro-ph]}

\bibitem{liang2010}
E.W. {Liang}, S.X. {Yi}, J.~{Zhang}, H.J. {L{\"u}}, B.B. {Zhang}, B.~{Zhang},
  {Constraining Gamma-ray Burst Initial Lorentz Factor with the Afterglow Onset
  Feature and Discovery of a Tight {\ensuremath{\Gamma}}$_{0}$-E
  $_{{\ensuremath{\gamma}},iso}$ Correlation}.
\newblock \apj \textbf{725}(2), 2209--2224 (2010).
\newblock \doi{10.1088/0004-637X/725/2/2209}.
\newblock {\href{https://arxiv.org/abs/0912.4800}{{arXiv:0912.4800}}}
  {[astro-ph.HE]}

\bibitem{ghirlanda2012}
G.~{Ghirlanda}, L.~{Nava}, G.~{Ghisellini}, A.~{Celotti}, D.~{Burlon},
  S.~{Covino}, A.~{Melandri}, {Gamma-ray bursts in the comoving frame}.
\newblock \mnras \textbf{420}(1), 483--494 (2012).
\newblock \doi{10.1111/j.1365-2966.2011.20053.x}.
\newblock {\href{https://arxiv.org/abs/1107.4096}{{arXiv:1107.4096}}}
  {[astro-ph.HE]}

\bibitem{Derishev}
E.V. Derishev, et~al., {Physical parameters and emission mechanism in gamma-ray
  bursts}.
\newblock A\&A \textbf{372} (2001)

\bibitem{Daigne}
F.~Daigne, et~al., {Reconciling observed gamma-ray burst prompt spectra with
  synchrotron radiation?}
\newblock A\&A \textbf{526} (2011)

\bibitem{2013ApJ...769...69B}
P.~{Beniamini}, T.~{Piran}, {Constraints on the Synchrotron Emission Mechanism
  in Gamma-Ray Bursts}.
\newblock \apj \textbf{769}(1), 69 (2013).
\newblock \doi{10.1088/0004-637X/769/1/69}.
\newblock {\href{https://arxiv.org/abs/1301.5575}{{arXiv:1301.5575}}}
  {[astro-ph.HE]}

\bibitem{Nakar2}
E.~{Nakar}, S.~{Ando}, R.~{Sari}, {Klein-Nishina Effects on Optically Thin
  Synchrotron and Synchrotron Self-Compton Spectrum}.
\newblock \apj \textbf{703}(1), 675--691 (2009).
\newblock \doi{10.1088/0004-637X/703/1/675}.
\newblock {\href{https://arxiv.org/abs/0903.2557}{{arXiv:0903.2557}}}
  {[astro-ph.HE]}

\bibitem{Uhm}
Z.L. {Uhm}, B.~{Zhang}, {Fast-cooling synchrotron radiation in a decaying
  magnetic field and {\ensuremath{\gamma}}-ray burst emission mechanism}.
\newblock Nature Physics \textbf{10}(5), 351--356 (2014).
\newblock \doi{10.1038/nphys2932}.
\newblock {\href{https://arxiv.org/abs/1303.2704}{{arXiv:1303.2704}}}
  {[astro-ph.HE]}

\bibitem{Med}
M.V. {Medvedev}, {The Theory of Spectral Evolution of the Gamma-Ray Burst
  Prompt Emission}.
\newblock \apj \textbf{637}(2), 869--872 (2006).
\newblock \doi{10.1086/498697}.
\newblock
  {\href{https://arxiv.org/abs/astro-ph/0510472}{{arXiv:astro-ph/0510472}}}
  {[astro-ph]}

\bibitem{band1993batse}
D.~Band, J.~Matteson, L.~Ford, B.~Schaefer, D.~Palmer, B.~Teegarden, T.~Cline,
  M.~Briggs, W.~Paciesas, G.~Pendleton, et~al., Batse observations of gamma-ray
  burst spectra. i-spectral diversity.
\newblock The Astrophysical Journal \textbf{413}, 281--292 (1993)

\bibitem{Oganesyan2018}
G.~{Oganesyan}, L.~{Nava}, G.~{Ghirlanda}, A.~{Celotti}, {Characterization of
  gamma-ray burst prompt emission spectra down to soft X-rays}.
\newblock \aap \textbf{616}, A138 (2018).
\newblock \doi{10.1051/0004-6361/201732172}.
\newblock {\href{https://arxiv.org/abs/1710.09383}{{arXiv:1710.09383}}}
  {[astro-ph.HE]}

\bibitem{Norris2010}
J.P. Norris, N.~Gehrels, J.D. Scargle, Threshold for extended emission in short
  gamma-ray bursts.
\newblock The Astrophysical Journal \textbf{717}(1), 411--419 (2010).
\newblock \doi{10.1088/0004-637x/717/1/411}.
\newblock \urlprefix\url{http://dx.doi.org/10.1088/0004-637X/717/1/411}

\bibitem{Bostanci2012}
Z.F. Bostanci, Y.~Kaneko, E.~G{\"o}{\u g}{\"u}{\c s}, Gamma-ray bursts with
  extended emission observed with batse.
\newblock Monthly Notices of the Royal Astronomical Society \textbf{428}(2),
  1623--1630 (2012).
\newblock \doi{10.1093/mnras/sts157}.
\newblock \urlprefix\url{http://dx.doi.org/10.1093/mnras/sts157}

\bibitem{Perley2009}
D.A. Perley, B.D. Metzger, J.~Granot, N.R. Butler, T.~Sakamoto,
  E.~Ramirez-Ruiz, A.J. Levan, J.S. Bloom, A.A. Miller, A.~Bunker, et~al., Grb
  080503: Implications of a naked short gamma-ray burst dominated by extended
  emission.
\newblock The Astrophysical Journal \textbf{696}(2), 1871--1885 (2009).
\newblock \doi{10.1088/0004-637x/696/2/1871}.
\newblock \urlprefix\url{http://dx.doi.org/10.1088/0004-637X/696/2/1871}

\bibitem{LVCFERMI170817}
B.~Abbott, R.~Abbott, T.~Abbott, F.~Acernese, K.~Ackley, C.~Adams, T.~Adams,
  P.~Addesso, R.~Adhikari, V.~Adya, C.~Affeldt, M.~Afrough, B.~Agarwal,
  M.~Agathos, K.~Agatsuma, N.~Aggarwal, O.~Aguiar, L.~Aiello, A.~Ain, P.~Ajith,
  O.~Smirnov, R.~Fender, P.~Woudt, {Gravitational Waves and Gamma-Rays from a
  Binary Neutron Star Merger: GW170817 and GRB 170817A}.
\newblock \apjl \textbf{848}(2), L13 (2017).
\newblock \doi{10.3847/2041-8213/aa920c}.
\newblock {\href{https://arxiv.org/abs/1710.05834}{{arXiv:1710.05834}}}
  {[astro-ph.HE]}

\bibitem{Shemi90}
A.~{Shemi}, T.~{Piran}, {The Appearance of Cosmic Fireballs}.
\newblock \apjl \textbf{365}, L55 (1990).
\newblock \doi{10.1086/185887}

\bibitem{Daigne98}
F.~Daigne, R.~Mochkovitch, {Gamma-ray bursts from internal shocks in a
  relativistic wind: temporal and spectral properties}.
\newblock Monthly Notices of the Royal Astronomical Society \textbf{296}(2),
  275--286 (1998).
\newblock \doi{10.1046/j.1365-8711.1998.01305.x}.
\newblock \urlprefix\url{https://doi.org/10.1046/j.1365-8711.1998.01305.x}.
\newblock
  {\href{https://arxiv.org/abs/https://academic.oup.com/mnras/article-pdf/296/2/275/2985446/296-2-275.pdf}{{https://academic.oup.com/mnras/article-pdf/296/2/275/2985446/296-2-275.pdf}}}

\bibitem{Golen83}
S.V. {Golenetskii}, E.P. {Mazets}, R.L. {Aptekar}, V.N. {Ilinskii},
  {Correlation between luminosity and temperature in {\ensuremath{\gamma}}-ray
  burst sources}.
\newblock \nat \textbf{306}(5942), 451--453 (1983).
\newblock \doi{10.1038/306451a0}

\bibitem{McKinney12}
J.C. McKinney, D.A. Uzdensky, {A reconnection switch to trigger gamma-ray burst
  jet dissipation}.
\newblock Monthly Notices of the Royal Astronomical Society \textbf{419}(1),
  573--607 (2011).
\newblock \doi{10.1111/j.1365-2966.2011.19721.x}.
\newblock \urlprefix\url{https://doi.org/10.1111/j.1365-2966.2011.19721.x}.
\newblock
  {\href{https://arxiv.org/abs/https://academic.oup.com/mnras/article-pdf/419/1/573/18706384/mnras0419-0573.pdf}{{https://academic.oup.com/mnras/article-pdf/419/1/573/18706384/mnras0419-0573.pdf}}}

\bibitem{gottlieb170817}
O.~{Gottlieb}, E.~{Nakar}, T.~{Piran}, K.~{Hotokezaka}, {A cocoon shock
  breakout as the origin of the {\ensuremath{\gamma}}-ray emission in
  GW170817}.
\newblock \mnras \textbf{479}(1), 588--600 (2018).
\newblock \doi{10.1093/mnras/sty1462}.
\newblock {\href{https://arxiv.org/abs/1710.05896}{{arXiv:1710.05896}}}
  {[astro-ph.HE]}

\bibitem{Beloborodov2020}
A.M. {Beloborodov}, C.~{Lundman}, Y.~{Levin}, {Relativistic Envelopes and
  Gamma-Rays from Neutron Star Mergers}.
\newblock \apj \textbf{897}(2), 141 (2020).
\newblock \doi{10.3847/1538-4357/ab86a0}

\bibitem{Hotokezaka2018}
K.~{Hotokezaka}, K.~{Kiuchi}, M.~{Shibata}, E.~{Nakar}, T.~{Piran},
  {Synchrotron Radiation from the Fast Tail of Dynamical Ejecta of Neutron Star
  Mergers}.
\newblock \apj \textbf{867}(2), 95 (2018).
\newblock \doi{10.3847/1538-4357/aadf92}.
\newblock {\href{https://arxiv.org/abs/1803.00599}{{arXiv:1803.00599}}}
  {[astro-ph.HE]}

\bibitem{Radice2018}
D.~{Radice}, A.~{Perego}, K.~{Hotokezaka}, S.A. {Fromm}, S.~{Bernuzzi}, L.F.
  {Roberts}, {Binary Neutron Star Mergers: Mass Ejection, Electromagnetic
  Counterparts, and Nucleosynthesis}.
\newblock \apj \textbf{869}(2), 130 (2018).
\newblock \doi{10.3847/1538-4357/aaf054}.
\newblock {\href{https://arxiv.org/abs/1809.11161}{{arXiv:1809.11161}}}
  {[astro-ph.HE]}

\bibitem{Covino16}
S.~{Covino}, D.~{Gotz}, {Polarization of prompt and afterglow emission of
  Gamma-Ray Bursts}.
\newblock Astronomical and Astrophysical Transactions \textbf{29}(2), 205--244
  (2016).
\newblock {\href{https://arxiv.org/abs/1605.03588}{{arXiv:1605.03588}}}
  {[astro-ph.HE]}

\bibitem{lazzati2004compton}
D.~{Lazzati}, E.~{Rossi}, G.~{Ghisellini}, M.J. {Rees}, {Compton drag as a
  mechanism for very high linear polarization in gamma-ray bursts}.
\newblock \mnras \textbf{347}(1), L1--L5 (2004).
\newblock \doi{10.1111/j.1365-2966.2004.07387.x}.
\newblock
  {\href{https://arxiv.org/abs/astro-ph/0309038}{{arXiv:astro-ph/0309038}}}
  {[astro-ph]}

\bibitem{waxman2003new}
E.~{Waxman}, {Astronomy: New direction for {\ensuremath{\gamma}}-rays}.
\newblock \nat \textbf{423}(6938), 388--389 (2003).
\newblock \doi{10.1038/423388a}.
\newblock
  {\href{https://arxiv.org/abs/astro-ph/0305414}{{arXiv:astro-ph/0305414}}}
  {[astro-ph]}

\bibitem{gill20a}
R.~{Gill}, J.~{Granot}, P.~{Kumar}, {Linear polarization in gamma-ray burst
  prompt emission}.
\newblock \mnras \textbf{491}(3), 3343--3373 (2020).
\newblock \doi{10.1093/mnras/stz2976}.
\newblock {\href{https://arxiv.org/abs/1811.11555}{{arXiv:1811.11555}}}
  {[astro-ph.HE]}

\bibitem{2022ApJ...936...12C}
T.~{Chattopadhyay}, S.~{Gupta}, S.~{Iyyani}, D.~{Saraogi}, V.~{Sharma},
  A.~{Tsvetkova}, A.~{Ratheesh}, R.~{Gupta}, N.P.S. {Mithun}, C.S. {Vaishnava},
  V.~{Prasad}, E.~{Aarthy}, A.~{Kumar}, A.R. {Rao}, S.~{Vadawale},
  V.~{Bhalerao}, D.~{Bhattacharya}, A.~{Vibhute}, D.~{Frederiks}, {Hard X-Ray
  Polarization Catalog for a Five-year Sample of Gamma-Ray Bursts Using
  AstroSat CZT Imager}.
\newblock \apj \textbf{936}(1), 12 (2022).
\newblock \doi{10.3847/1538-4357/ac82ef}.
\newblock {\href{https://arxiv.org/abs/2207.09605}{{arXiv:2207.09605}}}
  {[astro-ph.HE]}

\bibitem{beloborodov11a}
A.M. {Beloborodov}, {Radiative Transfer in Ultrarelativistic Outflows}.
\newblock \apj \textbf{737}(2), 68 (2011).
\newblock \doi{10.1088/0004-637X/737/2/68}.
\newblock {\href{https://arxiv.org/abs/1011.6005}{{arXiv:1011.6005}}}
  {[astro-ph.HE]}

\bibitem{Toma_2009}
K.~Toma, X.F. Wu, P.~MÃ©szÃ¡ros, An up-scattered cocoon emission model of
  gamma-ray burst high-energy lags.
\newblock The Astrophysical Journal \textbf{707}(2), 1404--1416 (2009).
\newblock \doi{10.1088/0004-637x/707/2/1404}.
\newblock \urlprefix\url{http://dx.doi.org/10.1088/0004-637X/707/2/1404}

\bibitem{ghisellini99a}
G.~{Ghisellini}, D.~{Lazzati}, {Polarization light curves and position angle
  variation of beamed gamma-ray bursts}.
\newblock \mnras \textbf{309}(1), L7--L11 (1999).
\newblock \doi{10.1046/j.1365-8711.1999.03025.x}.
\newblock
  {\href{https://arxiv.org/abs/astro-ph/9906471}{{arXiv:astro-ph/9906471}}}
  {[astro-ph]}

\bibitem{burgess20a}
J.M. {Burgess}, D.~{B{\'e}gu{\'e}}, J.~{Greiner}, D.~{Giannios}, A.~{Bacelj},
  F.~{Berlato}, {Gamma-ray bursts as cool synchrotron sources}.
\newblock Nature Astronomy \textbf{4}, 174--179 (2020).
\newblock \doi{10.1038/s41550-019-0911-z}.
\newblock {\href{https://arxiv.org/abs/1810.06965}{{arXiv:1810.06965}}}
  {[astro-ph.HE]}

\bibitem{medvedev99a}
M.V. {Medvedev}, A.~{Loeb}, {Generation of Magnetic Fields in the Relativistic
  Shock of Gamma-Ray Burst Sources}.
\newblock \apj \textbf{526}(2), 697--706 (1999).
\newblock \doi{10.1086/308038}.
\newblock
  {\href{https://arxiv.org/abs/astro-ph/9904363}{{arXiv:astro-ph/9904363}}}
  {[astro-ph]}

\bibitem{Burgess}
J.M. Burgess, et~al., {Time-Resolved GRB Polarization with POLAR and GBM}.
\newblock ArXiv e-prints  (2019).
\newblock {\href{https://arxiv.org/abs/1810.06965}{{arXiv:1810.06965}}}

\bibitem{Troja}
E.~{Troja}, V.M. {Lipunov}, C.G. {Mundell}, N.R. {Butler}, A.M. {Watson},
  S.~{Kobayashi}, S.B. {Cenko}, F.E. {Marshall}, R.~{Ricci}, A.~{Fruchter},
  M.H. {Wieringa}, E.S. {Gorbovskoy}, V.~{Kornilov}, A.~{Kutyrev}, W.H. {Lee},
  V.~{Toy}, N.V. {Tyurina}, N.M. {Budnev}, D.A.H. {Buckley}, J.~{Gonz{\'a}lez},
  O.~{Gress}, A.~{Horesh}, M.I. {Panasyuk}, J.X. {Prochaska},
  E.~{Ramirez-Ruiz}, R.~{Rebolo Lopez}, M.G. {Richer}, C.~{Roman-Zuniga},
  M.~{Serra-Ricart}, V.~{Yurkov}, N.~{Gehrels}, {Significant and variable
  linear polarization during the prompt optical flash of GRB 160625B.}
\newblock \nat \textbf{547}, 425--427 (2017).
\newblock \doi{10.1038/nature23289}

\bibitem{Vianello}
G.~{Vianello}, R.~{Gill}, J.~{Granot}, N.~{Omodei}, J.~{Cohen-Tanugi},
  F.~{Longo}, {The Bright and the Slow{\textemdash}GRBs 100724B and 160509A
  with High-energy Cutoffs at {\ensuremath{\lesssim}}100 MeV}.
\newblock \apj \textbf{864}(2), 163 (2018).
\newblock \doi{10.3847/1538-4357/aad6ea}.
\newblock {\href{https://arxiv.org/abs/1706.01481}{{arXiv:1706.01481}}}
  {[astro-ph.HE]}

\bibitem{gotz}
D.~{G{\"o}tz}, P.~{Laurent}, F.~{Lebrun}, F.~{Daigne},
  {\v{Z}}.~{Bo{\v{s}}njak}, {Variable Polarization Measured in the Prompt
  Emission of GRB 041219A Using IBIS on Board INTEGRAL}.
\newblock \apjl \textbf{695}(2), L208--L212 (2009).
\newblock \doi{10.1088/0004-637X/695/2/L208}.
\newblock {\href{https://arxiv.org/abs/0903.1712}{{arXiv:0903.1712}}}
  {[astro-ph.HE]}

\bibitem{yonetoku11a}
D.~{Yonetoku}, T.~{Murakami}, S.~{Gunji}, T.~{Mihara}, K.~{Toma},
  T.~{Sakashita}, Y.~{Morihara}, T.~{Takahashi}, N.~{Toukairin}, H.~{Fujimoto},
  Y.~{Kodama}, S.~{Kubo}, {IKAROS Demonstration Team}, {Detection of Gamma-Ray
  Polarization in Prompt Emission of GRB 100826A}.
\newblock \apjl \textbf{743}(2), L30 (2011).
\newblock \doi{10.1088/2041-8205/743/2/L30}.
\newblock {\href{https://arxiv.org/abs/1111.1779}{{arXiv:1111.1779}}}
  {[astro-ph.HE]}

\bibitem{zhang19a}
S.N. {Zhang}, M.~{Kole}, T.W. {Bao}, T.~{Batsch}, T.~{Bernasconi}, F.~{Cadoux},
  J.Y. {Chai}, Z.G. {Dai}, Y.W. {Dong}, N.~{Gauvin}, W.~{Hajdas}, M.X. {Lan},
  H.C. {Li}, L.~{Li}, Z.H. {Li}, J.T. {Liu}, X.~{Liu}, R.~{Marcinkowski},
  N.~{Produit}, S.~{Orsi}, M.~{Pohl}, D.~{Rybka}, H.L. {Shi}, L.M. {Song}, J.C.
  {Sun}, J.~{Szabelski}, T.~{Tymieniecka}, R.J. {Wang}, Y.H. {Wang}, X.~{Wen},
  B.B. {Wu}, X.~{Wu}, X.F. {Wu}, H.L. {Xiao}, S.L. {Xiong}, L.Y. {Zhang},
  L.~{Zhang}, X.F. {Zhang}, Y.J. {Zhang}, A.~{Zwolinska}, {Detailed
  polarization measurements of the prompt emission of five gamma-ray bursts}.
\newblock Nature Astronomy \textbf{3}, 258--264 (2019).
\newblock \doi{10.1038/s41550-018-0664-0}.
\newblock {\href{https://arxiv.org/abs/1901.04207}{{arXiv:1901.04207}}}
  {[astro-ph.HE]}

\bibitem{sharma19a}
V.~{Sharma}, S.~{Iyyani}, D.~{Bhattacharya}, T.~{Chattopadhyay}, A.R. {Rao},
  E.~{Aarthy}, S.V. {Vadawale}, N.P.S. {Mithun}, V.B. {Bhalerao}, F.~{Ryde},
  A.~{Pe'er}, {Time-varying Polarized Gamma-Rays from GRB 160821A: Evidence for
  Ordered Magnetic Fields}.
\newblock \apjl \textbf{882}(1), L10 (2019).
\newblock \doi{10.3847/2041-8213/ab3a48}.
\newblock {\href{https://arxiv.org/abs/1908.10885}{{arXiv:1908.10885}}}
  {[astro-ph.HE]}

\bibitem{lazzati09a}
D.~{Lazzati}, M.C. {Begelman}, {Polarization Signature of Gamma-Ray Bursts from
  Fragmented Fireballs}.
\newblock \apjl \textbf{700}(2), L141--L144 (2009).
\newblock \doi{10.1088/0004-637X/700/2/L141}.
\newblock {\href{https://arxiv.org/abs/0906.4337}{{arXiv:0906.4337}}}
  {[astro-ph.HE]}

\bibitem{Bala2023}
S.~Bala, S.~Mate, A.~Mehla, P.~Sastry, N.P.S. Mithun, S.~Palit, M.V. Chanda,
  D.~Saraogi, C.S. Vaishnava, G.~Waratkar, V.~Bhalerao, D.~Bhattacharya,
  S.~Tendulkar, S.~Vadawale, {Prospects of measuring gamma-ray burst
  polarization with the Daksha mission}.
\newblock Journal of Astronomical Telescopes, Instruments, and Systems
  \textbf{9}(4), 048,002 (2023).
\newblock \doi{10.1117/1.JATIS.9.4.048002}.
\newblock \urlprefix\url{https://doi.org/10.1117/1.JATIS.9.4.048002}

\bibitem{kaspi_beloborodov2017}
V.M. {Kaspi}, A.M. {Beloborodov}, {Magnetars}.
\newblock \araa \textbf{55}(1), 261--301 (2017).
\newblock \doi{10.1146/annurev-astro-081915-023329}.
\newblock {\href{https://arxiv.org/abs/1703.00068}{{arXiv:1703.00068}}}
  {[astro-ph.HE]}

\bibitem{hurley2005}
K.~{Hurley}, S.E. {Boggs}, D.M. {Smith}, R.C. {Duncan}, R.~{Lin},
  A.~{Zoglauer}, S.~{Krucker}, G.~{Hurford}, H.~{Hudson}, C.~{Wigger},
  W.~{Hajdas}, C.~{Thompson}, I.~{Mitrofanov}, A.~{Sanin}, W.~{Boynton},
  C.~{Fellows}, A.~{von Kienlin}, G.~{Lichti}, A.~{Rau}, T.~{Cline}, {An
  exceptionally bright flare from SGR 1806-20 and the origins of short-duration
  {\ensuremath{\gamma}}-ray bursts}.
\newblock \nat \textbf{434}(7037), 1098--1103 (2005).
\newblock \doi{10.1038/nature03519}.
\newblock
  {\href{https://arxiv.org/abs/astro-ph/0502329}{{arXiv:astro-ph/0502329}}}
  {[astro-ph]}

\bibitem{tavernaturolla2017}
R.~{Taverna}, R.~{Turolla}, {On the spectrum and polarization of magnetar flare
  emission}.
\newblock \mnras \textbf{469}(3), 3610--3628 (2017).
\newblock \doi{10.1093/mnras/stx1086}.
\newblock {\href{https://arxiv.org/abs/1705.01130}{{arXiv:1705.01130}}}
  {[astro-ph.HE]}

\bibitem{metzger2019}
B.D. {Metzger}, B.~{Margalit}, L.~{Sironi}, {Fast radio bursts as synchrotron
  maser emission from decelerating relativistic blast waves}.
\newblock \mnras \textbf{485}(3), 4091--4106 (2019).
\newblock \doi{10.1093/mnras/stz700}.
\newblock {\href{https://arxiv.org/abs/1902.01866}{{arXiv:1902.01866}}}
  {[astro-ph.HE]}

\bibitem{lyutikov2020}
M.~{Lyutikov}, S.~{Popov}, {Fast Radio Bursts from reconnection events in
  magnetar magnetospheres}.
\newblock arXiv e-prints arXiv:2005.05093 (2020).
\newblock {\href{https://arxiv.org/abs/2005.05093}{{arXiv:2005.05093}}}
  {[astro-ph.HE]}

\bibitem{pshirkov2010}
M.S. {Pshirkov}, K.A. {Postnov}, {Radio precursors to neutron star binary
  mergings}.
\newblock \apss \textbf{330}(1), 13--18 (2010).
\newblock \doi{10.1007/s10509-010-0395-x}.
\newblock {\href{https://arxiv.org/abs/1004.5115}{{arXiv:1004.5115}}}
  {[astro-ph.HE]}

\bibitem{totani2013}
T.~{Totani}, {Cosmological Fast Radio Bursts from Binary Neutron Star Mergers}.
\newblock \pasj \textbf{65}, L12 (2013).
\newblock \doi{10.1093/pasj/65.5.L12}.
\newblock {\href{https://arxiv.org/abs/1307.4985}{{arXiv:1307.4985}}}
  {[astro-ph.HE]}

\bibitem{mingarelli2015}
C.M.F. {Mingarelli}, J.~{Levin}, T.J.W. {Lazio}, {Fast Radio Bursts and Radio
  Transients from Black Hole Batteries}.
\newblock \apjl \textbf{814}(2), L20 (2015).
\newblock \doi{10.1088/2041-8205/814/2/L20}.
\newblock {\href{https://arxiv.org/abs/1511.02870}{{arXiv:1511.02870}}}
  {[astro-ph.HE]}

\bibitem{paschalidis2019}
V.~{Paschalidis}, M.~{Ruiz}, {Are fast radio bursts the most likely
  electromagnetic counterpart of neutron star mergers resulting in prompt
  collapse?}
\newblock \prd \textbf{100}(4), 043001 (2019).
\newblock \doi{10.1103/PhysRevD.100.043001}.
\newblock {\href{https://arxiv.org/abs/1808.04822}{{arXiv:1808.04822}}}
  {[astro-ph.HE]}

\bibitem{rowlinson2019}
A.~{Rowlinson}, G.E. {Anderson}, {Constraining coherent low-frequency radio
  flares from compact binary mergers}.
\newblock \mnras \textbf{489}(3), 3316--3333 (2019).
\newblock \doi{10.1093/mnras/stz2295}.
\newblock {\href{https://arxiv.org/abs/1905.02509}{{arXiv:1905.02509}}}
  {[astro-ph.HE]}

\bibitem{platts2019}
E.~{Platts}, A.~{Weltman}, A.~{Walters}, S.P. {Tendulkar}, J.E.B. {Gordin},
  S.~{Kandhai}, {A living theory catalogue for fast radio bursts}.
\newblock \physrep \textbf{821}, 1--27 (2019).
\newblock \doi{10.1016/j.physrep.2019.06.003}.
\newblock {\href{https://arxiv.org/abs/1810.05836}{{arXiv:1810.05836}}}
  {[astro-ph.HE]}

\bibitem{petroff2019}
E.~{Petroff}, J.W.T. {Hessels}, D.R. {Lorimer}, {Fast radio bursts}.
\newblock \aapr \textbf{27}(1), 4 (2019).
\newblock \doi{10.1007/s00159-019-0116-6}.
\newblock {\href{https://arxiv.org/abs/1904.07947}{{arXiv:1904.07947}}}
  {[astro-ph.HE]}

\bibitem{petroff2022}
E.~{Petroff}, J.W.T. {Hessels}, D.R. {Lorimer}, {Fast radio bursts at the dawn
  of the 2020s}.
\newblock \aapr \textbf{30}(1), 2 (2022).
\newblock \doi{10.1007/s00159-022-00139-w}.
\newblock {\href{https://arxiv.org/abs/2107.10113}{{arXiv:2107.10113}}}
  {[astro-ph.HE]}

\bibitem{cunningham2019}
V.~{Cunningham}, S.B. {Cenko}, E.~{Burns}, A.~{Goldstein}, A.~{Lien},
  D.~{Kocevski}, M.~{Briggs}, V.~{Connaughton}, M.C. {Miller}, J.~{Racusin},
  M.~{Stanbro}, {A Search for High-energy Counterparts to Fast Radio Bursts}.
\newblock \apj \textbf{879}(1), 40 (2019).
\newblock \doi{10.3847/1538-4357/ab2235}.
\newblock {\href{https://arxiv.org/abs/1905.06818}{{arXiv:1905.06818}}}
  {[astro-ph.HE]}

\bibitem{anumarlapudi2020}
A.~{Anumarlapudi}, V.~{Bhalerao}, S.P. {Tendulkar}, A.~{Balasubramanian},
  {Prompt X-Ray Emission from Fast Radio Bursts{\textemdash}Upper Limits with
  AstroSat}.
\newblock \apj \textbf{888}(1), 40 (2020).
\newblock \doi{10.3847/1538-4357/ab5363}.
\newblock {\href{https://arxiv.org/abs/1911.00537}{{arXiv:1911.00537}}}
  {[astro-ph.HE]}

\bibitem{guidorzi2020}
C.~{Guidorzi}, M.~{Marongiu}, R.~{Martone}, L.~{Nicastro}, S.L. {Xiong}, J.Y.
  {Liao}, G.~{Li}, S.N. {Zhang}, L.~{Amati}, F.~{Frontera}, M.~{Orlandini},
  P.~{Rosati}, E.~{Virgilli}, S.~{Zhang}, Q.C. {Bu}, C.~{Cai}, X.L. {Cao},
  Z.~{Chang}, G.~{Chen}, L.~{Chen}, T.X. {Chen}, Y.B. {Chen}, Y.P. {Chen},
  W.~{Cui}, W.W. {Cui}, J.K. {Deng}, Y.W. {Dong}, Y.Y. {Du}, M.X. {Fu}, G.H.
  {Gao}, H.~{Gao}, M.~{Gao}, M.Y. {Ge}, Y.D. {Gu}, J.~{Guan}, C.C. {Guo}, D.W.
  {Han}, Y.~{Huang}, J.~{Huo}, S.M. {Jia}, L.H. {Jiang}, W.C. {Jiang},
  J.~{Jin}, Y.J. {Jin}, L.D. {Kong}, B.~{Li}, C.K. {Li}, M.S. {Li}, T.P. {Li},
  W.~{Li}, X.~{Li}, X.B. {Li}, X.F. {Li}, Y.G. {Li}, Z.W. {Li}, X.H. {Liang},
  B.S. {Liu}, C.Z. {Liu}, G.Q. {Liu}, H.W. {Liu}, X.J. {Liu}, Y.N. {Liu},
  B.~{Lu}, F.J. {Lu}, X.F. {Lu}, Q.~{Luo}, T.~{Luo}, R.C. {Ma}, X.~{Ma},
  B.~{Meng}, Y.~{Nang}, J.Y. {Nie}, G.~{Ou}, J.L. {Qu}, N.~{Sai}, R.C. {Shang},
  L.M. {Song}, X.Y. {Song}, L.~{Sun}, Y.~{Tan}, L.~{Tao}, Y.L. {Tuo},
  C.~{Wang}, G.F. {Wang}, J.~{Wang}, W.S. {Wang}, Y.S. {Wang}, X.Y. {Wen}, B.Y.
  {Wu}, B.B. {Wu}, M.~{Wu}, G.C. {Xiao}, S.~{Xiao}, Y.P. {Xu}, J.W. {Yang},
  S.~{Yang}, Y.J. {Yang}, Q.B. {Yi}, Q.Q. {Yin}, Y.~{You}, A.M. {Zhang}, C.M.
  {Zhang}, F.~{Zhang}, H.M. {Zhang}, J.~{Zhang}, T.~{Zhang}, W.C. {Zhang},
  W.~{Zhang}, W.Z. {Zhang}, Y.~{Zhang}, Y.F. {Zhang}, Y.J. {Zhang}, Y.~{Zhang},
  Z.~{Zhang}, Z.~{Zhang}, Z.L. {Zhang}, H.S. {Zhang}, X.F. {Zhang}, S.J.
  {Zheng}, D.K. {Zhou}, J.F. {Zhou}, Y.X. {Zhu}, Y.~{Zhu}, R.L. {Zhuang}, {A
  search for prompt {\ensuremath{\gamma}}-ray counterparts to fast radio bursts
  in the Insight-HXMT data}.
\newblock \aap \textbf{637}, A69 (2020).
\newblock \doi{10.1051/0004-6361/202037797}.
\newblock {\href{https://arxiv.org/abs/2003.10889}{{arXiv:2003.10889}}}
  {[astro-ph.HE]}

\bibitem{curtin2022}
A.P. {Curtin}, S.P. {Tendulkar}, A.~{Josephy}, P.~{Chawla}, B.~{Andersen}, V.M.
  {Kaspi}, M.~{Bhardwaj}, T.~{Cassanelli}, A.~{Cook}, F.A. {Dong},
  E.~{Fonseca}, B.M. {Gaensler}, J.F. {Kaczmarek}, A.E. {Lanmnan}, C.~{Leung},
  A.B. {Pearlman}, E.~{Petroff}, Z.~{Pleunis}, M.~{Rafiei-Ravandi}, S.M.
  {Ransom}, K.~{Shin}, P.~{Scholz}, K.~{Smith}, I.~{Stairs}, {Limits on Fast
  Radio Burst-like Counterparts to Gamma-ray Bursts using CHIME/FRB}.
\newblock arXiv e-prints arXiv:2208.00803 (2022).
\newblock {\href{https://arxiv.org/abs/2208.00803}{{arXiv:2208.00803}}}
  {[astro-ph.HE]}

\bibitem{principe2022}
G.~{Principe}, N.~{Omodei}, F.~{Longo}, L.~{Di Venere}, N.~{Di Lalla},
  {Fermi-LAT Collaboration}, in \emph{37th International Cosmic Ray Conference.
  12-23 July 2021. Berlin} (2022), p. 624

\bibitem{chime_sgr1935_2020}
{CHIME/FRB Collaboration}, B.C. {Andersen}, K.M. {Bandura}, M.~{Bhardwaj},
  A.~{Bij}, M.M. {Boyce}, P.J. {Boyle}, C.~{Brar}, T.~{Cassanelli},
  P.~{Chawla}, T.~{Chen}, J.F. {Cliche}, A.~{Cook}, D.~{Cubranic}, A.P.
  {Curtin}, N.T. {Denman}, M.~{Dobbs}, F.Q. {Dong}, M.~{Fandino}, E.~{Fonseca},
  B.M. {Gaensler}, U.~{Giri}, D.C. {Good}, M.~{Halpern}, A.S. {Hill}, G.F.
  {Hinshaw}, C.~{H{\"o}fer}, A.~{Josephy}, J.W. {Kania}, V.M. {Kaspi}, T.L.
  {Landecker}, C.~{Leung}, D.Z. {Li}, H.H. {Lin}, K.W. {Masui}, R.~{McKinven},
  J.~{Mena-Parra}, M.~{Merryfield}, B.W. {Meyers}, D.~{Michilli},
  N.~{Milutinovic}, A.~{Mirhosseini}, M.~{M{\"u}nchmeyer}, A.~{Naidu}, L.B.
  {Newburgh}, C.~{Ng}, C.~{Patel}, U.L. {Pen}, T.~{Pinsonneault-Marotte},
  Z.~{Pleunis}, B.M. {Quine}, M.~{Rafiei-Ravandi}, M.~{Rahman}, S.M. {Ransom},
  A.~{Renard}, P.~{Sanghavi}, P.~{Scholz}, J.R. {Shaw}, K.~{Shin}, S.R.
  {Siegel}, S.~{Singh}, R.J. {Smegal}, K.M. {Smith}, I.H. {Stairs}, C.M. {Tan},
  S.P. {Tendulkar}, I.~{Tretyakov}, K.~{Vanderlinde}, H.~{Wang}, D.~{Wulf},
  A.V. {Zwaniga}, {A bright millisecond-duration radio burst from a Galactic
  magnetar}.
\newblock \nat \textbf{587}(7832), 54--58 (2020).
\newblock \doi{10.1038/s41586-020-2863-y}.
\newblock {\href{https://arxiv.org/abs/2005.10324}{{arXiv:2005.10324}}}
  {[astro-ph.HE]}

\bibitem{bochenek2020b}
C.D. {Bochenek}, V.~{Ravi}, K.V. {Belov}, G.~{Hallinan}, J.~{Kocz}, S.R.
  {Kulkarni}, D.L. {McKenna}, {A fast radio burst associated with a Galactic
  magnetar}.
\newblock \nat \textbf{587}(7832), 59--62 (2020).
\newblock \doi{10.1038/s41586-020-2872-x}.
\newblock {\href{https://arxiv.org/abs/2005.10828}{{arXiv:2005.10828}}}
  {[astro-ph.HE]}

\bibitem{mereghetti2020}
S.~{Mereghetti}, V.~{Savchenko}, C.~{Ferrigno}, D.~{G{\"o}tz}, M.~{Rigoselli},
  A.~{Tiengo}, A.~{Bazzano}, E.~{Bozzo}, A.~{Coleiro}, T.J.L. {Courvoisier},
  M.~{Doyle}, A.~{Goldwurm}, L.~{Hanlon}, E.~{Jourdain}, A.~{von Kienlin},
  A.~{Lutovinov}, A.~{Martin-Carrillo}, S.~{Molkov}, L.~{Natalucci},
  F.~{Onori}, F.~{Panessa}, J.~{Rodi}, J.~{Rodriguez},
  C.~{S{\'a}nchez-Fern{\'a}ndez}, R.~{Sunyaev}, P.~{Ubertini}, {INTEGRAL
  Discovery of a Burst with Associated Radio Emission from the Magnetar SGR
  1935+2154}.
\newblock \apjl \textbf{898}(2), L29 (2020).
\newblock \doi{10.3847/2041-8213/aba2cf}.
\newblock {\href{https://arxiv.org/abs/2005.06335}{{arXiv:2005.06335}}}
  {[astro-ph.HE]}

\bibitem{chime2018}
{CHIME/FRB Collaboration}, M.~{Amiri}, K.~{Bandura}, P.~{Berger},
  M.~{Bhardwaj}, M.M. {Boyce}, P.J. {Boyle}, C.~{Brar}, M.~{Burhanpurkar},
  P.~{Chawla}, J.~{Chowdhury}, J.F. {Cliche}, M.D. {Cranmer}, D.~{Cubranic},
  M.~{Deng}, N.~{Denman}, M.~{Dobbs}, M.~{Fandino}, E.~{Fonseca}, B.M.
  {Gaensler}, U.~{Giri}, A.J. {Gilbert}, D.C. {Good}, S.~{Guliani},
  M.~{Halpern}, G.~{Hinshaw}, C.~{H{\"o}fer}, A.~{Josephy}, V.M. {Kaspi}, T.L.
  {Landecker}, D.~{Lang}, H.~{Liao}, K.W. {Masui}, J.~{Mena-Parra}, A.~{Naidu},
  L.B. {Newburgh}, C.~{Ng}, C.~{Patel}, U.L. {Pen}, T.~{Pinsonneault-Marotte},
  Z.~{Pleunis}, M.~{Rafiei Ravandi}, S.M. {Ransom}, A.~{Renard}, P.~{Scholz},
  K.~{Sigurdson}, S.R. {Siegel}, K.M. {Smith}, I.H. {Stairs}, S.P. {Tendulkar},
  K.~{Vand erlinde}, D.V. {Wiebe}, {The CHIME Fast Radio Burst Project: System
  Overview}.
\newblock \apj \textbf{863}(1), 48 (2018).
\newblock \doi{10.3847/1538-4357/aad188}.
\newblock {\href{https://arxiv.org/abs/1803.11235}{{arXiv:1803.11235}}}
  {[astro-ph.IM]}

\bibitem{bannister2017}
K.W. {Bannister}, R.M. {Shannon}, J.P. {Macquart}, C.~{Flynn}, P.G. {Edwards},
  M.~{O'Neill}, S.~{Os{\l}owski}, M.~{Bailes}, B.~{Zackay}, N.~{Clarke}, L.R.
  {D'Addario}, R.~{Dodson}, P.J. {Hall}, A.~{Jameson}, D.~{Jones},
  R.~{Navarro}, J.T. {Trinh}, J.~{Allison}, C.S. {Anderson}, M.~{Bell}, A.P.
  {Chippendale}, J.D. {Collier}, G.~{Heald}, I.~{Heywood}, A.W. {Hotan},
  K.~{Lee-Waddell}, J.P. {Madrid}, J.~{Marvil}, D.~{McConnell}, A.~{Popping},
  M.A. {Voronkov}, M.T. {Whiting}, G.R. {Allen}, D.C.J. {Bock}, D.P.
  {Brodrick}, F.~{Cooray}, D.R. {DeBoer}, P.J. {Diamond}, R.~{Ekers}, R.G.
  {Gough}, G.A. {Hampson}, L.~{Harvey-Smith}, S.G. {Hay}, D.B. {Hayman}, C.A.
  {Jackson}, S.~{Johnston}, B.S. {Koribalski}, N.M. {McClure-Griffiths},
  P.~{Mirtschin}, A.~{Ng}, R.P. {Norris}, S.E. {Pearce}, C.J. {Phillips}, D.N.
  {Roxby}, E.R. {Troup}, T.~{Westmeier}, {The Detection of an Extremely Bright
  Fast Radio Burst in a Phased Array Feed Survey}.
\newblock \apjl \textbf{841}(1), L12 (2017).
\newblock \doi{10.3847/2041-8213/aa71ff}.
\newblock {\href{https://arxiv.org/abs/1705.07581}{{arXiv:1705.07581}}}
  {[astro-ph.HE]}

\bibitem{bochenek2020a}
C.D. {Bochenek}, D.L. {McKenna}, K.V. {Belov}, J.~{Kocz}, S.R. {Kulkarni},
  J.~{Lamb}, V.~{Ravi}, D.~{Woody}, {STARE2: Detecting Fast Radio Bursts in the
  Milky Way}.
\newblock \pasp \textbf{132}(1009), 034,202 (2020).
\newblock \doi{10.1088/1538-3873/ab63b3}.
\newblock {\href{https://arxiv.org/abs/2001.05077}{{arXiv:2001.05077}}}
  {[astro-ph.HE]}

\bibitem{lin2022}
H.H. {Lin}, K.y. {Lin}, C.T. {Li}, Y.H. {Tseng}, H.~{Jiang}, J.H. {Wang}, J.C.
  {Cheng}, U.L. {Pen}, M.T. {Chen}, P.~{Chen}, Y.~{Chen}, T.~{Goto},
  T.~{Hashimoto}, Y.J. {Hwang}, S.K. {King}, D.~{Kubo}, C.Y. {Kuo}, A.~{Mills},
  J.~{Nam}, P.~{Oshiro}, C.S. {Shen}, H.C. {Tseng}, S.H. {Wang}, V.F.S. {Wu},
  G.~{Bower}, S.H. {Chang}, P.A. {Chen}, Y.C. {Chen}, Y.K. {Chiang},
  A.~{Fedynitch}, N.~{Gusinskaia}, S.C.C. {Ho}, T.Y.Y. {Hsiao}, C.P. {Hu}, Y.D.
  {Huang}, J.M. {J{\'a}uregui Garc{\'\i}a}, S.J. {Kim}, C.Y. {Kuo}, D.F.J.
  {Ling}, A.Y.L. {On}, J.B. {Peterson}, B.J. {R. Raquel}, S.C. {Su}, Y.~{Uno},
  C.K.W. {Wu}, S.~{Yamasaki}, H.M. {Zhu}, {BURSTT: Bustling Universe Radio
  Survey Telescope in Taiwan}.
\newblock \pasp \textbf{134}(1039), 094106 (2022).
\newblock \doi{10.1088/1538-3873/ac8f71}.
\newblock {\href{https://arxiv.org/abs/2206.08983}{{arXiv:2206.08983}}}
  {[astro-ph.IM]}

\bibitem{1983ApJ...270..711W}
N.E. {White}, J.H. {Swank}, S.S. {Holt}, {Accretion powered X-ray pulsars.}
\newblock \apj \textbf{270}, 711--734 (1983).
\newblock \doi{10.1086/161162}

\bibitem{1989PASJ...41....1N}
F.~{Nagase}, {Accretion-powered X-ray pulsars.}
\newblock \pasj \textbf{41}, 1 (1989)

\bibitem{1997ApJS..113..367B}
L.~{Bildsten}, D.~{Chakrabarty}, J.~{Chiu}, M.H. {Finger}, D.T. {Koh}, R.W.
  {Nelson}, T.A. {Prince}, B.C. {Rubin}, D.M. {Scott}, M.~{Stollberg}, B.A.
  {Vaughan}, C.A. {Wilson}, R.B. {Wilson}, {Observations of Accreting Pulsars}.
\newblock \apjs \textbf{113}(2), 367--408 (1997).
\newblock \doi{10.1086/313060}.
\newblock
  {\href{https://arxiv.org/abs/astro-ph/9707125}{{arXiv:astro-ph/9707125}}}
  {[astro-ph]}

\bibitem{2020ApJ...896...90M}
C.~{Malacaria}, P.~{Jenke}, O.J. {Roberts}, C.A. {Wilson-Hodge}, W.H.
  {Cleveland}, B.~{Mailyan}, {GBM Accreting Pulsars Program Team}, {The Ups and
  Downs of Accreting X-Ray Pulsars: Decade-long Observations with the Fermi
  Gamma-Ray Burst Monitor}.
\newblock \apj \textbf{896}(1), 90 (2020).
\newblock \doi{10.3847/1538-4357/ab855c}.
\newblock {\href{https://arxiv.org/abs/2004.00051}{{arXiv:2004.00051}}}
  {[astro-ph.HE]}

\bibitem{2012ApJ...759..124J}
P.A. {Jenke}, M.H. {Finger}, C.A. {Wilson-Hodge}, A.~{Camero-Arranz}, {Orbital
  Decay and Evidence of Disk Formation in the X-Ray Binary Pulsar OAO
  1657-415}.
\newblock \apj \textbf{759}(2), 124 (2012).
\newblock \doi{10.1088/0004-637X/759/2/124}.
\newblock {\href{https://arxiv.org/abs/1112.5190}{{arXiv:1112.5190}}}
  {[astro-ph.HE]}

\bibitem{2023MNRAS.520.1411M}
H.~{Manikantan}, B.~{Paul}, K.~{Roy}, V.~{Rana}, {Changes in the distribution
  of circum-binary material around the HMXB GX 301-2 during a rapid spin-up
  episode of the neutron star}.
\newblock \mnras \textbf{520}(1), 1411--1416 (2023).
\newblock \doi{10.1093/mnras/stad037}.
\newblock {\href{https://arxiv.org/abs/2301.02815}{{arXiv:2301.02815}}}
  {[astro-ph.HE]}

\bibitem{2011BASI...39..429P}
B.~{Paul}, S.~{Naik}, {Transient High Mass X-ray Binaries}.
\newblock Bulletin of the Astronomical Society of India \textbf{39}(3),
  429--449 (2011).
\newblock \doi{10.48550/arXiv.1110.4446}.
\newblock {\href{https://arxiv.org/abs/1110.4446}{{arXiv:1110.4446}}}
  {[astro-ph.HE]}

\bibitem{2017PASJ...69..100S}
M.~{Sugizaki}, T.~{Mihara}, M.~{Nakajima}, K.~{Makishima}, {Correlation between
  the luminosity and spin-period changes during outbursts of 12 Be binary
  pulsars observed by the MAXI/GSC and the Fermi/GBM}.
\newblock \pasj \textbf{69}(6), 100 (2017).
\newblock \doi{10.1093/pasj/psx119}.
\newblock {\href{https://arxiv.org/abs/1709.07579}{{arXiv:1709.07579}}}
  {[astro-ph.HE]}

\bibitem{ZeldovichNovikov1966}
Y.B. Zel'dovich, I.D. Novikov, The hypothesis of cores retarded during
  expansion and the hot cosmological model.
\newblock Soviet Astronomy \textbf{10}, 602 (1967)

\bibitem{Hawking1971}
S.~Hawking, Gravitationally collapsed objects of very low mass.
\newblock Monthly Notices of the Royal Astronomical Society \textbf{152}(1),
  75--78 (1971)

\bibitem{Carr2020}
B.~Carr, F.~K{\"u}hnel, Primordial black holes as dark matter: recent
  developments.
\newblock Annual Review of Nuclear and Particle Science \textbf{70}, 355--394
  (2020)

\bibitem{Niikura2019}
H.~Niikura, M.~Takada, N.~Yasuda, R.H. Lupton, T.~Sumi, S.~More, T.~Kurita,
  S.~Sugiyama, A.~More, M.~Oguri, et~al., Microlensing constraints on
  primordial black holes with subaru/hsc andromeda observations.
\newblock Nature Astronomy \textbf{3}(6), 524--534 (2019)

\bibitem{JungKim2020}
S.~Jung, T.~Kim, Gamma-ray burst lensing parallax: closing the primordial black
  hole dark matter mass window.
\newblock Physical Review Research \textbf{2}(1), 013,113 (2020)

\bibitem{Gawade2023}
P.~{Gawade}, S.~{More}, V.~{Bhalerao}, {On the feasibility of primordial black
  hole abundance constraints using lensing parallax of GRBs}.
\newblock arXiv e-prints arXiv:2308.01775 (2023).
\newblock \doi{10.48550/arXiv.2308.01775}.
\newblock {\href{https://arxiv.org/abs/2308.01775}{{arXiv:2308.01775}}}
  {[astro-ph.CO]}

\bibitem{2002ApJS..138..149H}
B.A. {Harmon}, G.J. {Fishman}, C.A. {Wilson}, W.S. {Paciesas}, S.N. {Zhang},
  M.H. {Finger}, T.M. {Koshut}, M.L. {McCollough}, C.R. {Robinson}, B.C.
  {Rubin}, {The Burst and Transient Source Experiment Earth Occultation
  Technique}.
\newblock \apjs \textbf{138}(1), 149--183 (2002).
\newblock \doi{10.1086/324018}.
\newblock
  {\href{https://arxiv.org/abs/astro-ph/0109069}{{arXiv:astro-ph/0109069}}}
  {[astro-ph]}

\bibitem{2021JApA...42...64S}
A.~{Singhal}, R.~{Srinivasan}, V.~{Bhalerao}, D.~{Bhattacharya}, A.R. {Rao},
  S.~{Vadawale}, {Using collimated CZTI as all-sky X-ray detector based on
  Earth occultation technique}.
\newblock Journal of Astrophysics and Astronomy \textbf{42}(2), 64 (2021).
\newblock \doi{10.1007/s12036-021-09743-1}.
\newblock {\href{https://arxiv.org/abs/2105.09527}{{arXiv:2105.09527}}}
  {[astro-ph.IM]}

\bibitem{2018ApJS..235....4O}
K.~{Oh}, M.~{Koss}, C.B. {Markwardt}, K.~{Schawinski}, W.H. {Baumgartner}, S.D.
  {Barthelmy}, S.B. {Cenko}, N.~{Gehrels}, R.~{Mushotzky}, A.~{Petulante},
  C.~{Ricci}, A.~{Lien}, B.~{Trakhtenbrot}, {The 105-Month Swift-BAT All-sky
  Hard X-Ray Survey}.
\newblock \apjs \textbf{235}(1), 4 (2018).
\newblock \doi{10.3847/1538-4365/aaa7fd}.
\newblock {\href{https://arxiv.org/abs/1801.01882}{{arXiv:1801.01882}}}
  {[astro-ph.HE]}

\bibitem{2009A&A...493..501M}
A.~{Moretti}, C.~{Pagani}, G.~{Cusumano}, S.~{Campana}, M.~{Perri}, A.~{Abbey},
  M.~{Ajello}, A.P. {Beardmore}, D.~{Burrows}, G.~{Chincarini}, O.~{Godet},
  C.~{Guidorzi}, J.E. {Hill}, J.~{Kennea}, J.~{Nousek}, J.P. {Osborne},
  G.~{Tagliaferri}, {A new measurement of the cosmic X-ray background}.
\newblock \aap \textbf{493}(2), 501--509 (2009).
\newblock \doi{10.1051/0004-6361:200811197}

\bibitem{2008ASPC..401..323S}
F.~{Senziani}, G.~{Skinner}, P.~{Jean}, M.~{Hernanz}, in \emph{RS Ophiuchi
  (2006) and the Recurrent Nova Phenomenon}, \emph{Astronomical Society of the
  Pacific Conference Series}, vol. 401, ed. by A.~{Evans}, M.F. {Bode}, T.J.
  {O'Brien}, M.J. {Darnley} (2008), p. 323

\bibitem{2021ApJ...910..134G}
A.C. {Gordon}, E.~{Aydi}, K.L. {Page}, K.L. {Li}, L.~{Chomiuk}, K.V.
  {Sokolovsky}, K.~{Mukai}, J.~{Seitz}, {Surveying the X-Ray Behavior of Novae
  as They Emit {\ensuremath{\gamma}}-Rays}.
\newblock \apj \textbf{910}(2), 134 (2021).
\newblock \doi{10.3847/1538-4357/abe547}.
\newblock {\href{https://arxiv.org/abs/2010.15930}{{arXiv:2010.15930}}}
  {[astro-ph.HE]}

\bibitem{2008A&A...485..223S}
F.~{Senziani}, G.K. {Skinner}, P.~{Jean}, M.~{Hernanz}, {Detectability of
  gamma-ray emission from classical novae with Swift/BAT}.
\newblock \aap \textbf{485}(1), 223--231 (2008).
\newblock \doi{10.1051/0004-6361:200809863}.
\newblock {\href{https://arxiv.org/abs/0804.4791}{{arXiv:0804.4791}}}
  {[astro-ph]}

\bibitem{2006ApJ...652..629B}
M.F. {Bode}, T.J. {O'Brien}, J.P. {Osborne}, K.L. {Page}, F.~{Senziani}, G.K.
  {Skinner}, S.~{Starrfield}, J.U. {Ness}, J.J. {Drake}, G.~{Schwarz}, A.P.
  {Beardmore}, M.J. {Darnley}, S.P.S. {Eyres}, A.~{Evans}, N.~{Gehrels}, M.R.
  {Goad}, P.~{Jean}, J.~{Krautter}, G.~{Novara}, {Swift Observations of the
  2006 Outburst of the Recurrent Nova RS Ophiuchi. I. Early X-Ray Emission from
  the Shocked Ejecta and Red Giant Wind}.
\newblock \apj \textbf{652}(1), 629--635 (2006).
\newblock \doi{10.1086/507980}.
\newblock
  {\href{https://arxiv.org/abs/astro-ph/0604618}{{arXiv:astro-ph/0604618}}}
  {[astro-ph]}

\bibitem{2015ATel.7244....1T}
M.~{Tuerler}, C.~{Ferrigno}, D.~{Eckert}, K.~{Watanabe}, E.~{Kuulkers},
  {INTEGRAL hard X-ray observation of the nova GK Per during its 2015
  outburst}.
\newblock The Astronomer's Telegram \textbf{7244}, 1 (2015)

\bibitem{2019ApJ...872...86N}
T.~{Nelson}, K.~{Mukai}, K.L. {Li}, I.~{Vurm}, B.D. {Metzger}, L.~{Chomiuk},
  J.L. {Sokoloski}, J.D. {Linford}, T.~{Bohlsen}, P.~{Luckas}, {NuSTAR
  Detection of X-Rays Concurrent with Gamma-Rays in the Nova V5855 Sgr}.
\newblock \apj \textbf{872}(1), 86 (2019).
\newblock \doi{10.3847/1538-4357/aafb6d}.
\newblock {\href{https://arxiv.org/abs/1901.00030}{{arXiv:1901.00030}}}
  {[astro-ph.HE]}

\bibitem{2020MNRAS.497.2569S}
K.V. {Sokolovsky}, K.~{Mukai}, L.~{Chomiuk}, R.~{Lopes de Oliveira}, E.~{Aydi},
  K.L. {Li}, E.~{Steinberg}, I.~{Vurm}, B.D. {Metzger}, A.~{Kawash}, J.D.
  {Linford}, A.J. {Mioduszewski}, T.~{Nelson}, J.U. {Ness}, K.L. {Page}, M.P.
  {Rupen}, J.L. {Sokoloski}, J.~{Strader}, {X-ray spectroscopy of the
  {\ensuremath{\gamma}}-ray brightest nova V906 Car (ASASSN-18fv)}.
\newblock \mnras \textbf{497}(3), 2569--2585 (2020).
\newblock \doi{10.1093/mnras/staa2104}.
\newblock {\href{https://arxiv.org/abs/2007.07885}{{arXiv:2007.07885}}}
  {[astro-ph.HE]}

\bibitem{2022MNRAS.514.2239S}
K.V. {Sokolovsky}, K.L. {Li}, R.~{Lopes de Oliveira}, J.U. {Ness}, K.~{Mukai},
  L.~{Chomiuk}, E.~{Aydi}, E.~{Steinberg}, I.~{Vurm}, B.D. {Metzger}, A.N.
  {Babul}, A.~{Kawash}, J.D. {Linford}, T.~{Nelson}, K.L. {Page}, M.P. {Rupen},
  J.L. {Sokoloski}, J.~{Strader}, D.~{Kilkenny}, {The first nova eruption in a
  novalike variable: YZ Ret as seen in X-rays and {\ensuremath{\gamma}}-rays}.
\newblock \mnras \textbf{514}(2), 2239--2258 (2022).
\newblock \doi{10.1093/mnras/stac1440}.
\newblock {\href{https://arxiv.org/abs/2108.03241}{{arXiv:2108.03241}}}
  {[astro-ph.HE]}

\bibitem{2022MNRAS.514.1557P}
K.L. {Page}, A.P. {Beardmore}, J.P. {Osborne}, U.~{Munari}, J.U. {Ness}, P.A.
  {Evans}, M.F. {Bode}, M.J. {Darnley}, J.J. {Drake}, N.P.M. {Kuin}, T.J.
  {O'Brien}, M.~{Orio}, S.N. {Shore}, S.~{Starrfield}, C.E. {Woodward}, {The
  2021 outburst of the recurrent nova RS Ophiuchi observed in X-rays by the
  Neil Gehrels Swift Observatory: a comparative study}.
\newblock \mnras \textbf{514}(2), 1557--1574 (2022).
\newblock \doi{10.1093/mnras/stac1295}.
\newblock {\href{https://arxiv.org/abs/2205.03232}{{arXiv:2205.03232}}}
  {[astro-ph.HE]}

\bibitem{2016AstL...42...69G}
S.A. {Grebenev}, A.V. {Prosvetov}, R.A. {Burenin}, R.A. {Krivonos}, A.V.
  {Mescheryakov}, {X-ray nova MAXI J1828-249. Evolution of the broadband
  spectrum during its 2013-2014 outburst}.
\newblock Astronomy Letters \textbf{42}(2), 69--81 (2016).
\newblock \doi{10.1134/S1063773716020031}.
\newblock {\href{https://arxiv.org/abs/1604.00158}{{arXiv:1604.00158}}}
  {[astro-ph.HE]}

\bibitem{2017AstL...43..167M}
I.A. {Mereminskiy}, E.V. {Filippova}, R.A. {Krivonos}, S.A. {Grebenev}, R.A.
  {Burenin}, R.A. {Sunyaev}, {The outburst of the X-ray nova GRS 1739-278 in
  September 2016}.
\newblock Astronomy Letters \textbf{43}(3), 167--174 (2017).
\newblock \doi{10.1134/S1063773717030057}.
\newblock {\href{https://arxiv.org/abs/1610.08102}{{arXiv:1610.08102}}}
  {[astro-ph.HE]}

\bibitem{2018AstL...44..378M}
I.A. {Mereminskiy}, S.A. {Grebenev}, A.V. {Prosvetov}, A.N. {Semena},
  {Low-Frequency Quasi-Periodic Oscillations in the X-ray Nova MAXI J1535-571
  at the Initial Stage of Its 2017 Outburst}.
\newblock Astronomy Letters \textbf{44}(6), 378--389 (2018).
\newblock \doi{10.1134/S106377371806004X}.
\newblock {\href{https://arxiv.org/abs/1806.06025}{{arXiv:1806.06025}}}
  {[astro-ph.HE]}

\bibitem{tgf_fishman94}
G.J. {Fishman}, P.N. {Bhat}, R.~{Mallozzi}, J.M. {Horack}, T.~{Koshut},
  C.~{Kouveliotou}, G.N. {Pendleton}, C.A. {Meegan}, R.B. {Wilson}, W.S.
  {Paciesas}, S.J. {Goodman}, H.J. {Christian}, {Discovery of Intense Gamma-Ray
  Flashes of Atmospheric Origin}.
\newblock Science \textbf{264}(5163), 1313--1316 (1994).
\newblock \doi{10.1126/science.264.5163.1313}

\bibitem{tgf_tavani11}
M.~{Tavani}, M.~{Marisaldi}, C.~{Labanti}, F.~{Fuschino}, A.~{Argan},
  A.~{Trois}, P.~{Giommi}, S.~{Colafrancesco}, C.~{Pittori}, F.~{Palma},
  M.~{Trifoglio}, F.~{Gianotti}, A.~{Bulgarelli}, V.~{Vittorini},
  F.~{Verrecchia}, L.~{Salotti}, G.~{Barbiellini}, P.~{Caraveo}, P.W.
  {Cattaneo}, A.~{Chen}, T.~{Contessi}, E.~{Costa}, F.~{D'Ammando}, E.~{Del
  Monte}, G.~{de Paris}, G.~{Di Cocco}, G.~{di Persio}, I.~{Donnarumma},
  Y.~{Evangelista}, M.~{Feroci}, A.~{Ferrari}, M.~{Galli}, A.~{Giuliani},
  M.~{Giusti}, I.~{Lapshov}, F.~{Lazzarotto}, P.~{Lipari}, F.~{Longo},
  S.~{Mereghetti}, E.~{Morelli}, E.~{Moretti}, A.~{Morselli}, L.~{Pacciani},
  A.~{Pellizzoni}, F.~{Perotti}, G.~{Piano}, P.~{Picozza}, M.~{Pilia},
  G.~{Pucella}, M.~{Prest}, M.~{Rapisarda}, A.~{Rappoldi}, E.~{Rossi},
  A.~{Rubini}, S.~{Sabatini}, E.~{Scalise}, P.~{Soffitta}, E.~{Striani},
  E.~{Vallazza}, S.~{Vercellone}, A.~{Zambra}, D.~{Zanello}, {Terrestrial
  Gamma-Ray Flashes as Powerful Particle Accelerators}.
\newblock \prl \textbf{106}(1), 018501 (2011).
\newblock \doi{10.1103/PhysRevLett.106.018501}

\bibitem{tgf_carlson07}
B.E. {Carlson}, N.G. {Lehtinen}, U.S. {Inan}, {Constraints on terrestrial gamma
  ray flash production from satellite observation}.
\newblock \grl \textbf{34}(8), L08809 (2007).
\newblock \doi{10.1029/2006GL029229}

\bibitem{tgf_dwyer12}
J.R. {Dwyer}, D.M. {Smith}, S.A. {Cummer}, {High-Energy Atmospheric Physics:
  Terrestrial Gamma-Ray Flashes and Related Phenomena}.
\newblock \ssr \textbf{173}(1-4), 133--196 (2012).
\newblock \doi{10.1007/s11214-012-9894-0}

\bibitem{2018EP&S...70..101M}
Y.~{Miyoshi}, I.~{Shinohara}, T.~{Takashima}, K.~{Asamura}, N.~{Higashio},
  T.~{Mitani}, S.~{Kasahara}, S.~{Yokota}, Y.~{Kazama}, S.Y. {Wang}, S.W.Y.
  {Tam}, P.T.P. {Ho}, Y.~{Kasahara}, Y.~{Kasaba}, S.~{Yagitani}, A.~{Matsuoka},
  H.~{Kojima}, Y.~{Katoh}, K.~{Shiokawa}, K.~{Seki}, {Geospace exploration
  project ERG}.
\newblock Earth, Planets and Space \textbf{70}(1), 101 (2018).
\newblock \doi{10.1186/s40623-018-0862-0}

\bibitem{2019JGRD..12414024O}
N.~{{\O}stgaard}, T.~{Neubert}, V.~{Reglero}, K.~{Ullaland}, S.~{Yang},
  G.~{Genov}, M.~{Marisaldi}, A.~{Mezentsev}, P.~{Kochkin}, N.~{Lehtinen},
  D.~{Sarria}, B.H. {Qureshi}, A.~{Solberg}, C.~{Maiorana}, K.~{Albrechtsen},
  C.~{Budtz-J{\o}rgensen}, I.~{Kuvvetli}, F.~{Christiansen}, O.~{Chanrion},
  M.~{Heumesser}, J.~{Navarro-Gonzalez}, P.~{Connell}, C.~{Eyles},
  H.~{Christian}, S.~{Al-nussirat}, {First 10 Months of TGF Observations by
  ASIM}.
\newblock Journal of Geophysical Research (Atmospheres) \textbf{124}(24),
  14,024--14,036 (2019).
\newblock \doi{10.1029/2019JD031214}

\bibitem{2018JGRA..123.4381R}
O.J. {Roberts}, G.~{Fitzpatrick}, M.~{Stanbro}, S.~{McBreen}, M.S. {Briggs},
  R.H. {Holzworth}, J.E. {Grove}, A.~{Chekhtman}, E.S. {Cramer}, B.G.
  {Mailyan}, {The First Fermi-GBM Terrestrial Gamma Ray Flash Catalog}.
\newblock Journal of Geophysical Research (Space Physics) \textbf{123}(5),
  4381--4401 (2018).
\newblock \doi{10.1029/2017JA024837}

\bibitem{tgf_bagheri19}
M.~{Bagheri}, J.R. {Dwyer}, M.L. {McConnell}, {On the Linear Polarization of
  TGFs and X-Rays From Natural and Rocket-Triggered Lightning and Its
  Association With Source Geometry}.
\newblock Journal of Geophysical Research (Space Physics) \textbf{124}(11),
  9166--9183 (2019).
\newblock \doi{10.1029/2019JA026570}

\bibitem{hariharan19}
B.~{Hariharan}, S.R. {Dugad}, S.K. {Gupta}, Y.~{Hayashi}, S.S.R. {Inbanathan},
  P.~{Jagadeesan}, A.~{Jain}, S.~{Kawakami}, P.K. {Mohanty}, B.S. {Rao},
  {Modeling of rigidity dependent CORSIKA simulations for GRAPES-3}.
\newblock Experimental Astronomy \textbf{48}(2-3), 111--120 (2019).
\newblock \doi{10.1007/s10686-019-09640-0}.
\newblock {\href{https://arxiv.org/abs/1908.05948}{{arXiv:1908.05948}}}
  {[astro-ph.IM]}

\bibitem{vichare18}
G.~{Vichare}, A.~{Bhaskar}, G.~{Datar}, A.~{Raghav}, K.U. {Nair},
  C.~{Selvaraj}, M.~{Ananthi}, A.K. {Sinha}, M.~{Paranjape}, T.~{Gawade}, C.P.
  {Anil Kumar}, C.~{Panneerselvam}, S.~{Sathishkumar}, S.~{Gurubaran},
  {Equatorial secondary cosmic ray observatory to study space weather and
  terrestrial events}.
\newblock Advances in Space Research \textbf{61}(10), 2555--2568 (2018).
\newblock \doi{10.1016/j.asr.2018.03.006}

\bibitem{uma09}
K.N. {Uma}, T.N. {Rao}, {Characteristics of Vertical Velocity Cores in
  Different Convective Systems Observed over Gadanki, India}.
\newblock Monthly Weather Review \textbf{137}(3), 954 (2009).
\newblock \doi{10.1175/2008MWR2677.1}

\bibitem{subrahmanyam22}
K.V. Subrahmanyam, K.K. Kumar, C-band polarimetric doppler weather radar
  observations during an extreme precipitation event and associated dynamics
  over peninsular india.
\newblock Natural Hazards \textbf{114}(2), 1307--1322 (2022).
\newblock \doi{10.1007/s11069-022-05426-4}.
\newblock \urlprefix\url{https://doi.org/10.1007/s11069-022-05426-4}

\bibitem{2017LRSP...14....2B}
A.O. {Benz}, {Flare Observations}.
\newblock Living Reviews in Solar Physics \textbf{14}(1), 2 (2017).
\newblock \doi{10.1007/s41116-016-0004-3}

\bibitem{2008A&ARv..16..155K}
S.~{Krucker}, M.~{Battaglia}, P.J. {Cargill}, L.~{Fletcher}, H.S. {Hudson},
  A.L. {MacKinnon}, S.~{Masuda}, L.~{Sui}, M.~{Tomczak}, A.L. {Veronig},
  L.~{Vlahos}, S.M. {White}, {Hard X-ray emission from the solar corona}.
\newblock \aapr \textbf{16}, 155--208 (2008).
\newblock \doi{10.1007/s00159-008-0014-9}

\bibitem{2002SoPh..210....3L}
R.P. {Lin}, B.R. {Dennis}, G.J. {Hurford}, D.M. {Smith}, A.~{Zehnder}, P.R.
  {Harvey}, D.W. {Curtis}, D.~{Pankow}, P.~{Turin}, M.~{Bester},
  A.~{Csillaghy}, M.~{Lewis}, N.~{Madden}, H.F. {van Beek}, M.~{Appleby},
  T.~{Raudorf}, J.~{McTiernan}, R.~{Ramaty}, E.~{Schmahl}, R.~{Schwartz},
  S.~{Krucker}, R.~{Abiad}, T.~{Quinn}, P.~{Berg}, M.~{Hashii}, R.~{Sterling},
  R.~{Jackson}, R.~{Pratt}, R.D. {Campbell}, D.~{Malone}, D.~{Landis}, C.P.
  {Barrington-Leigh}, S.~{Slassi-Sennou}, C.~{Cork}, D.~{Clark}, D.~{Amato},
  L.~{Orwig}, R.~{Boyle}, I.S. {Banks}, K.~{Shirey}, A.K. {Tolbert},
  D.~{Zarro}, F.~{Snow}, K.~{Thomsen}, R.~{Henneck}, A.~{McHedlishvili},
  P.~{Ming}, M.~{Fivian}, J.~{Jordan}, R.~{Wanner}, J.~{Crubb}, J.~{Preble},
  M.~{Matranga}, A.~{Benz}, H.~{Hudson}, R.C. {Canfield}, G.D. {Holman},
  C.~{Crannell}, T.~{Kosugi}, A.G. {Emslie}, N.~{Vilmer}, J.C. {Brown},
  C.~{Johns-Krull}, M.~{Aschwanden}, T.~{Metcalf}, A.~{Conway}, {The Reuven
  Ramaty High-Energy Solar Spectroscopic Imager (RHESSI)}.
\newblock \solphys \textbf{210}(1), 3--32 (2002).
\newblock \doi{10.1023/A:1022428818870}

\bibitem{2008ApJ...677..704H}
I.G. {Hannah}, S.~{Christe}, S.~{Krucker}, G.J. {Hurford}, H.S. {Hudson}, R.P.
  {Lin}, {RHESSI Microflare Statistics. II. X-Ray Imaging, Spectroscopy, and
  Energy Distributions}.
\newblock \apj \textbf{677}(1), 704--718 (2008).
\newblock \doi{10.1086/529012}.
\newblock {\href{https://arxiv.org/abs/0712.2544}{{arXiv:0712.2544}}}
  {[astro-ph]}

\bibitem{2020A&A...642A..15K}
S.~{Krucker}, G.J. {Hurford}, O.~{Grimm}, S.~{K{\"o}gl}, H.P.
  {Gr{\"o}belbauer}, L.~{Etesi}, D.~{Casadei}, A.~{Csillaghy}, A.O. {Benz},
  N.G. {Arnold}, F.~{Molendini}, P.~{Orleanski}, D.~{Schori}, H.~{Xiao}, M.e.a.
  {Kuhar}, {The Spectrometer/Telescope for Imaging X-rays (STIX)}.
\newblock \aap \textbf{642}, A15 (2020).
\newblock \doi{10.1051/0004-6361/201937362}

\bibitem{2017CSci..113..610S}
S.~{Seetha}, S.~{Megala}, {Aditya-L1 mission}.
\newblock Current Science \textbf{113}(4), 610 (2017).
\newblock \doi{10.18520/cs/v113/i04/610-612}

\bibitem{2020ApJ...891..126K}
S.~{Katsuda}, M.~{Ohno}, K.~{Mori}, T.~{Beppu}, Y.~{Kanemaru}, M.S. {Tashiro},
  Y.~{Terada}, K.~{Sato}, K.~{Morita}, H.~{Sagara}, F.~{Ogawa}, H.~{Takahashi},
  H.~{Murakami}, M.~{Nobukawa}, H.~{Tsunemi}, K.~{Hayashida}, H.~{Matsumoto},
  H.~{Noda}, H.~{Nakajima}, Y.~{Ezoe}, Y.~{Tsuboi}, Y.~{Maeda}, T.~{Yokoyama},
  N.~{Narukage}, {Inverse First Ionization Potential Effects in Giant Solar
  Flares Found from Earth X-Ray Albedo with Suzaku/XIS}.
\newblock \apj \textbf{891}(2), 126 (2020).
\newblock \doi{10.3847/1538-4357/ab7207}.
\newblock {\href{https://arxiv.org/abs/2001.10643}{{arXiv:2001.10643}}}
  {[astro-ph.SR]}

\bibitem{numpy}
S.~van~der Walt, S.C. Colbert, G.~Varoquaux, {The NumPy Array: A Structure for
  Efficient Numerical Computation}.
\newblock Computing in Science {\&} Engineering \textbf{13}(2), 22--30 (2011).
\newblock \doi{10.1109/MCSE.2011.37}.
\newblock
  \urlprefix\url{http://ieeexplore.ieee.org/lpdocs/epic03/wrapper.htm?arnumber=5725236}

\bibitem{matplotlib}
J.D. Hunter, {Matplotlib: A 2D Graphics Environment}.
\newblock Computing in Science {\&} Engineering \textbf{9}(3), 90--95 (2007).
\newblock \doi{10.1109/MCSE.2007.55}.
\newblock
  \urlprefix\url{http://scitation.aip.org/content/aip/journal/cise/9/3/10.1109/MCSE.2007.55}

\bibitem{2013A&A...558A..33A}
{Astropy Collaboration}, T.P. {Robitaille}, E.J. {Tollerud}, P.~{Greenfield},
  M.~{Droettboom}, E.~{Bray}, T.~{Aldcroft}, M.~{Davis}, A.~{Ginsburg}, A.M.
  {Price-Whelan}, W.E. {Kerzendorf}, A.~{Conley}, N.~{Crighton}, K.~{Barbary},
  D.~{Muna}, H.~{Ferguson}, F.~{Grollier}, M.M. {Parikh}, P.H. {Nair}, H.M.
  {Unther}, C.~{Deil}, J.~{Woillez}, S.~{Conseil}, R.~{Kramer}, J.E.H.
  {Turner}, L.~{Singer}, R.~{Fox}, B.A. {Weaver}, V.~{Zabalza}, Z.I. {Edwards},
  K.~{Azalee Bostroem}, D.J. {Burke}, A.R. {Casey}, S.M. {Crawford},
  N.~{Dencheva}, J.~{Ely}, T.~{Jenness}, K.~{Labrie}, P.L. {Lim},
  F.~{Pierfederici}, A.~{Pontzen}, A.~{Ptak}, B.~{Refsdal}, M.~{Servillat},
  O.~{Streicher}, {Astropy: A community Python package for astronomy}.
\newblock \aap \textbf{558}, A33 (2013).
\newblock \doi{10.1051/0004-6361/201322068}.
\newblock {\href{https://arxiv.org/abs/1307.6212}{{arXiv:1307.6212}}}
  {[astro-ph.IM]}

\bibitem{2018AJ....156..123A}
{Astropy Collaboration}, A.M. {Price-Whelan}, B.M. {Sip{\H{o}}cz}, H.M.
  {G{\"u}nther}, P.L. {Lim}, S.M. {Crawford}, S.~{Conseil}, D.L. {Shupe}, M.W.
  {Craig}, N.~{Dencheva}, A.~{Ginsburg}, J.T. {VanderPlas}, L.D. {Bradley},
  D.~{P{\'e}rez-Su{\'a}rez}, M.~{de Val-Borro}, T.L. {Aldcroft}, K.L. {Cruz},
  T.P. {Robitaille}, E.J. {Tollerud}, C.~{Ardelean}, T.~{Babej}, Y.P. {Bach},
  M.~{Bachetti}, A.V. {Bakanov}, S.P. {Bamford}, G.~{Barentsen}, P.~{Barmby},
  A.~{Baumbach}, K.L. {Berry}, F.~{Biscani}, M.~{Boquien}, K.A. {Bostroem},
  L.G. {Bouma}, G.B. {Brammer}, E.M. {Bray}, H.~{Breytenbach},
  H.~{Buddelmeijer}, D.J. {Burke}, G.~{Calderone}, J.L. {Cano Rodr{\'\i}guez},
  M.~{Cara}, J.V.M. {Cardoso}, S.~{Cheedella}, Y.~{Copin}, L.~{Corrales},
  D.~{Crichton}, D.~{D'Avella}, C.~{Deil}, {\'E}.~{Depagne}, J.P. {Dietrich},
  A.~{Donath}, M.~{Droettboom}, N.~{Earl}, T.~{Erben}, S.~{Fabbro}, L.A.
  {Ferreira}, T.~{Finethy}, R.T. {Fox}, L.H. {Garrison}, S.L.J. {Gibbons}, D.A.
  {Goldstein}, R.~{Gommers}, J.P. {Greco}, P.~{Greenfield}, A.M. {Groener},
  F.~{Grollier}, A.~{Hagen}, P.~{Hirst}, D.~{Homeier}, A.J. {Horton},
  G.~{Hosseinzadeh}, L.~{Hu}, J.S. {Hunkeler}, {\v{Z}}.~{Ivezi{\'c}},
  A.~{Jain}, T.~{Jenness}, G.~{Kanarek}, S.~{Kendrew}, N.S. {Kern}, W.E.
  {Kerzendorf}, A.~{Khvalko}, J.~{King}, D.~{Kirkby}, A.M. {Kulkarni},
  A.~{Kumar}, A.~{Lee}, D.~{Lenz}, S.P. {Littlefair}, Z.~{Ma}, D.M. {Macleod},
  M.~{Mastropietro}, C.~{McCully}, S.~{Montagnac}, B.M. {Morris}, M.~{Mueller},
  S.J. {Mumford}, D.~{Muna}, N.A. {Murphy}, S.~{Nelson}, G.H. {Nguyen}, J.P.
  {Ninan}, M.~{N{\"o}the}, S.~{Ogaz}, S.~{Oh}, J.K. {Parejko}, N.~{Parley},
  S.~{Pascual}, R.~{Patil}, A.A. {Patil}, A.L. {Plunkett}, J.X. {Prochaska},
  T.~{Rastogi}, V.~{Reddy Janga}, J.~{Sabater}, P.~{Sakurikar}, M.~{Seifert},
  L.E. {Sherbert}, H.~{Sherwood-Taylor}, A.Y. {Shih}, J.~{Sick}, M.T.
  {Silbiger}, S.~{Singanamalla}, L.P. {Singer}, P.H. {Sladen}, K.A. {Sooley},
  S.~{Sornarajah}, O.~{Streicher}, P.~{Teuben}, S.W. {Thomas}, G.R. {Tremblay},
  J.E.H. {Turner}, V.~{Terr{\'o}n}, M.H. {van Kerkwijk}, A.~{de la Vega}, L.L.
  {Watkins}, B.A. {Weaver}, J.B. {Whitmore}, J.~{Woillez}, V.~{Zabalza},
  {Astropy Contributors}, {The Astropy Project: Building an Open-science
  Project and Status of the v2.0 Core Package}.
\newblock \aj \textbf{156}(3), 123 (2018).
\newblock \doi{10.3847/1538-3881/aabc4f}.
\newblock {\href{https://arxiv.org/abs/1801.02634}{{arXiv:1801.02634}}}
  {[astro-ph.IM]}

\bibitem{ghb+05}
K.M. Gorski, E.~Hivon, A.J. Banday, B.D. Wandelt, F.K. Hansen, M.~Reinecke,
  M.~Bartelmann, {HEALPix: A Framework for High‐Resolution Discretization and
  Fast Analysis of Data Distributed on the Sphere}.
\newblock The Astrophysical Journal \textbf{622}(2), 759--771 (2005).
\newblock \doi{10.1086/427976}.
\newblock \urlprefix\url{http://adsabs.harvard.edu/abs/2005ApJ...622..759G}

\bibitem{Agostinelli2003}
S.~Agostinelli, J.~Allison, K.~Amako, J.~Apostolakis, H.~Araujo, P.~Arce,
  M.~Asai, D.~Axen, S.~Banerjee, G.~Barrand, F.~Behner, L.~Bellagamba,
  J.~Boudreau, L.~Broglia, A.~Brunengo, H.~Burkhardt, S.~Chauvie, J.~Chuma,
  R.~Chytracek, G.~Cooperman, G.~Cosmo, P.~Degtyarenko, A.~Dell'Acqua,
  G.~Depaola, D.~Dietrich, R.~Enami, A.~Feliciello, C.~Ferguson, H.~Fesefeldt,
  G.~Folger, F.~Foppiano, A.~Forti, S.~Garelli, S.~Giani, R.~Giannitrapani,
  D.~Gibin, J.J. {Gomez Cadenas}, I.~Gonzalez, G.~{Gracia Abril}, G.~Greeniaus,
  W.~Greiner, V.~Grichine, A.~Grossheim, S.~Guatelli, P.~Gumplinger,
  R.~Hamatsu, K.~Hashimoto, H.~Hasui, A.~Heikkinen, A.~Howard, V.~Ivanchenko,
  A.~Johnson, F.W. Jones, J.~Kallenbach, N.~Kanaya, M.~Kawabata, Y.~Kawabata,
  M.~Kawaguti, S.~Kelner, P.~Kent, A.~Kimura, T.~Kodama, R.~Kokoulin,
  M.~Kossov, H.~Kurashige, E.~Lamanna, T.~Lampen, V.~Lara, V.~Lefebure, F.~Lei,
  M.~Liendl, W.~Lockman, F.~Longo, S.~Magni, M.~Maire, E.~Medernach,
  K.~Minamimoto, P.~{Mora de Freitas}, Y.~Morita, K.~Murakami, M.~Nagamatu,
  R.~Nartallo, P.~Nieminen, T.~Nishimura, K.~Ohtsubo, M.~Okamura, S.~O'Neale,
  Y.~Oohata, K.~Paech, J.~Perl, A.~Pfeiffer, M.G. Pia, F.~Ranjard, A.~Rybin,
  S.~Sadilov, E.~di~Salvo, G.~Santin, T.~Sasaki, N.~Savvas, Y.~Sawada,
  S.~Scherer, S.~Sei, V.~Sirotenko, D.~Smith, N.~Starkov, H.~Stoecker,
  J.~Sulkimo, M.~Takahata, S.~Tanaka, E.~Tcherniaev, E.~{Safai Tehrani},
  M.~Tropeano, P.~Truscott, H.~Uno, L.~Urban, P.~Urban, M.~Verderi, A.~Walkden,
  W.~Wander, H.~Weber, J.P. Wellisch, T.~Wenaus, D.C. Williams, D.~Wright,
  T.~Yamada, H.~Yoshida, D.~Zschiesche, {GEANT4 - A simulation toolkit}.
\newblock Nucl. Instruments Methods Phys. Res. Sect. A Accel. Spectrometers,
  Detect. Assoc. Equip. \textbf{506}(3), 250--303 (2003).
\newblock \doi{10.1016/S0168-9002(03)01368-8}

\end{thebibliography}


\end{document}